\documentclass[11pt,a4paper]{article}
\usepackage[makeroom]{cancel}
\usepackage{latexsym}
\usepackage{longtable}
\usepackage{epsfig}
\usepackage{graphicx,bbm,psfrag}
\usepackage{amssymb}
\usepackage{amsmath,amsthm,booktabs,mathtools}
\usepackage{bm}
\usepackage{color}
\usepackage{dutchcal}
\usepackage{hyperref}
\hypersetup{
    colorlinks,
    citecolor=red,
    linkcolor=blue,
}

\newtheorem{statement}{Statement}[section]
\newcommand{\st}[1]{
\bc
\fbox{\begin{minipage}[t]{0.9\textwidth}
\begin{statement}
#1
\end{statement}
\end{minipage}}
\ec
}
\numberwithin{equation}{section}

\newcommand{\bc}{\begin{center}}
\newcommand{\ec}{\end{center}}
\def\ba#1{\begin{array}{#1}\displaystyle}
\newcommand{\ea}{\end{array}}
\newcommand{\z}{\\[0.3cm] \displaystyle}
\newcommand{\beq}{\begin{equation}}
\newcommand{\eeq}{\end{equation}}
\newcommand{\beqa}{\begin{eqnarray}}
\newcommand{\eeqa}{\end{eqnarray}}
\newcommand{\no}{\nonumber}
\newcommand{\n}{\nonumber\\}
\newcommand{\bi}{\begin{itemize}}
\newcommand{\ei}{\end{itemize}}
\def\mato#1{\left(\ba{#1}} 
\def\matf{\ea\right)}

\def\lt#1{\left#1}
\def\rt#1{\right#1}
\def\t#1{\tilde{#1}}
\def\h#1{\hat{#1}}
\def\b#1{\bar{#1}}
\def\frc#1#2{\frac{#1}{#2}}

\newcommand{\prin}{\underline{\mathrm{P}}}
\newcommand{\p}{\partial}
\newcommand{\id}{{\rm id}}
\newcommand{\Pexp}{{\mathsf T}\exp}
\newcommand{\vac}{{\rm vac}}
\newcommand{\bra}{\langle}
\newcommand{\ket}{\rangle}

\newcommand{\Z}{{\mathbb{Z}}}

\newcommand{\R}{{\mathbb{R}}}
\newcommand{\C}{{\mathbb{C}}}
\newcommand{\Tr}{{\rm Tr}}

\newcommand{\rz}{{\rm z}}

\newcommand{\ep}{\epsilon}

\newcommand{\ri}{{\rm i}}

\newcommand{\dd}{{\rm d}}
\newcommand{\sd}{{\hat{\rm d}}}
\newcommand{\1}{{\bf 1}}
\DeclareMathOperator{\sgn}{sgn}
\DeclareMathOperator{\ad}{ad}

\DeclareMathOperator{\End}{{\rm End}}
\DeclareMathOperator{\Aut}{{\rm Aut}}
\DeclareMathOperator{\spn}{{\rm span}}

\begin{document}

\begin{center}
{\Large {\sc Twist fields in many-body physics}}

\vspace{1cm}

{\large Benjamin Doyon
}

\vspace{0.2cm}
Department of Mathematics, King’s College London, Strand, London WC2R 2LS, U.K.
\ec

\medskip

The notion of twist fields has played a fundamental role in many-body physics. It is used to construct the so-called disorder parameter for the study of phase transitions in the classical Ising model of statistical mechanics, it is involved in the Jordan-Wigner transformation in quantum chains and bosonisation in quantum field theory, and it is related to measures of entanglement in many-body quantum systems. I provide a pedagogical introduction to the notion of twist field and the concepts at its roots, and review some of its applications, focussing on 1+1 dimension. This includes: locality and extensivity, internal symmetries, semi-locality, the standard exponential form and height fields, path integral defects and Riemann surfaces, topological invariance, and twist families. Additional topics touched upon include renormalisation and form factors in relativistic quantum field theory, tau functions of integrable PDEs, thermodynamic and hydrodynamic principles, and branch-point twist fields for entanglement entropy. One-dimensional quantum systems such as chains (e.g.~quantum Heisenberg model) and field theory (e.g.~quantum sine-Gordon model) are the main focus, but I also explain how the notion applies to equilibrium statistical mechanics (e.g.~classical Ising lattice model), and how some aspects can be adapted to one-dimensional classical dynamical systems (e.g.~classical Toda chain).

\tableofcontents

\section{Introduction}

Locality is a fundamental idea of many-body physics. In its most basic description, it says that cause-effect correlations between things (probes, observers) far apart from each other are very weak. But, locality has to be expressed more accurately. Very often, in quantum many-body physics, studies of locality concentrate on Lieb-Robinson bounds \cite{brattelioperator1,brattelioperator2,hastings2010localityquantumsystems,ranard2021aspects,chen2023speed}.
In models of relativistic particles and QFT, algebraic formulations are based on ``local nets'' \cite{haag1992principle,thomas1998standard}. In both cases, the basic statement is that local observables commute at space-like distances, outside some ``light cone'' -- that is, they are independent when not causally connected.

But, of course, any locality concept is only as good as the physical applications it has. The main object of these notes, {\em twist fields}, are certain type of observables that ``twist'' our notion of locality. Yet, they have far-reaching physical applications.

As far as I am aware, the first appearance of twist fields was in \cite{kadanoff1971determination}, in the context of phase transitions in classical statistical mechanics \cite{cardy1996scaling}. The idea was to characterise the high-temperature, disordered phase of the Ising lattice model in a way similar to the characterisation of its low-temperature, ordered phase. In the latter, the spin-reversal symmetry is spontaneously broken and the magnetisation acquires a non-zero value at zero magnetic field. The magnetisation is an ``order parameter'', whose study near second-order phase transition is at the basis of much development of conformal field theory (CFT) \cite{francesco2012conformal}. In order to characterise the disordered phase, there was a need for a ``disordered parameter''. This was found in \cite{kadanoff1971determination}, but, instead of being a local observable like the local spin, it was a non-local observable, involving modifications of the Ising interactions on a contiguous ``string'' of edges. It arises from the high-to-low-temperature Kramers-Wannier duality transformation of the local spin. Despite being non-local, the observable has certain topological invariance properties -- the string can be taken with any shape, as long as it extends to infinity (in infinite volume). It is my understanding that its study \cite{luther1975calculation,zuber1977quantum,schroer1978order,karowski1978exact} within quantum field theory (QFT) \cite{itzykson2006quantum,peskin2018introduction,zinn2021quantum}, whose Euclidean formulation describes near-critical temperatures, led to the first development of the concept of twist fields. The mapping between classical statistical mechanics in two dimensions, and quantum dynamics in 1+1 dimensions (one dimension of space, one of time), then gives twist fields within QFT on Minkowski spacetime, for relativistic particles and quantum critical phenomena \cite{itzykson2006quantum,sachdevbook}. Perhaps the most prominent two strands of this development were the idea of semi-locality in integrable QFT, well explained in F.A.~Smirnov's \cite{smirnov1992form} and G.~Mussardo's \cite{mussardo2010statistical} books, and the construction of twisted modules in conformal field theory (CFT), which, I believe, arose independently from the physics works on twist fields, and which are, for instance, pedagogically explained in Lepowsky and Li's book \cite{lepowsky2012introduction}.

But the notion of ``twist field'' is much more general, and goes beyond the study of phase transitions. One of its most potent recent applications is in the study of quantum entanglement. For this purpose, the branch-point twist field was introduced in integrable and non-integrable QFT \cite{cardy2008form,doyon2009bipartite}, inspired by previous geometric constructions of the entanglement entropy in QFT and CFT \cite{holzhey1994geometric,calabrese2004entanglement}, and was later defined away from field theory, in quantum spin chains \cite{castro2011permutation}. It was used for many other measures of entanglement, see the reviews \cite{Pasquale_Calabrese_2009,laflorencie2016quantum,castro2025symmetry}.

However, it appears as though there is no all-encompassing, pedagogical overview of twist fields in many-body physics yet. Here within ``many-body physics'' I include interacting chains, gases and field theories. In these notes, I attempt to fill this gap. For this purpose, I point out that locality and extensivity are crucial concepts. I describe these in a general way, loosely based on a description of locality in QFT from my notes \cite{doyon2009notes}, which I refer to as ``locality from the dynamics'', and on my more recent work \cite{doyon2017thermalization,doyon2022hydrodynamic}, ``locality from the state''. I then use these concepts in order to define a notion of semi-locality and twist fields. This, I believe, slighly generalises what was done in the past. I discuss a number of applications, including the Jordan-Wigner transformation, and branch-point twist fields for entanglement entropy. It turns out that some aspects of twist fields make sense beyond quantum dynamical systems and classical statistical mechanics: they apply also to classical dynamical systems, something which was not emphasised before. I briefly explain how this works. Thus, one can find twist fields with interesting applications in: the Ising, Heisenberg and other quantum spin chains; the sine-Gordon model of relativistic QFT, the Lieb-Liniger quantum gas, and other interacting QFT models and gases; the Ising lattice model and other models of classical statistical mechanics; and the classical Toda chain and other classical dynamical chains, gases of particles and field theories.

For terminological consistency, I still use ``twist fields'': although many constructions and applications are not in field theory, the original inspiration for twist fields and concepts of locality and semi-locality is from QFT.

I keep the level of the discussion non-rigorous. The language is somehwat rigorous-looking, with limits and Hilbert spaces, and the mathematical symbols I use are indeed exactly what I mean to use. However, what makes the discussion non-rigorous is the lack of {\em analysis}. In what topology is the limit? What function space am I taking? I give a few general directions only. Restricting myself to the structural properties allows me to express the fundamental concepts, while not getting lost into technicalities. But it is to be kept in mind that such technicalities are important, and may sometimes be crucial for aspects of the many-body physics itself.

The literature on subjects and results connected to twist fields is immense. It is impossible to cite all of it, let alone to fully overview it. My choice is guided by my experience of working on the topic, and is  subjective and non-exhaustive.

These notes are organised as follows. In Section \ref{sectmodels} I specify the set of models and physical setups I have in mind, some of which I will use as examples. In Section \ref{sectloc} I introduce the theoretical basis for the understanding of twist fields: notions of locality and extensivity in many-body quantum and classical physics and field theory, and of symmetry. Section \ref{secttwist} is the main part of these notes, with a pedagogical introduction to twist fields, their constructions and properties. In Section \ref{sectappli}, I overview some of their applications in various contexts, giving simple examples. Finally, I conclude in Section \ref{sectconclu}. I review the path integral formulation in Appendix \ref{apppath}, which play an important role in the formulation of twist fields, and provide a special twist field construction in Appendix \ref{appgen} and other supporting calculaions in Appendix \ref{appunwinding}.

\section{Models and examples of twist fields}\label{sectmodels}

A local observable only probes or influences a small region of space. However, this cannot probe, or produce, more topological effects, such as ``dislocation-like'' phenomena. Dislocations happen when a symmetry is applied along an extensive region, and disconnects some fundamental degrees of freedom there. The na\"ive picture conjured by the term ``dislocation'' is associated to translation symmetry, but other symmetries can be used, such as the internal symmetry of phase multiplication for describing a vortex in quantum mechanics. Because it is associated to a symmetry, the exact shape of the dislocation is not that important -- it is the point where it starts that matters. We thus need observables that have local support regions, from which a long tail emanates, looking for dislocations. These are twist fields. The concept applies both in space (statistical physics) and spacetime (quantum physics) -- the latter being the main focus here.

In order illustrate this, I first exhibit families of models in various contexts of many-body physics, and give in each case a symmetry group, and examples of associated twist fields. It turns out, perhaps surprisingly, that all these twist fields come out of a single, common framework, which is the one I will explain in the following sections. These are models of relativistic and non-relativistic quantum field theory, quantum and classical chains, and classical statistical mechanics.

Throughout these notes, I concentrate, in the quantum case, on Bosonic observables, however the extension to Fermionic observables is immediate (touched upon in Subsec.~\ref{ssectjordan} and \ref{ssectfree}).

\medskip

{\bf 1.} A relativistic QFT has Hilbert space $\mathcal H$ spanned by asymptotic states, parametrised by rapidities of asymptotic particles $\theta_i\in\R$ and particle types $a_i$ (taking values in some set that depends on the model),
\beq\label{qftstates}
	|\theta_1,\ldots,\theta_n\ket_{a_1,\ldots,a_n}.
\eeq
By convention one takes $\theta_1>\cdots>\theta_n$ for in-states, and the opposite order for out-states. We may take the example of a real bosonic field $\phi(x)$ and its canonical conjugate $\pi(x)$,
\beq
	[\phi(x),\pi(x')] = \ri \,\delta(x-x'),\quad
	[\phi(x),\phi(x')]= [\pi(x),\pi(x')] = 0,
\eeq
and Hamiltonian
\beq\label{HQFT}
	H = \int \dd x\,h(x),\quad h(x) = \frc12 \Big((\p_x \phi(x))^2 + \pi(x)^2\Big) + V(\phi(x)^2).
\eeq
The spectrum of asymptotic particle types, and the action of $\phi(x),\,\pi(x)$ on asymptotic states \eqref{qftstates}, are non-trivial, and depend on the interaction. The case $V(\phi^2) = g\cos(\beta\phi)$ is the well-known sine-Gordon model, where the spectrum is constructed explicitly \cite{smirnov1992form,mussardo2010statistical,korepin1997quantum}. The model \eqref{HQFT} has $\Z_2$ symmetry group, with symmetry $\sigma$ given by
\beq
	\sigma(\phi(x))=-\phi(x), \quad  \sigma(\pi(x))=-\pi(x).
\eeq
The twist field associated to the $\Z_2$ symmetry is\footnote{I use $\pi$ for $3.14159265...$, and $\pi(x)$ for the field.}:
\beq\label{Tbosonintro}
	\mathcal T(x)
	=\exp\Big[\pi \int_{x}^\infty \dd x'\,\phi(x')\pi(x')\Big].
\eeq
See Subsec.~\ref{ssectultra}. The sine-Gordon model also has a non-compact, discrete $\Z$ symmetry group $\phi\to\phi+2\pi n/\beta,\,n\in\Z$. The twist field associated to this is
\beq\label{TZintro}
	\mathcal T_n(x) = \exp\Big[\frc{2\pi \ri n}\beta \int_x^\infty \dd x'\,\pi(x')\Big].
\eeq
See also Subsec.~\ref{ssectultra}. The dual to this symmetry is a compact, continuous $U(1)$ symmetry, formally generated by $\phi(\infty)-\phi(-\infty)$ and made explicit in the massive Thirring model, dual to the sine-Gordon model \cite{korepin1997quantum}. The family of twist fields associated to this dual $U(1)$ symmetry group is
\beq\label{Tvertexintro}
	\mathcal T_\lambda(x) = e^{-\ri \lambda \phi(x)}.
\eeq
See Subsec.~\ref{ssectexp} and \ref{ssectjordan}.

\medskip

{\bf 2.} We may consider instead a quantun spin chain of spin $s$, with Hilbert space $\mathcal H = \bigotimes_{x\in\Z} \C^{2s+1}$. In the case $s=1/2$, operators are formed out of the identity $\1_x$ and Pauli matrices $\sigma_x^i$, $i=1,2,3$ acting on sites $x\in\Z$, and we may consider the Hamiltonian
\beq\label{heisenberggen}
	H = \sum_{x\in\Z} h(x),\quad h(x) = V_1(\vec\sigma_{x+1}\cdot\vec\sigma_x) + 
	V_2(\vec\sigma_{x+2}\cdot\vec\sigma_x) + \ldots + \mu\sigma_x^3.
\eeq
It preserves the total spin in the z direction, the extensive conserved quantity formally defined as $S^3 = \sum_{x\in\Z} \sigma_x^3$,
\beq
	[H,S^3]=0,
\eeq
and at $\mu=0$ it preserves $S^i=\sum_{x\in\Z}\sigma_x^i$ for $i=1,2,3$. The case $V_1(a) = a$, $V_{n\geq 2}(a)=0$ is the Heisenberg spin chain. The family of twist fields associated to the symmetry generated by $\vec\lambda\cdot\vec S$ is
\beq\label{Tspinintro}
	\mathcal T_\lambda(x) = \prod_{x'\geq x} e^{\ri \vec\lambda\cdot\vec\sigma_{x'}}.
\eeq
These are twist fields only for $\vec\lambda = (0,0,\lambda)$ if $\mu\neq 0$ (as otherwise $\vec\lambda\cdot\vec S$ does not generate a symmetry). See Subsec.~\ref{ssectultra} and \ref{ssectexp}.

\medskip

{\bf 3.} A common set of models are Bosonic Galilean quantum gases. Their formulation in second quantisation is often the most convenient, with canonical complex Bosonic fields $\psi(x)$,
\beq
	[\psi(x),\psi^\dag(x')] = \delta(x-x')
\eeq
and Hamiltonian, for instance, given by
\beq
	H = \int \dd x\,h(x),\quad h(x) = -\frc12 \psi^\dag(x)\p_x^2 \psi(x)  + V(|\psi(x)|^2).
\eeq
The Hilbert space $\mathcal H$ is spanned by $n$-particle states $\psi^\dag(x_1)\cdots\psi^\dag(x_n)|0\ket$ at positions $x_1,\ldots,x_n$ for $n=0,1,2,\ldots$, with pseudovacuum $|0\ket$ defined as $\psi(x)|0\ket = 0$, and the first-quantised $n$-particle wave functions associated to the vector $|v\ket$ is $\bra v|\psi^\dag(x_1)\cdots\psi^\dag(x_n)|0\ket$. The model has $U(1)$ symmetry generated by the number operator $\int \dd x\,\psi^\dag(x)\psi(x)$ as
\beq
	\sigma_\lambda(\psi(x)) = e^{-\ri \lambda}\psi(x),\quad
	\sigma_\lambda = e^{\ri \lambda \ad \int \dd x\,\psi^\dag(x)\psi(x)}.
\eeq
The family of twist fields associated to this $U(1)$ symmetry is
\beq\label{Tfermionintro}
	\mathcal T_\lambda(x)
	=\exp\Big[\ri \lambda\int_{x}^\infty \dd x'\,\psi^\dag(x')\psi(x')\Big].
\eeq
See  Subsec.~\ref{ssectultra} and \ref{ssectexp}.

\medskip

{\bf 4.} The classical counterpart would be a classical gas. A related set of models are classical chains of particle-like degrees of freedom with classical positions $q_x$ and momenta $p_x$ satisfying the Poisson bracket $\{q_x,p_{x'}\}= \delta_{x,x'}$, with Hamiltonian
\beq\label{Hclassical}
	H = \sum_{x\in\Z} h(x),\quad h(x) = \frc{p_x^2}2 + f(q_{x+1}-q_x).
\eeq
For appropriate function $f(a)$, such as in the Toda chain $f(a) = e^{-a}$, the ``total stretch'' is an extensive conserved quantity:
\beq\label{totalstretch}
	Q = q_{\infty}-q_{-\infty} = \sum_{x\in\Z} q(x),\quad
	q(x) = q_{x+1}-q_x.
\eeq
This generates the symmetry group that is dual to the non-compact $\R$ symmetry group of ``internal translations'' $q_x\to q_x+a,\,a\in\R$. The family of twist fields associated to the total stretch is
\beq
	\mathcal T_\lambda(x) = e^{-\lambda q_x},
\eeq
while that associated to the internal translations is
\beq
	\mathcal T_a(x) = \exp\Big[-a \sum_{x'\geq x} p_{x'}\Big].
\eeq
See Subsec.~\ref{ssectcon} and \ref{ssectjordan}.

\medskip

{\bf 5.} Finally, still in the classical realm but forgoing the dynamics, we may look at statistical lattice models. On the square lattice formed out of the vertices $\Z^2$, we could define the model by its Boltzmann weight
\beq
	\exp\Big[-\sum_{\text{edge}\, ({\vec x, \vec x'})\in\text{lattice}(\Z^2)} f(\vec s_{\vec x}\cdot\vec s_{\vec x'})\Big]
\eeq
where $\vec s_{\vec x}\in \R^d$ with, say, constraint $|\vec s_{\vec x}|=1$ and measure $\prod_{\vec x\in\Z^2} \dd^d s_{\vec x}$. This has $O(d)$ symmetry group
\beq
	\sigma_R(\vec s_{\vec x})= R\vec s_{\vec x}
\eeq
for $R\in O(d)$. Then the family of twist fields associated to this symmetry group is $\mathcal T_R(z)$, for $z\in\Z^2$, defined (at least partially) by the fact that $\bra \mathcal T_R(z)\cdots\ket/\bra \mathcal T_R(z)\ket$ is the expectation value of $\cdots$ with modified Boltzmann weight
\beq\label{spintwistintro}
	\exp\Big[-\sum_{\text{edge}\, e=({\vec x, \vec x'})\in\text{lattice}(\Z^2)} f\big((R^{\chi_{C_z}(e)}\vec s_{\vec x})\cdot \vec s_{\vec x'}\big)\Big]
\eeq
where $\chi_C$ is the characteristic function for the set $C$ of edges, and the set $C_z$ is
\beq
	C_z = \{((x',y),(x',y+1)): x'\geq x \},\quad z= (x,y).
\eeq
See Subsec.~\ref{ssecttwistpath} and \ref{ssectcon}.

\section{Locality and extensivity in many-body physics}\label{sectloc}

A fundamental aspect of many-body physics is the concept of {\em locality}, and its close cousin {\em extensivity}.

Intuitively, a local observable of a many-body system is something that only affects, or probes, a small region of space. This is important, as in many-body physics, it is difficult to influence, or observe, all degrees of freedom simultaneously, because there are too many of them. Normally, an observable relates to few degrees of freedom in some region of space, and correlations between these and others in another region of space. For instance, neutron scattering experiments give us the dynamical structure factor, the Fourier transform of space-time two-point correlations of the local observables representing the local effects of neutrons (on spin, atomic positions, etc).

But on the theoretical side, there are many subtleties with this informal description, and in order to understand the concept of twist field in a general enough fashion, it is important to understand them. I concentrate on one-dimensional systems for simplicity, but most of this discussion can be extended beyond this. I think of Hamiltonian systems such as quantum or classical chains, gases or fields, but also many of the concepts may be adapted to circuits, or stochastic or Linbladian systems.

First, the notion of ``small" requires a topology, for instance on the lattice sites $\Z$ on which degrees of freedom lie, or on continuous space $\R$. What tells us that a lattice site is near to another one? What tells us that these lattice sites form a ``one-dimensional'' system? A natural candidate is the topology induced by the system's {\em interaction}, or dynamics. Another candidate is, instead, given by the {\em state} in which the system finds itself. Claiming that two sites do not interact, or instead that correlations decay at large distances, as they should in many physically sensible states such as at equilibrium, suggest notions of neighbourhoods. Thus, locality may be {\em with respect to the dynamics}, or {\em with respect to the state}, two different notions, useful in their own way.

Second, ``small" also requires that there is a separation between a {\em microscopic} and a {\em macroscopic} length. A small region is, intuitively, a finite number of nearby lattice sites in a chain, or a finite interval, or even perhaps a single point, in a system that lives on the continuum. But, for instance, in a quantum chain of 10 sites, if any observable supported on ``a few'' sites is local, then all observables are local! So, the ``thermodynamic limit" must be taken, where the volume is made large. The volume may be taken as a macroscopic length scale, or at infinite volume, there may be a large, macroscopic scale of variations, such as that of a confining potential. Any macroscopic length scale is such that the number of independent degrees of freedom lying between two spatial points at a macroscopic distance from each other, is large (mathematically, tend to infinity).

Locality becomes a useful notion when looking for large-scale, universal emergent behaviours, such as hydrodynamic behaviours out of equilibrium, or quantum or thermodynamic critical correlations and fluctuations at equilibrium. The separation between {\em microscopic and macroscopic length scales} is therefore at the heart of locality. This is familiar from thermodynamics, where we have dual notions of {\em extensive and intensive} observables. But also, it is at the heart of twist fields, as these {\em ``appear'' to be non-local, observing a macroscopic length of the system}. This was essential in order to characterise the disordered phase of the Ising model. But they do satisfy a type of locality, and this is crucial for their formal and physical properties.

We will see how one can define an {\em extensive observable} in two alternative way: in the most physical way as {\em an observable that scales extensively with a macroscopic lenght scale $\ell$}, and in a more mathematical, but powerful way as an {\em equivalence class of local observables under total derivatives}.

I now discuss locality and extensivity in a pedagogical fashion, concentrating on the main intuitive concepts and avoiding mathematical details. In all my discussions, except otherwise stated, the system lies on an infinite volume. The microscopic length is implicitely finite, and controls the large-separation limits that are taken in order to define locality. In the discussion of extensivity the macroscopic length scale $\ell$ appears, which is the volume over which the extensive quantity lies.

The most useful notions for the construction of twist field are those based on the algebra of observables, Subsecs.~\ref{ssectdyn} and \ref{ssectextdyn}. These will be the basis for the exchange relations discussed in Subsec.~\ref{ssectexch}. Notions based on states, Subsec.~\ref{ssectstate} and \ref{ssectextstate}, are useful in applications, especilally for evaluating twist fields correlations, as described in Section \ref{sectappli}.


\subsection{Locality from the dynamics}\label{ssectdyn}

In relativistic quantum field theory (QFT), the notion of locality plays an essential role from the outset. From standard textbooks, one ascertains the ``quantum locality" of operators via commutation relations. One says that $o(x)$ is local if\footnote{As mentioned, throughout I concentrate on Bosonic observables. For Fermionic observables, the commutator would be replaced by the anti-commutator.}
\beq\label{QFTlocality}
	[o(x,t),o(x',t')] = 0\quad \mbox{at space-like distances }
	|x-x'| > c|t-t'|\qquad\mbox{(QFT)}
\eeq
where $c$ is the speed of light. In quantum systems, if observables commute, then they are quantum mechanically ``independent", and hence this is the statement that local observables far apart from each other are independent observables.

In the $C^*$-algebra formulation of quantum spin chains with short-range interactions on infinite volume, the Lieb-Robinson bound gives a similar statement to \eqref{QFTlocality} {\em for observables $o(x)$ supported on finite numbers of sites around $x$} \cite{brattelioperator1,brattelioperator2} (extensions of this are reviewed in \cite{chen2023speed}). There, the commutator is not zero, but bounded by a decaying exponential in $|x-x'|-c|t-t'|$. The full $C^*$ algebra contains more than just operators supported on finite numbers of site -- it is a ``completion'' of these, with respect to the topology induced by the operator norm. There is also a commutation statement for this completion: simply $\lim_{|x-x'|\to\infty}[o(x),o(x')]=0$.


Similar statements can be made in classical systems, using the Poisson bracket $\ri [\cdot,\cdot]\to \{\cdot,\cdot\}$. Now, observables are functions on phase space, and if they have vanishing Poisson brackets, then again they can be said to be ``independent".


Thus, {\em commutation relations tell us about independence}, and this appears to be a good way of assessing locality, independently of the specific context (QFT, spin chains, quantum or classical gases, etc). So, we may simply consider the following setup:

\subsubsection*{Setup}
We have a Lie algebra of observables $\mathfrak A$, with a commutator or Poisson bracket operation (we will use the commutator notation $[\cdot,\cdot]$ from the quantum setup)\footnote{Here and throughout these notes, an ``observable'' is taken in its most general sense: in the quantum case, it not necessarily Hermitian (and there isn't necessarily a Hermiticity structure).}. There may be underlying this an associative product with Leibniz's rule, but for now this is not essential. We have a notion that allows us to say that a sequence of observables tend to 0 -- for instance, a norm. We have a one-parameter group of (linear) automorphisms $\iota_x:\mathfrak A\to \mathfrak A$ of the algebra of observables, 
\beq\label{oxalgebra}
	\iota_x o = o(x),\quad \iota_x[a,b] = [\iota_x a,\iota_x b],\quad \iota_0 = \1,\quad \iota_x\circ \iota_{x'} = \iota_{x+x'}.
\eeq
The parameter $x$ is deemed a ``spatial position", so $\iota_x$'s are spatial translations, discrete or continuous; below we will use the continuous notation, and consider the spatial set to be $\R$, for simplicity. Finally, we identify one particular observable, $h$, which we deem to be the {\em energy density}. In this setup, it is this particular observable -- or more precisely the subspace $\spn\{h(x):x\in\R\}$ -- that fixes our notion of locality and determines all our local observables.

We say that {\em the dynamics is local} if
\beq\label{localh}
	[h(x),h(x')]\to 0 \quad \mbox{fast enough as $|x-x'|\to\infty$}.
\eeq
A good notion of ``fast enough" is exponential decay, but in these notes I will keep such issues of analysis out of sight. Eq.~\eqref{localh} means that the energy densities at well-separated positions are independent of each other as observables. This brings us towards a topology: it loosely says that $x$ and $x'$ are far enough -- in different neighbourhoods -- if $[h(x),h(x')]$ is small enough. Related to it is the notion of support: we may say that $h(x)$ is ``supported around'' $x$. In many cases, the commutator is exactly zero for $|x-x'|>d$ for some $d>0$; in these cases, the system has ``finite interaction range" $d$.
In quantum and classical field theory and gases, one often has $d=0$.

Then, we construct the subspace of {\em all local observables} $\mathfrak L_h\subset\mathfrak A$ by looking for all observables $o$ such that
\beq\label{local}
	o\in \mathfrak L_h \quad : \quad [h(x),o(x')]\to 0 \quad \mbox{fast enough as $|x-x'|\to\infty$.}
\eeq
Note that if $o\in\mathfrak L_h$, then $o(x)\in\mathfrak L_h$, so $\iota_x \mathfrak L_h = \mathfrak L_h$ for all $x$. That is, the energy density is our ``base observable" to define other local observables. In systems with finite interaction range, we may ask for the commutator to vanish  at all large enough distances; the distance at which vanishing occurs may depend on $o$. In QFT, this distance is usually taken to be 0, such as in \eqref{QFTlocality} at $t=t'=0$. In such cases, one often reserves the word ``local" for these, while observables whose commutators with $h(x)$ vanish only asymptotically are ``quasi-local". Here, I keep the word ``local" for the more general concept, forgoing the specific requirements on the large-separation vanishing.

The set of local observables as defined in \eqref{local} is usually too wide for useful applications. Important subspaces of local observables are {\em families of mutually local observables}. These are subspaces $\mathfrak L\subset \mathfrak L_h$ such that
\beq\label{localset}
	[o(x),o'(x')] \to 0 \quad \mbox{fast enough as $|x-x'|\to\infty$}\quad \forall\ o,o'\in\mathfrak L.
\eeq
Again $\iota_x\mathfrak L = \mathfrak L\;\forall \,x$. With $o=o'$ in \eqref{localset} we say that $o$ is self-local (instead of mutually local with itself). We will see in Sec.~\ref{secttwist} how twist fields are local observables that go beyond this subspace -- we will consider larger subspaces of ``mutually semi-local observables''.

The first equation in \eqref{oxalgebra} means that the observable $o(x)$ is {\em homogeneous}. For instance, the observable $o'(x) = xo(x)$ may still have decaying commutation relations as above, but it is not homogeneous, $\iota_x o'(x') = o'(x'+x) - xo(x'+x)\neq o'(x'+x)$. In this work, by local observables I mean homogeneous local observables, but many concepts can be extended to inhomogeneous observables.

\st{
In this construction, we start with what we deem to be an energy density $h$, we asks for it to be local in the sense of observable independence at large distances -- thus defining our topology -- and we then define from it local observables, within which we may restrict to mutually local observables.
}

\subsubsection*{The case of relativistic QFT}

In relativistic QFT, there is a momentum operator $P$  which generates space translations,
\beq
	e^{-\ri Px}\,o\, e^{-\ri Px} = \iota_x o = o(x).
\eeq
Further, there is a Hamiltonian operator $H$ for time evolution, and the Hilbert space $\mathcal H$ is spanned by asymptotic states, Eq.~\eqref{qftstates}. On these states, $P$ and $H$ act in a specifc way dictated by relativistic invariance,
\beqa
	P|\theta_1,\ldots,\theta_n\ket_{a_1,\ldots,a_n} &=& \sum_{i=1}^n m_{a_i}\sinh(\theta_i)|\theta_1,\ldots,\theta_b\ket_{a_1,\ldots,a_n}\n
	H|\theta_1,\ldots,\theta_n\ket_{a_1,\ldots,a_n} &=& \sum_{i=1}^n m_{a_i}\cosh(\theta_i)|\theta_1,\ldots,\theta_b\ket_{a_1,\ldots,a_n},
	\label{PHqft}
\eeqa
where $m_a$ is the mass of particle of type $a$.

But, even though  $\mathcal H$ is given and $P, H$ are fully specified as above, {\em the model of QFT is not defined until a Hamiltonian density $h$ is specified}, such that $H = \int \dd x\,h(x)$, as an operator on this Hilbert space, with the property \eqref{localh}. Knowing the Hilbert space of asymptotic particles, and the Hamiltonian, is not enough: the locality structure must be specified as well. There are many $h$ that give rise to $H$ acting as \eqref{PHqft}: any interacting model of QFT with one bosonic (fermionic) particle type has the same $\mathcal H$, $P$ and $H$ as the free bosonic (fermionic) field theory, but different models have different energy densities $h$. The question of characterising all QFT's with a given particle spectrum, is that of finding all Hamiltonian densities $h$ such that \eqref{localh} holds. Once this is set, one looks for local operators satisfying \eqref{local}, and their matrix elements. In fact, for relativistic QFT, we also ask for a local momentum density $p$ to be amongst those, and as well as all elements of the energy-momentum tensor $T^{\mu\nu}$, with $T^{00} = h,\,T^{01} = p$, which is conserved $\p_\mu T^{\mu\nu} = 0$ and symmetric $T^{\mu\nu}= T^{\nu\mu}$. From this, all physical aspects of the theory can be deduced, including the scattering matrix through the LSZ reduction formula. The above formal description of locality is a generalisation of such principles to a wider family of quantum and classical many-body systems.


\subsubsection*{Consequences}

By the Jacobi identity (or, if there is an underlying product, by Leibniz's rule) {\em the set of observables $\mathfrak L_h$, satisfying \eqref{local}, forms a subalgebra}:
\beq
	a,b\quad \mbox{local}\quad \Rightarrow \quad
	[a,b]\quad \mbox{local},\quad ab\quad \mbox{local}.
\eeq
Indeed take the observable $[a,b]$, and its translate $[a,b](x') = [a(x'),b(x')]$, and evaluate the following using \eqref{local}:
\beq
	[h(x),[a(x'),b(x')]] = [[h(x),a(x')],b(x')]
	+
	[a(x'),[h(x),b(x')]] \to 0 \quad \mbox{fast enough as $|x-x'|\to\infty$,}
\eeq
or using Leibniz's ruls
\beq
	[h(x),a(x')b(x')] = [h(x),a(x')]b(x') + a(x')[h(x),b(x')]
	\to 0 \quad \mbox{fast enough as $|x-x'|\to\infty$.}
\eeq

The requirement \eqref{local} physically makes sense, as it allows us to show, using series expansions, if the decay is indeed fast enough and at least for $t$ in some neighbourhood of 0, that {\em time evolution
\beq
	o(x,t) = e^{\ri t[H,\cdot]} o(x)
\eeq
generated by the Hamiltonian}\footnote{As mentioned, here, we use the integral symbol, but this covers also systems on discrete space: we may cleverly take $h(x) = h(\lfloor x\rfloor)$, so that this becomes $H = \sum_\Z h(x)$. Below we mostly use the continuous notation, keeping in mind that this covers applications to chains as well.}
\beq\label{H}
	H = \int \dd x\,h(x)
\eeq
{\em is well defined}. Indeed, although $H$, an integral over all of $\R$, is usually not part of $\mathfrak A$, the first time derivative it generates, $\dot o(x) = \ri[H,o(x)]$, restricts to a finite neighbourhood by locality:
\beq\label{Ho}
	[H,o(x)] = \int \dd x'\,[h(x'),o(x)]
	= \int_{x'\sim x} \dd x'\,[h(x'),o(x)].
\eeq
Here I use the symbol $x'\sim x$ to indicate the condition that $x'$ be ``near to" $x$, in the sense of \eqref{localh} and \eqref{local} -- thus if we restrict $x'-x$ to lie on a finite interval $I\subset \R$, by putting the characteristic function $\chi_{I+x}(x')$, the error term vanishes fast enough as $I$ is made large\footnote{The relation $\sim$ is an equivalence relation, because $\chi_{I+x}(x')\chi_{I'+x'}(x'') = \chi_{I+x}(x')\chi_{I''+x}(x'')$ where $I'' = [\min(I)+\min(I'),\max(I)+\max(I')]$, etc. Note that, formally, the relation $x\sim x': \exists\, \ep>0 \,|\,[h(x),h(x')]<\ep$ is {\em not} an equivalence relation; here, instead, we use the fact that there is a limit taken in \eqref{localh}, \eqref{local}.}. The result is local:
\beqa
	[h(x'),[H,o(x)]] 
	&=& \int_{x''\sim x} \dd x''\,[h(x'),[h(x''),o(x)]]\n
	&=& \int_{x''\sim x} \dd x''\,\big([[h(x'),h(x'')],o(x)] + [h(x''),[h(x'),o(x)]]\big)\n
	&=& \int_{x'\sim x''\sim x} \dd x''\,[[h(x'),h(x'')],o(x)] + 
	\int_{x''\sim x\sim x'} \dd x''\,[h(x''),[h(x'),o(x)]]\n
	&\to& 0 \qquad (|x'-x|\to\infty).\label{odotlocal}
\eeqa
Similarly, the $n$th term in the series
\beq\label{oxt}
	o(x,t) = e^{\ri t[H,\cdot]} o(x) = \sum_{n=0}^\infty \frc{(\ri t)^n}{n!} \int_{x_i\sim x} \dd x_1\cdots \dd x_n\,
	[h(x_1),[\ldots,[h(x_n),o(x)]\ldots]]
\eeq
indeed restricts to an integral over a finite region of $\R^n$ possibly up to small, decaying errors, and by a calculation as in \eqref{odotlocal}, is local. Thus, if the approach to 0 is fast enough in \eqref{local}, or at least in the sense of formal series, the time-evolved observable $o(x,t)$ is also local:
\beq
	[h(x),o(x',t)] \to 0 \qquad (|x-x'|\to\infty),
\eeq
and similarly for mutual locality. Therefore
\beq\label{oxtlocal}
	o \quad\mbox{local, mutually local}\quad \Rightarrow\quad
	o(x,t) \quad\mbox{local, mutually local}.
\eeq
Formal expressions such as \eqref{H}, and expressions such as \eqref{Ho} and \eqref{oxt}, are discussed at more length in the context of extensive observables in Subsec.~\ref{ssectextstate} and especially \ref{ssectextdyn}.

In fact, time-evolved operators are often local for all $t\in\R$, for instance thanks to a Lieb-Robinson bound, see the discussions in \cite{brattelioperator1,brattelioperator2,hastings2010localityquantumsystems,ranard2021aspects,chen2023speed}.

\subsubsection*{Existence of local currents}

One of the most interesting statement in many-body local physics is that about the existence of currents associated to conserved densities. In order to show this, let us show a general statement about local observables whose total value on all of space vanishes. Suppose $o(x)$ is a local observable that satisfies the equation
\beq\label{ovanish}
	\int \dd x\,o(x) = 0.
\eeq
As for the Hamiltonian, the integral over all of $\R$ does not necessarily make sense. But again, we may understand this in terms of the algebra of observables, considering commutators: $[\int \dd x\,o(x),a]=0$ for any local observable $a$. Then, one can show that {\em there must exist a local observable $o'$ such that}
\beq\label{oop}
	o(x) = \p_x o'(x).
\eeq
Indeed, consider the following observable $o'(x)$, which has a ``linear tail'' emanating from $x$ that we can put either to the right or to the left:
\beq\label{opdef}
	o'(x) = \int_{-\infty}^x \dd x'\,o(x') = -\int_{x}^\infty \dd x'\,o(x')
	=:\p_x^{-1}o(x).
\eeq
The equivalence follows from \eqref{ovanish}. Then \eqref{oop} holds, and $o'(x)$ is local: by the first way of writing $o'(x)$ we have
\beq
	[h(x),o'(x')] \to 0 \qquad (|x-x'|\to\infty)
\eeq
while by the second way
\beq
	[h(x),o'(x')] \to 0 \qquad (|x-x'|\to\infty).
\eeq
In fact, the observable \eqref{opdef} may not have been in $\mathfrak A$. But one can {\em formally} adjoin it to a space of observables; this makes  sense as a Lie algebra if this is a space of mutually local observables of which $o$ is part, because we can always choose the right or left tail of $o'(x)$ to make sure that the commutator exists, and vanishes at large separations. Once this is done, then it is a local observable. See \cite{doyon2017thermalization} for a construction similar to that, where mathematically rigorous sense is given via a Hilbert space of observables (that we discuss below in Subsec.~\ref{ssectextstate}).

Now with $H$ generating time translations, suppose
\beq\label{Q}
	Q=\int \dd x\,q(x)
\eeq
is a conserved quantity: formally
\beq
	[H,Q]=0\ \Rightarrow\ \int \dd x\,\dot q(x)=0.
\eeq
Then there is an associated local current
\beq
	j(x) = -\p_x^{-1}\dot q(x).
\eeq
Thus we have shown that local densities of conserved quantities must have an associated local current and satisfy a conservation law,
\beq\label{conscurrent}
	\dot q + \p_x j = 0\quad\mbox{i.e.}\quad \p_\mu j^\mu = 0
\eeq
where in the second equation we define $j^0 = q,\,j^1 = j$ with $\p_0 = \p_t,\,\p_1 = \p_x$.

These aspects are discussed again in Subsec.~\ref{ssectsymmetry} in the context of symmetries.

\subsection{Locality from the state}\label{ssectstate}

In equilibrium thermodynamic states and in ground states of typical many-body systems with short-range interactions, connected correlation functions, or Ursell functions, of local observables decay at large spatial separations:
\beq\label{decayequil}
	\bra a(x)b(x')\ket^{\rm c}:= \bra a(x)b(x')\ket - \bra a(x)\ket\bra b(x')\ket \to 0\quad \mbox{as } |x-x'|\to\infty\qquad \mbox{(equilibrium states).}
\eeq
Decay of correlations -- the asymptotic clustering of correlation functions -- indicates that, within this state, observables $a(x)$ and $b(x')$ that are far from each other, become statistically uncorrelated\footnote{Here and below, by a slight abuse of notation, $\bra\cdots\ket^{\rm c}$ is the connected correlation function where all factors in the apparent product are considered separate variables. When a clarity requires it, I will put commas, e.g.~$\bra a(x),b(x')\ket^{\rm c}$.}.

In thermodynamics such a decay plays a fundamental role: if the decay is strong enough, then the corresponding susceptibility is finite. Take, for instance, a Gibbs state for a system in volume $L$, with inverse temperature $\beta$ and a chemical poential $\mu$ associated to a conserved charge $Q$ -- which may be a number of particle, a total spin, etc. Both the Hamiltonian $H$ and charge $Q$ are supported on $[0,L]$, and this is what determines the volume of the system (we take it finite only for this paragraph). The susceptibility is the variation of the average volume density of $Q$ under a variation of the chemical potential:
\beq\label{thremostate}
	\frc{\p}{\p\mu} \frc{\Tr\Big( e^{-\beta(H-\mu Q)}Q/L\Big)}{
	\Tr\Big( e^{-\beta(H-\mu Q)}\Big)}
	=
	\beta L^{-1}\bra QQ\ket^{\rm c}.
\eeq
If the conserved charge has the form \eqref{Q}, $Q = \int_0^L \dd x\,q(x)$, and if the correlation decay is strong enough, one can show, using spatial translation invariance of the state, that (see e.g.~\cite{doyon2017thermalization})
\beq\label{infvolume}
	\lim_{L\to\infty}
	L^{-1}\bra QQ\ket^{\rm c}
	=
	\int \dd x\,\bra q(x)q(0)\ket^{\rm c}.
\eeq
If the correlation decay is not strong enough, the susceptibility may diverge, such as at critical points \cite{cardy1996scaling,sachdevbook}.

In the $C^*$-algebra formulation of quantum statistical mechanics, correlation decay is also a fundamental property. In this context, a state $\bra\cdots\ket = \omega(\cdots)$ is seen as a normalised positive linear map
\beq
	\omega : \mathfrak A \to \C
\eeq
an such maps form a convex set, because we can add them with positive coefficients summing to 1. Convex sets have extremal points, which are those that cannot be written as convex linear combinations of more than one states. It turns out that extremal points satisfy \eqref{decayequil}, and thus correlation decay is an indication that the system is in a ``single thermodynamic state''. This is natural, at least in one direction of the implication: if $\omega_1$ and $\omega_2$ satifsy \eqref{decayequil}, then their weighted sums, e.g.~$\omega = \frc12(\omega_1+\omega_2)$, clearly don't, because the term $\bra a(x)\ket\bra b(x')\ket = \omega(a(x'))\omega(b(x'))$ is non-linear $\omega$. Such weighted sums may be used to represent coexistence of thermodynamic states (for instance, a liquid-vapour mixture). In fact, in the $C^*$-algebra context, there is a close relation between decay of commutators $[a(x),b(x')]\to0$ ($|x-x'|\to\infty$), and decay of correlations in states: the former imply the latter in so-called ``factor'' states. See \cite{brattelioperator1,brattelioperator2}.

Thus, {\em decay of correlations are important properties of thermodyamic state}, and this appears to be a good way of assessing locality. So, we may simply consider the following setup:

\subsubsection*{Setup}

Much like in the previous Subsection, we have an algebra of observables $\mathfrak A$, now with an associative product. We again have a notion that allows us to say that a sequence of observables tend to 0, and one-parameter group of automorphisms $\iota_x:\mathfrak A\to \mathfrak A$ of the algebra of observables, 
\beq
	\iota_x o = o(x),\quad \iota_x(ab) = \iota_x (a)\iota_x (b)
	,\quad \iota_0 = \1,
\eeq
with $x$ deemed a spatial position. Finally, we have a normalised, positive linear map on $\mathfrak A$,
\beq
	\bra\cdots\ket: \mathfrak A \to \C,
\eeq
a {\em state}. In this setup, it is this state that fixes our notion of locality and determines all our local observables.

We say that $\mathfrak L\subset\mathfrak A$ is a subalgebra of local observables if it is a subalgebra, and if the state asymptotically clusters for all elements of this subalgebra:
\beq\label{localstate}
	\bra o(x)o'(x')\ket^{\rm c} \to 0 \quad \mbox{fast enough as $|x-x'|\to\infty$}\quad \forall\ o,o'\in\mathfrak L.
\eeq
That is, local observables decorrelate from each other at far positions. Now the topology, loosely, says that $x$ and $x'$ are far enough if $\bra o(x)o'(x')\ket^{\rm c}$ is small enough for all $o,o'\in\mathfrak L$.

Here the algebraic structure does not play as fundamental a role as in the definition based on commutators or Poisson brackets -- it is there so we can define the product and therefore correlation functions. But it simplifies the discussion to ask for local observables to form not just a subspace, but also a subalgebra, and this requirement is often satisfied in applications. Again, we note translation invariance $\iota_x\mathfrak L = \mathfrak L$ for all $x$.

It is sometimes insightful to emphasise the non-algebraic basis of this locality concept by construting a Hilbert space, in a simple modification of the Gelfand-Naimark-Segal construction \cite{brattelioperator1}. The pre-inner product is the sesquilinear form
\beq\label{gns}
	(a,b) = \bra a^\dag b\ket^{\rm c}
\eeq
and from this the null-space $\mathcal N = \{a\in \mathfrak A:(a,a)=0\}$ is moded out (linearly), and the Hilbert space is the completion with respect to the inner-product, $\mathcal H_{\rm GNS} = \overline{\mathfrak A / \mathcal N}$. Then local observables are $\mathcal L\subset \mathcal H_{\rm GNS}$ such that $(a(x),b(x'))\to0$ as $|x-x'|\to\infty$.

But the fact that local observables do form an algebra (that is, that we can multiply them) is useful, for instance the decay of two-point correlations controls the decay of higher-point correlations \cite{ampelogiannis2025clustering}\footnote{This is a slightly stronger version of the result proven there.}:
\beq\label{localstatemulti}
	\bra o_1(x_1),o_2(x_2),\ldots,o_n(x_n)\ket^{\rm c} \to 0\quad
	\mbox{as}\ \max_{s\subset \{1,\ldots,n\}} {\rm dist}(\{x_i:i\in s\},\{x_j:j\in \{1,\ldots,n\}\setminus s) \to \infty
\eeq
for $o_1,\ldots,o_n\in\mathfrak L$.

\st{
In this construction, we start with a state $\bra\cdots\ket$, and we ask for it to be clustering fast enough for all local observables -- thus defining our topology.
}

\subsubsection*{Consequences}

The most immediate consequence of locality is the ability to construct a {\em flow on states} using local observables. Indeed, if $w\in\mathfrak L$, then we may construct the one-parameter flow, say with $\beta\in\R$,
\beq\label{flowstate}
	\frc{\dd}{\dd\beta} \bra o\ket_\beta
	= - \int \dd x\,\bra w(x)\,o\ket^{\rm c}_\beta,\quad
	\bra \cdots\ket_0 = \bra\cdots\ket.
\eeq 
This defines the state $\bra\cdots\ket_\beta$ on the subalgebra $\mathfrak L$, as it defines its application on every local observable. This flow preserves the set of local observables, because, by using the definition of the connected correlation functions, we have
\beq
	\frc{\dd}{\dd\beta} \bra o(x)o'(x')\ket_\beta^{\rm c}
	= - \int \dd x''\,\bra w(x'')o(x)o'(x')\ket^{\rm c}_\beta
\eeq
and therefore, by \eqref{localstatemulti}, if the decay at large separations is strong enough,
\beq
	\frc{\dd}{\dd\beta} \bra o(x)o'(x')\ket_\beta \to 0\quad (|x-x'|\to\infty).
\eeq
That is, if we can exchange limit and derivative, $\mathfrak L$ is an algebra of local observables not only with respect to $\bra \cdots\ket$, but also with respect to all of $\bra\cdots\ket_\beta$. The flow can be solved formally, in the form of the following series, which parallels \eqref{oxt} in the state setup:
\beq\label{oxw}
	\bra o\ket_\beta = \frc{\bra e^{-\beta W} o\ket}{\bra e^{-\beta W}\ket} = \sum_{n=0}^\infty \frc{(-\beta)^n}{n!} \int \dd x_1\cdots \dd x_n\,
	\bra w(x_1)\cdots w(x_n)\,o(0)\ket^{\rm c}
\eeq
where
\beq
	W = \int \dd x\,w(x).
\eeq
Again, $W$ as written, with an integral over $\R$, does not need to be part of $\mathfrak A$: it is its exponential within a state, as in \eqref{oxw}, that needs to make sense, and this is so, for instance, by the series expansion in \eqref{oxw}.

Choosing $\bra\cdots\ket$ to be the trace state $\Tr$ and $W=H-\mu N$, we recover the thermodynamic state in \eqref{thremostate}. A more general construction of such flows was introduced in \cite{doyon2017thermalization} in the context of studying long-time relaxation in homogeneous systems, and was extended in \cite{buvca2023unified}.

\subsubsection*{Existence of local currents}

Like in the previous Subsection, it is simple to argue that if
\beq\label{ovanishstate}
	\int \dd x\,o(x) = 0,
\eeq
in the sense that $\int \dd x\,\bra o(x)\,a\ket^{\rm c} = 0$ for any local observable $a$, then
\beq\label{oopstate}
	o(x) = \p_x o'(x),\quad o'(x) =\p_x^{-1}o(x)
\eeq
where $o'(x)$ is local. Here, one may adjoin $o'$ to the subspace $\mathfrak L$, and all multi-point connected correlation functions $\bra o_1(x_1),o_2(x_2),\ldots,o_n(x_n)\ket^{\rm c}$ still make sense and satisfy \eqref{localstatemulti}, because we can always put the tails towards the right or left as required, and use locality, to guarantee that the connected correlation function is finite, and decays when distances are large. But this adjoining does not necessarily make sense as an algebra.

Likewise, if there is a time flow $o(x)\mapsto o(x,t)$, and if $Q=\int \dd x\,q(x)$ is conserved in the sense that
\beq
	\int \dd x\,\dot q(x,t) = 0,
\eeq
then there is an associated local current
\beq
	j(x) = -\p_x^{-1}\dot q(x)
\eeq
and a conservation law,
\beq
	\dot q + \p_x j = 0.
\eeq
This construction was made mathematically accurate for the inner product \eqref{gns} in the context of $C^*$-algebra formulation of quantum spin chains in \cite{doyon2022hydrodynamic}.

\subsection{Extensivity from the state: pseudolocal charges}\label{ssectextstate}

I use the setup of Subsec.~\ref{ssectstate}: locality is defined with respect to a state $\bra\cdots\ket$, which satifies an asymptotic clustering condition \eqref{localstate}. For simplicity, I assume that the state satisfies translation invariance, so in particular
\beq\label{translation}
	\bra o_1(x_1+x)\cdots o_n(x_n+x)\ket^{\rm c} =
	\bra o_1(x_1)\cdots o_n(x_n)\ket^{\rm c}.
\eeq

\subsubsection*{Linear scaling with the macroscopic length: ``pseudolocal'' observables}

Recall that the Hamiltonian $H$, and the conserved quantities $Q$, were written as integrals over $\R$, Eqs.~\eqref{H}, \eqref{Q}. So, consider the following, which for now is not necessarily a conserved quantity:
\beq
	Q = \int \dd x\,q(x).
\eeq
As I mentioned, this usually does not make sense in the algebra of observables $\mathfrak A$, let alone as a local observable $\mathfrak L$. Naturally, we could make the integration region finite,
\beq\label{Qelldensity}
	Q_\ell = \int_{-\ell/2}^{\ell/2} \dd x\,q(x)\in\mathfrak L \subset \mathfrak A.
\eeq
Now this exists within the algebra of local observables, in both constuctions of Subsec.~\ref{ssectdyn} and \ref{ssectstate}. Here, in contrast to the discussion at the beginning of Subsec.~\ref{ssectstate}, the system still lies on an infinite volume, and the state is homogeneous, Eq.~\eqref{translation},  but we restrict the observable to the interval $[-\ell/2,\ell/2]$; that is, $\ell$ is our macroscopic length.

Clearly, by translation invariance,
\beq
	\bra Q_\ell\ket = \ell \bra q(0)\ket.
\eeq
But also, we note the following two properties: by \eqref{localstate}, a relation such as \eqref{infvolume} holds:
\beq\label{Qell2}
	\bra Q_\ell^2\ket^{\rm c}
	\sim \ell \int\dd x\,\bra q(x)q(0)\ket^{\rm c}\quad\mbox{as }\ell\to\infty,
\eeq
and for any local $o$,
\beq
	\lim_{\ell\to\infty} \bra Q_\ell\, o(x)\ket^{\rm c}
	=
	\lim_{\ell\to\infty} \bra Q_\ell\, o(0)\ket^{\rm c}
	=
	\int\dd x\,\bra q(x)\,o\ket^{\rm c}.
\eeq

An {\em extensive observable} with respect to the state $\bra\cdots\ket$ is defined by formalising these properties: it is a one-parameter family of local observables
\beq
	Q_\ell\in\mathfrak L,\quad \ell>0
\eeq
which may or may not be written as in \eqref{Qelldensity}, such that
\beq\label{cond1}
	\bra Q_\ell^\dag Q_\ell\ket^{\rm c} \leq \gamma \ell\quad\forall\ \ell>0
\eeq
for some $\gamma>0$, and such that the following limit exists and is independent of the position $x$, for every local $o$:
\beq\label{cond2}
	\lim_{\ell\to\infty} \bra Q_\ell^\dag\, o(x)\ket^{\rm c}
	=
	\lim_{\ell\to\infty} \bra Q_\ell^\dag\, o(0)\ket^{\rm c}
	=: \mathsf Q(o).
\eeq
We note that non-zero extensivity is required: if $\bra Q_\ell^\dag Q_\ell\ket^{\rm c}\ll \ell$ then one can show, by using the Cauchy-Schwartz inequality, that $\mathsf Q(o)=0$. If $Q_\ell$ is written as in \eqref{Qelldensity}, then $q(x)$ is its {\em local density} and
\beq\label{mapQlocal}
	\mathsf Q(o) = \int \dd x\,\bra q^\dag(x)o(0)\ket^{\rm c}.
\eeq
In this case, we note the ``asymptotic derivative property''
\beq\label{derstate}
	\mathsf Q(o(x)o'(x')) \to \bra o'\ket \mathsf Q(o) + \bra o\ket \mathsf Q(o'),\quad  |x-x'|\to\infty.
\eeq

Condition \eqref{cond1} imposes  {\em linear scaling} of the second cumulant of $Q_\ell$, while \eqref{cond2} imposes spatial homogeneity. They have been given a precise mathematical description in \cite{doyon2017thermalization}, and are a formalisation of the concept originally referred to as {\em pseudolocal charge} (see the review \cite{ilievski2016quasilocal}). In this original description, the quantity $\bra Q_\ell^\dag Q_\ell\ket^{\rm c}$ was taken in the (normalised) trace state $\bra\cdots\ket = \Tr(\cdots)$, and, setting $\bra Q_\ell\ket = 0$ by an appropriate shift, this becomes the square of the (normalised) Hilbert-Schmidt norm of $Q_\ell$. Here, the concept is expressed in its full generality, as introduced in \cite{doyon2017thermalization}, with a norm that is, in general, state-dependent. Thus {\em the notion of an extensive observable is state-dependent}. It is the limit $\ell\to\infty$ that is referred to as {\em extensive observable}, and here, this limit is identified with the map $\mathsf Q:\mathfrak L \to \C$.

I discuss higher cumulants below, but first I will discuss the general theory from the above definition, only considering two-point functions. Many important concepts arise from this mathematically precise theory, in particular that of the equivalence class $\int q$, that will be used in wider contexts.

\subsubsection*{Extensivity as a Hilbert space and stationary state manifold}

It turns out that extensive observables form a Hilbert space. A theorem established in \cite{doyon2017thermalization} identifies this Hilbert space with $\mathcal H'$ constructed similarly to the GNS Hilbert space $\mathcal H_{\rm GNS}$, but from a modification of the inner product \eqref{gns}. In this construction, we do not need any macroscopic length; everything is defined directly using local observables.

Starting with all observables in $\mathfrak L$, the new pre-inner product is
\beq\label{innerprime}
	(a,b)' =  \int \dd x\,( a(x),b(0)) = \int \dd x\,\bra a^\dag(x)b(0)\ket^{\rm c}.
\eeq
Again from this the null-space $\mathcal N' = \{a\in \mathfrak L:(a,a)'=0\}$ is moded out (linearly), and the Hilbert space is the completion with respect to the inner-product, $\mathcal H' = \overline{\mathfrak L / \mathcal N'}$. A theorem in \cite{doyon2017thermalization} establishes the equivalence between $\mathcal H'$ and the space of extensive observables: for every $\mathsf Q$ (defined as a limit $\ell\to\infty$ as above), there is $Q\in \mathcal H'$ such that $\mathsf Q(o) = (Q,o)'$ for all local $o$, and {\em vice versa}. Note how we use the capital letter $Q$. This is to emphasise that this is an element of $\mathcal H'$, which differs from local observables $\mathfrak L$ in two ways: the moding out (taking equivalence classes), and the completion (adjoining Cauchy sequences). Then, clearly, $\mathsf Q$ can be extended, by continuity, to all $o\in\mathcal H'$, and $\mathsf Q:\mathcal H'\to \C$ is a continuous linear map.

The viewpoint of extensive observables via the Hilbert space $\mathcal H'$ is more mathematical, but rather insightful. In particular, we emphasise that an extensive observable is {\em an equivalence class}, or Cauchy sequence thereof. For a given local $o$, let us denote this equivalence class suggestively using the integral symbol:
\beq
	\int o = \{o + a: a\in\mathcal N'\} \in \mathcal H'.
\eeq
What is $\mathcal N'$? It is all local $a$'s such that $\int \dd x\,\bra a^\dag(x)a(0)\ket^{\rm c}=0$, and by Cauchy-Schwartz, this implies $\int \dd x\,\bra a^\dag(x)o(0)\ket^{\rm c}=0$ for all local $o$. Thus, recalling \eqref{ovanishstate} and \eqref{oopstate},
\beq\label{Nder}
	\mathcal N' = \{a\in\mathfrak L:\int \dd x\,a(x) = 0\}
	= \{\p b:b\in\mathfrak L\}
\eeq
where $(\p b)(x) = \p_xb(x)$. Naturally, we may add any total derivative  $a(x)\to a(x)+ \p_x o(x)$ in $(a,b)'$, and the result stays the same, because the boundary terms at infinity vanish due to asymptotic clustering \eqref{localstate}; that is, $\int \dd x\,a(x)$ is unchanged. But importantly, \eqref{Nder} says that, by the explanations around \eqref{ovanishstate} and \eqref{oopstate}, total derivatives are the only things in $\mathcal N'$, the only local observables we can add to $a(x)$ to keep $(a,b)'$ unchanged. An extensive observable associated to a local density, is identified with {\em the equivalence class of its densities, implementing the idea that a local density is defined only up to total derivatives of local observables},
\beq\label{intoder}
	\int o = \{o + \p a: a\in\mathfrak L\}.
\eeq

The full Hilbert space $\mathcal H'$ of (i.e.~in bijection with) extensive observables includes the completion of these with respect to the inner product \eqref{innerprime}. What is the physical meaning of this completion? This includes {\em Cauchy sequences $\int o_n$}, where $o_n$ are local, but where the limit of the sequence $\lim_{n\to\infty} \int o_n$ may not be writable as $\int o'$ some local density $o'$. Thus, we go beyond local densities, and this is why this was originally referred to as ``pseudolocal''. We only ask for the observable to have ``extensivity properties'' with respect to the macroscopic parameter $\ell$, and this gives an extended notion of its density, obtained via the map $\mathsf Q$ {\em acting on} local observables (that is, in the dual space of that of local observables).

Putting in the time evolution, one obtains a one-parameter strongly continuous group $\tau_t:\mathcal H'\to\mathcal H'$ as shown in \cite{doyon2022hydrodynamic}, and one can restricts to {\em extensive conserved quantities}, extensive observables that are time-independent,
\beq
	\mathcal Q' = \{Q\in\mathcal H': \tau_t Q = Q\}.
\eeq
It turns out that it is, indeed, the space $\mathcal H'$ that controls the large-time relaxation \cite{doyon2017thermalization}, and asymptotics of correlation functions in space-time \cite{doyon2022hydrodynamic}, as expected from thermodynamics and hydrodynamics. Thus, a crucial problem in order to study the long-time dynamics of a given model is to determine the full set of extensive conserved quantities, {\em including those that do not possess a local density}, of the form $\lim_{n\to\infty} \int o_n$ where it may be that none of the $\int o_n$ is conserved.

It is worth illustrating the relation between $\mathcal H'$ and the thermodynamic states discussed above. Once we have extensive conserved quantities for a given state $\bra\cdots\ket$, we may use this to construct the {\em stationary state manifold}. Suppose we have a family of states $\bra\cdots\ket_\beta$, and the associated family of Hilbert spaces $\mathcal H_\beta'$ and extensive conserved quantities $\mathcal Q'_\beta$. Then these determine the possible stationary states via flows as in \eqref{flowstate}, which are more generally
\beq\label{flowformal}
	\frc{\dd}{\dd\beta} \bra o\ket_\beta
	= - (W_\beta,o)_\beta,\quad W_\beta\in\mathcal Q_\beta'.
\eeq
By this construction, the states $\bra\cdots\ket_\beta$ should satisfy the clustering property, thus be good, extremal stationary states. In this way, one constructs a {\em manifold} of Gibbs and generalised Gibbs states, with $\mathcal H'$ the tangent space at the ``point'' $\bra \cdots\ket$ on the manifold \cite{doyon2017thermalization}. It is shown in \cite{doyon2017thermalization,buvca2023unified}, under certain assumptions about spatial clustering at long times, that long-time relaxation goes to such states.

\subsubsection*{Cumulants and large deviations}

In the above, I have concentrated on extensivity from the viewpoint of two-point correlation functions only. There is a strong mathematical theory behind this, which I have overviewed. But it is natural to use the higher-point clustering property \eqref{localstatemulti}.

Taking again the sequence of $Q_\ell$ defined in \eqref{Qelldensity}, it is simple to see that the equivalent of \eqref{Qell2} holds for higher number of points as well:
\beq\label{Qelln}
	\bra Q_\ell^n\ket^{\rm c}
	\sim \ell \int\dd x_1\cdots\dd x_{n-1}\,\bra q(x_1)\cdots q(x_{n-1})q(0)\ket^{\rm c}\quad\mbox{as }\ell\to\infty.
\eeq
The important point is that {\em the scaling is proportional to the macroscopic length $\ell$, for all cumulant orders $n\geq 1$}. The quantities
\beq\label{scaledcumu}
	c_n:=\int\dd x_1\cdots\dd x_{n-1}\,\bra q(x_1)\cdots q(x_{n-1})q(0)\ket^{\rm c}
	=
	\lim_{\ell\to\infty}\frc{\bra Q_\ell^n\ket^{\rm c}}{\ell}
\eeq
are called the {\em scaled cumulants} of $Q_\ell$.

It is natural to enquiry if we may obtain a refined formal definition of extensive observables, using \eqref{Qelln}, instead of just \eqref{cond1},
\beq
	\bra Q_\ell^n\ket^{\rm c} \leq \gamma_n\ell,\quad n=2,3,\ldots
\eeq
along with homogeneity such as in \eqref{cond2}. We may want to define a sequence of linear maps generalising that in \eqref{cond2},
\beq
	\mathsf Q^{n}(o) := \lim_{\ell\to\infty}\bra Q_\ell^{n}\,o(0)\ket^{\rm c},\quad n=1,2,3,\ldots
\eeq
The associated mathematical theory has not been developed, as far as I am aware.

Much like in Eq.~\eqref{oxw}, the scaled cumulants allow us to express, as a power series, the ``free energy shift'' incurred due to $Q_\ell$,
\beq
	\log \bra e^{\lambda Q_\ell}\ket
	=
	\sum_{n=1}^\infty \frc{\lambda^n }{n!}
	\bra Q_\ell^n\ket^{\rm c}
	\sim
	\ell 
	\sum_{n=1}^\infty \frc{c_n\lambda^n }{n!}.
\eeq
That is, the insertion of the exponential of an extensive observable within the state $\bra\cdots\ket$ must shift its partition function in an ``exponential-extensive'' way, which we express via the {\em asymptotic relation}
\beq\label{deltaf}
	\bra e^{\lambda Q_\ell}\ket \asymp
	e^{\ell F(\lambda)},\quad
	F(\lambda) := 
	\sum_{n=1}^\infty \frc{c_n\lambda^n }{n!}
\eeq
with (for our purpose) the meaning $a_\ell \asymp b_\ell \Leftrightarrow \lim_{\ell\to\infty}\log a_\ell/\ell = \lim_{\ell\to\infty} \log b_\ell/\ell$. The quantity $F(\lambda)$ as defined here is called the {\em scaled cumulant generating function} (SCGF).

In \eqref{deltaf}, there is a subtle exchange of limits: the asymptotic expansion as $\ell\to\infty$, and the power series expansion in $\lambda$. The exchange is not always valid, and it may happen that even though scaled cumulants \eqref{scaledcumu} exist and are finite, $\log \bra e^{\lambda Q_\ell}\ket$ does not asymptote to $\ell F(\lambda)$ with $F(\lambda)$ the SCGF.

If $Q$ is an extensive conserved quantities, then the quantity $F(\lambda)$ is related to thermodynamic functions associated to the state $\bra\cdots\ket$ and its modification by $Q$. Suppose there is a space of conserved quantities $\spn(Q_1,\ldots,Q_n)$, and $Q = Q_i$ for some $i$. Consider the manifold of states $\bra\cdots\ket_{\underline\beta}$ where $\underline\beta = (\beta_1,\ldots,\beta_n)$, as defined via \eqref{flowstate}. One can define the specific free energy as the function $f(\beta_1,\ldots\beta_n)$ which satisfies ($Q_j = \int \dd x\,q_j(x)$)
\beq
	\frc{\p f}{\p\beta_j} = \bra q_j\ket_{\underline\beta}.
\eeq
If the state $\bra \cdots\ket$ is put on a finite volume $L$ (with appropriate boundary conditions, say periodic), then one can evaluate its parition function $Z_L(\underline{\beta})$ and the above agrees with the usual definition $f = -\lim_{L\to\infty} \frc1L \log Z_L(\underline{\beta})$,
\beq
	Z_L = \Tr \Big(e^{-\sum_j \beta_j Q_j^{(L)}}\Big) 
	\asymp e^{-Lf(\underline\beta)}.
\eeq
Then one can show \cite{doyon2020fluctuations} that $F(\lambda)$ is a difference of free energies
\beq\label{Fff}
	F(\lambda) = f(\beta_1,\ldots,\beta_n) -  f(\beta_1,\ldots,\beta_i-\lambda,\ldots,\beta_n).
\eeq
Intuitively, this means that it is possible factorise the partition function, and its modification by the insertion of $e^{Q_\ell}$, into a product of partition functions around positions $x \in L\Z$ for some $1\ll L\ll \ell$, so that
\beq\label{factZ}
	\bra e^{\lambda Q_\ell}\ket
	\asymp
	\prod_x \frc{Z_{L}(\beta_1,\ldots,\beta_i - \chi_{[-\ell/2,\ell/2]}(x)\lambda,\ldots,\beta_n)}{Z_L(\beta_1,\ldots,\beta_n)}
\eeq
where $\chi_A(x)$ is the indicator function for the set $A$.

One may also study averages of exponential of total currents on a large time domain, instead of total charges on a large spatial region, Eq.~\eqref{ttdefTUgencontpathtime} below; see the brief discussion in Subsec \ref{ssectld}.

The importance of the exponential scaling \eqref{deltaf} lies in its relation to the {\em large deviation theory} for the distribution of $Q_\ell$. I take the quantum language, but the same can be said for classical systems. In a state with normalised density matrix $\rho$, the probability of measuring the value $\ell z$ for the operator $Q_\ell$ is
\beq
	P(Q_\ell = \ell z) = \Tr \Big(\rho\mathbb P_{\ell z}^{Q_\ell}\Big)
\eeq
where $\mathbb P_{\ell z}^{Q_\ell}$ is the projector onto the eigenspace of $Q_\ell$ with eighenvalue $\ell z$. Then, we have
\beq
	\bra e^{\lambda Q_\ell}\ket = \int \dd\mu(z)\, P(Q_\ell = \ell z) e^{\ell \lambda  z}
\eeq
where $\dd\mu(z)$ is the spectral measure. It is expected that, in many-body systems and for integrals of local observables like $Q_\ell$, the measure on $z$ becomes, in an appropriate sense, absolutely continuous with respect to the Lebesgue measure as $\ell\to\infty$. Then, \eqref{deltaf} naturally suggests that
\beq
	P(Q_\ell = \ell z) \asymp e^{-\ell I(z)}
\eeq
and by a saddle-point analysis, we find that $I(z)$ and $F(\lambda)$ are {\em Legendre-Frenchel transforms of each other}:
\beq
	F(\lambda) = \lambda z_*(\lambda) - I(z_*(\lambda)),\quad
	I'(z_*(\lambda)) = \lambda.
\eeq
The function $I(z)$ is the {\em large-deviation function}, characterising the exponentially suppressed probability that $Q_\ell$ takes a value that is extensively different from its most likely value, $\ell z_*(0) = \bra Q_\ell\ket$, as $\ell\to\infty$. See \cite{touchette2011basic} for an introduction to large-deviation theory.

It is in this sense that the scaling \eqref{Qelln} is called the {\em large-deviation scaling} of extensive observables.

\subsection{Extensivity from the dynamics: transformations} \label{ssectextdyn}

I now discuss the notion of extensive observables in the setup of Subsec.~\ref{ssectdyn}, with local observables defined with respect to a dynamics implemented via a local hamiltonian density $h$. Here, we consider a family of mutually local observables $\mathfrak L$, via \eqref{localset}.

In this setup, it is the algebraic structure that takes center stage. Consider again $Q_\ell$ as defined in Eq.~\eqref{Qelldensity}. Now we note that the large-$\ell$ limit exists in the following sense:
\beq\label{extensiveell}
	\lim_{\ell\to\infty} [Q_\ell,o(x)] = \int \dd x'\,[q(x'),o(x)].
\eeq
This is of course implicitly what I meant in \eqref{Ho}. Importantly, {\em this is still a local observable}, as shown in \eqref{odotlocal}. Therefore, instead of defining an extensive observable with a local density via an equivalence class \eqref{intoder} and its associated map $\mathsf Q$, Eq.~\eqref{mapQlocal}, it is here defined by its action on local observables: it is the limit $\ell\to\infty$ of local observables $Q_\ell$ that give rise, under adjoint action, to a non-zero, homogeneous {\em transformation of local observables}, which we may denote
\beq\label{adintlocal}
	\ad \int q  : \mathfrak L \to \mathfrak L,\quad
	\ad \int q\  (o)= \int \dd x\,[q(x),o(0)] .
\eeq
Denoting $\mathrm Q = \ad \int q$ this gives the action
\beq\label{transfoextensive}
	\mathrm Q(o) = \int \dd x\,[q(x),o(0)] .
\eeq

This is of course the natural framework to construct the flow of transformations associated to time evolution, via
\beq\label{Hdotad}
	\dot o = \ri\, \mathrm H(o),\quad \mathrm H = \ad \int h,
\eeq
and the ``global transformation'' generated by any $\mathrm Q$ is that solving
\beq\label{global}
	\frc{\p o_\lambda}{\p\lambda} = \ri\,\mathrm Q (o_\lambda)\quad \Leftrightarrow\quad
	o_\lambda = e^{\ri \lambda \mathrm Q}(o).
\eeq
The factor of $\ri$ is not essential -- this simply preserves any hermitian structure of the algebra of observables, if there is such a structure (this structure does not play a fundamental role in our discussion).

Clearly, the same conclusion as that of \eqref{intoder} holds: if $\ad \int q = \ad \int (q+a)$, then
\beq\label{aoart}
	\int \dd x\,[a(x),o(0)] = 0 \quad \forall\ o\in\mathfrak L
\eeq
and hence, by the conclusion reached in Eq.~\eqref{oop}, it must be that $a = \p b$ for some $b\in\mathfrak L$. That is, using the anti-derivative defined in Eq.~\eqref{opdef},
\beq\label{Adder}
	\ad\int a = 0 \Rightarrow a = \p b\quad\mbox{for local }
	b=\p^{-1} a\in\mathfrak L.
\eeq
Therefore, in $\ad \int q$, the operation $\ad$ really acts on $\int q$ as defined by the equivalence class up to adding total derivatives, Eq.~\eqref{intoder} (but here interpreted in the setup of Subsec.~\ref{ssectdyn}). We may write this as $\ad: \mathfrak L/\mathcal N' \to \End(\mathfrak L)$, with $\mathcal N'$ given by \eqref{Nder}, and as such, $\ad$ has trivial kernel.

Instead of the asymptotic derivative property \eqref{derstate} for the map $\mathsf Q$ associated to $\int q$ and defined by \eqref{mapQlocal}, we directly have a derivative property of the transformation $\mathrm Q = \ad \int q$ associated to $\int q$. By the Jacobi identity,
\beq
	\mathrm Q([o,o']) = 
	[\mathrm Q(o) , o']
	+
	[o , \mathrm Q(o')],
\eeq
and similarly on $oo'$ by the Leibniz rule
\beq\label{autoQ}
	\mathrm Q(oo') = 
	\mathrm Q(o) o'
	+
	o \mathrm Q(o').
\eeq
Thus, recalling that locality is preserved, e.g.~\eqref{oxt}, the {\em global transformation} generated by $\mathrm Q$ via \eqref{global} is an automorphism of the algebra of mutually local observables,
\beq\label{autoglobal}
	e^{\ri \mathrm Q} \in \Aut(\mathfrak L).
\eeq
 What's more, {\em these objects themselves form an algebra}. Indeed, it is clear that, as $\ad \int q$ is a transformation of $\mathfrak L$, we can ``multiply'' such transformations by simply acting with them from the rightmost to the left most. The important quantity is their commutator. With $\mathrm Q = \ad \int q$ and $\mathrm Q' = \ad \int q'$, a simple calculation gives us (here the commutator is with respect to the composition $\circ$)
\beq\label{QQal}
	[\mathrm Q,\mathrm Q']
	=
	\ad \int \mathrm Q(q')
	=
	-\ad \int \mathrm Q' (q).
\eeq
That is, the set of transformations of the space of local observables obtained by commutators with total integrals of local observables, forms a Lie algebra. This is with respect to the commutator of two transformations that is induced by the multiplication defined by composition of transformations. This comes from the fact that  the commutator of two total integrals still is a total integral, something  formally obtained by shifting integration variables, e.g.:
\beq
	\Big[\int \dd x\,q(x)\ ,\ \int \dd x'\,q'(x')\Big]
	=
	\int \dd x\,\int \dd x'\,[q(x+x'),q'(x)]
	=
	\int \dd x\,\iota_x\mathrm Q(q')
	=
	\int \dd x\,\mathrm Q(q')(x).
\eeq

At this point it is useful to check the consistency of our general theory. Because $\ad\int$ is linear, Eq.~\eqref{QQal} along with \eqref{Adder} imply that
\beq\label{Qqa}
	\mathrm Q(q') + \mathrm Q'(q) = \p a
\eeq
for some local $a$. Indeed, it is a simple matter to verify that it takes the form
\beq\label{aexplicit}
	a = \int_0^\infty \dd x\int_{-\infty}^0 \dd y\,\big([q(x),q'(y)] -
	[q(y),q'(x)]\big)
	=
	\int_{x>0,\,y<0\atop x\sim y\sim0}\dd x \dd y\,\big([q(x),q'(y)] -
	[q(y),q'(x)]\big)
\eeq
where we used \eqref{localset} for the last equality, making it manifest that $a$ is local and supported around the position 0.

With the algebraic structure on extensive observables at hand, we may define extensive conserved quantities as those transformations $\mathrm Q = \ad\int q$ such that
\beq\label{HQ}
	[\mathrm H,\mathrm Q]=0,
\eeq
where $\mathrm H = \ad \int h$ is the extensive observable associated with the Hamiltonian density $h$.

\subsection{Symmetries, Noether's theorem, height fields}
\label{ssectsymmetry}

It is worth making clear a notion that we will use extensively: that of {\em symmetry}. Recall the notion of an algebra $\mathfrak L_h$ of local observable Eq.~\eqref{local} associated to a Hamiltonian density $h$. We define a symmetry in general with respect to some subalgebra $\mathfrak L\subset \mathfrak L_h$ of local observables {\em that are not necessarily mutually local} -- as this will be used for mutually semi-local observables below.

\subsubsection*{Symmetries and internal symmetries}

First, a homogeneous automorphism of $\mathfrak L$ is a bijective linear map $\sigma:\mathfrak L\to\mathfrak L$,
\beq
	\sigma(co + c'o') = c\sigma(o)+c'\sigma(o'),\quad \sigma\circ\sigma^{-1}= \1,\quad c,c'\in\C,\ o,o'\in\mathfrak L
\eeq
with the following two properties
\beq
	\sigma(oo') = \sigma(o)\sigma(o'),\quad \sigma\circ\iota_x = \iota_x \circ\sigma\quad \mbox{i.e. }\sigma(o(x))  = (\sigma(o))(x).
\eeq
Note that ``homogeneity'' corresponds to the last property of the above equation.

For us, a {\em symmetry} is a homogeneous automorphism that  commutes with the Hamiltonian extensive observable, the time evolution generator \eqref{Hdotad},
\beq\label{symmetry}
	\sigma\circ \mathrm H = \mathrm H \circ \sigma
	\quad \Leftrightarrow\quad \int \ad (\sigma(h)-h) = 0
	\qquad \mbox{(symmetry)}.
\eeq
Clearly, if $\sigma$ is a symmetry, then $\sigma^{-1}$ also commutes with the Hamiltonian, so is also a symmetry. So, in our definition, a symmetry is an automorphism of the algebra of local observables that commutes with both time evolution and space translations.

A crucial notion will be that of {\em internal symmetry}. This is a symmetry $\sigma$ for which the second equation of \eqref{symmetry} is trivially solved:
\beq\label{internal}
	\sigma(h) = h
	\qquad \mbox{(internal symmetry)}.
\eeq

\subsubsection*{Noether's theorem}

A cool application of the above along with the theory of Subsec.~\ref{ssectextdyn} is a somewhat non-conventional {\em derivation of half of Noether's theorem}, in just a few steps.

Suppose that $\mathrm Q$ is an extensive conserved quantity, Eq.~\eqref{HQ}. Clearly, the global transformation generated by $\mathrm Q$, Eq.~\eqref{global}, preserves the dynamics, so it is a symmetry:
\beq
	\sigma_\lambda = e^{\ri \lambda\mathrm Q},\quad
	e^{\ri \lambda \mathrm Q}\circ \mathrm H\circ e^{-\ri\lambda \mathrm Q}
	=
	e^{\ri \lambda [\mathrm Q,\cdot]} \mathrm H = \mathrm H
\eeq
(where the commutator is with respect to the composition of maps). These form a one-parameter (abelian, connected) group,
\beq
	\sigma_\lambda \circ\sigma_{\lambda'} = \sigma_{\lambda+\lambda'}.
\eeq
If we have many extensive conserved quantities, then each generate their one-parameter group of symmetries. By the Baker-Campbell-Hausdorff formula along with the algebra of extensive observables \eqref{QQal}, their compositions is also of this type. So, we in general have a vector of extensive conserved quantities $\vec{\mathrm Q}$, with symmetries implemented as
\beq
	\sigma_{\vec\lambda} = e^{\ri\vec\lambda\cdot\vec{\mathrm Q}}
\eeq
where $\vec\lambda$ parametrise the group element, and with Lie algebra associated to the Lie group represented by the Lie algebra \eqref{QQal} of extensive conserved quantities:
\beq\label{groupalg}
	[\mathrm Q_i,\mathrm Q_j] = -\ri\sum_k f_{ij}^k\mathrm Q_k.
\eeq
 We call such a symmetry group, generated by one or more extensive observables, a {\em continuous local symmetry group}, with ``local'' meaning that this comes from extensive observables (that are built, after all, from local densities). As \eqref{HQ} along with \eqref{QQal} implies $\ad \int \mathrm H(q_i)=0$, this means, by \eqref{Adder}, that for $\dot q_i := \ri\,\mathrm H(q_i)$, we must have
\beq\label{conslawext}
	\dot q_i + \p_x j_i = 0\quad \mbox{i.e.}\quad
	\p_\mu j_i^\mu = 0
\eeq
for $j_i = -\p^{-1}\dot q_i$ a local current, that is, re-instating space-time dependence,
\beq\label{conteq}
	\p_t q_i(x,t) + \p_x j_i(x,t) = 0.
\eeq
This is the same argument as that leading to \eqref{conscurrent}, just more formallly expressed.

The half that's missing of Noether's theorem is the idea that {\em any one-parameter group of symmetries} is generated by $\ad \int q$ for some local $q$, or perhaps by the limit $\lim_{n\to\infty} \ad \int q_n$ with convergence in an appropriate sense (paralleling the formal construction in Eqs.~\eqref{cond1}, \eqref{cond2} of extensive observables), or perhaps by such objects in an {\em extended algebra of observables} that include $\mathfrak L$. Such a general theory has not been developed as far as I know, except for standard results in QFT \cite{zinn2021quantum,peskin2018introduction}.

Thus we have found that:

\st{Any continuous local symmetry group gives rise to one or more local conservation laws.}

\subsubsection*{Explicit currents for continuous ultra-local symmetry groups}

Suppose that the symmetries are internal, Eq.~\eqref{internal}:
\beq
	e^{\ri\vec\lambda\cdot\vec{\mathrm Q}}(h) = h\quad \forall\, \vec\lambda.
\eeq
This implies
\beq\label{Qih0}
	\mathrm Q_i(h) = 0.
\eeq
In this case, it turns out that it is often possible to choose conserved densities that satisfy the equivalent of \eqref{groupalg} at the level of densities:
\beq\label{groupQq}
	\mathrm Q_i(q_j) = -\mathrm Q_j(q_i)
	= -\ri\sum_k f_{ij}^k q_k.
\eeq
Note that this along with \eqref{QQal} implies \eqref{groupalg}.

I will refer to such a connected Lie group of ultra-local symmetries a {\em continuous ultra-local symmetry group}. Here ``ultra-local'' means that not only it is generated by extensive observables, but these extensive observables annihilate the Hamiltonian density, and, more generally, their densities transform into each other under the transformation generated by the group as per \eqref{groupQq} -- their densities are ``ultra-local''. 

The generators of a continuous ultra-local symmetry group satisfy \eqref{Qih0}. By the general relation \eqref{Qqa}, this means $\mathrm H(q_i) = \p a_i$ for local observables $a_i$ given by \eqref{aexplicit}. Hence, we find that Eq.~\eqref{Qih0} imply \eqref{conteq}
with an {\em explicit, general form for the current}, that is explicitly local (without the need to adjoin new observables) and expressed purely in terms of the conserved densities and the Hamiltonian density:
\beq\label{jiexplicit}
	j_i(x,t) = -\ri\int_x^\infty \dd y\int_{-\infty}^x \dd z\,\big([q_i(y,t),h(z,t)] -
	[q_i(z,t),h(y,t)]\big).
\eeq

\subsubsection*{Height fields}

Let $q$ and $j$ satisfy the continuity equation \eqref{conteq}
\beq
	\p_t q(x,t) + \p_x j(x,t) = 0.
\eeq
If $q$ and $j$ were ordinary functions of two variables, then this would imply, by the Poincar\'e lemma, the existence of a function $\varphi(x,t)$ such that $\p_x \varphi = q$, $\p_t \varphi = -j$. Here, we may simply define the {\em height field}, the observable
\beq\label{heightfield}
	\varphi(x,t) = -\int_x^\infty \dd y\,q(y,t).
\eeq
We may also use the definition $\int_{-\infty}^x \dd y\,q(y,t)$ instead, but the former is more convenient for my purposes in relation to my choice of definition of twist fields below, and the results are equivalent. If the total charge $Q=\int \dd x\,q(x)$ vanishes, $Q=0$, then $\varphi(x) = \p_x^{-1} q(x)$, Eq.~\eqref{opdef}, which is a local observables. But this extends the notion to the cases where $Q$ does not vanish.

The height field is homogeneous, $\iota_{x'} \varphi(x,t) = \varphi(x+x',t)$. But also, if $q$ is ultra-local (i.e. $\mathrm Q = \ad\int q$ is one of the generators of a continuous ultra-local symmetry group), then
\beqa
	[\varphi(x),h(x')] &=&
	-\int_x^{\infty} \dd y\,[q(y),h(x')]
	\to \lt\{\ba{ll}
	0 & (x'\ll x) \z
	-\int_{-\infty}^\infty \dd y\,[q(y),h(x')] & (x'\gg x)
	\ea\rt.\n
	&=&
	\lt\{\ba{ll}
	0 & (x'\gg x) \z
	-\mathrm Q(h)(x') & (x'\ll x)
	\ea\rt.\ 
	=\ 
	0 \quad(|x'-x|\to\infty) 
\eeqa
and therefore, by the definition \eqref{local}, the height field $\varphi$ {\em is a local observable}. By a similar calculation
\beq
	[\varphi(x),\varphi(x')] \to 
	\lt\{\ba{ll}
	-\int_x^{\infty}\dd y\,\mathrm Q(q)(y) & (x'\ll x) \z
	\int_{x'}^{\infty}\dd y\,\mathrm Q(q)(y) & (x'\gg x)
	\ea\rt.
	=
	\sgn(x'-x) \p^{-1}\mathrm Q(q)(\max(x,x')) \quad
	(|x-x'|\to\infty). 
\eeq
We used $[\mathrm Q,\mathrm Q]=0\Rightarrow \ad\int \mathrm Q(q)=0\Rightarrow \p^{-1}\mathrm Q(q)$ is local -- and explicitly given by \eqref{aexplicit}. But if $q$ is ultra-local, then \eqref{groupQq} holds, and therefore
\beq
	\mathrm Q(q) = 0,
\eeq
so by the definition \eqref{localset} the height field is self-local.

Oppositely, given a self-local observable $\varphi(x)$, if $\lim_{\ell\to\infty}(\varphi(\ell/2)-\varphi(-\ell/2))$ is an extensive observable, then $\varphi(x)$ is a height field, and $q(x,t)=\p_x \varphi(x,t)$ is an ultra-local conserved density with current $-\p_t \varphi(x,t)$. Here, it is {\em crucial} that
\beq
	\lim_{\ell\to\infty}Q_\ell,\quad Q_\ell = (\varphi(\ell/2)-\varphi(-\ell/2))
\eeq
be extensive, otherwise, if it is sub-extensive, the density $\p_x \varphi(x)$ gives rise to a trivial conserved quantity $\int \dd x\,q(x)=0$. Non-trivial extensivity may be in the algebraic sense $\mathrm Q\neq 0$ explained around Eqs.~\eqref{extensiveell}, \eqref{adintlocal},
\beq\label{extensivehf}
	\lim_{\ell\to\infty} Q_\ell\ \mbox{is extensive:}\ 
	\lim_{\ell\to\infty} [\varphi(\ell/2)-\varphi(-\ell/2),o]\ \mbox{is a local observable, generically non-zero},
\eeq
or in the sense of states $\mathsf Q\neq 0$, Eqs.~\eqref{cond1}, \eqref{cond2}. A local observable $\varphi$ such that $\lim_{\ell\to\infty}(\varphi(\ell/2)-\varphi(-\ell/2))$ is extensive is often referred to as a {\em topological observable}, because the difference between its values at points far apart, which is non-trivial and may be large, represents a topological property of the state or the system, independent of the path that connects these points (this is related to topological invariance of twist fields discussed below, Subsec.~\ref{ssecttopo}). An example is the sine-Gordon field $\varphi(x) = \phi(x)$, from the Hamiltonian Eq.~\eqref{HQFT}, and the position $q_x$ in the Toda chain, Eq.~\eqref{heightfieldtoda}, see Subsec.~\ref{ssectjordan}.

More generally, with height fields
\beq\label{hf}
	\varphi_i = -\int_x^{\infty}\dd x'\,q_i(x')
\eeq
and using \eqref{groupQq}, we have
\beq
	[\varphi_i(x),\varphi_j(x')] \to 
	\ri \sum_k f_{ij}^k \varphi_k(\max(x,x'))\quad (|x-x'|\to\infty).
\eeq 
We may say that they satisfy a close asymptotic algebra.

That is,
\st{
The height fields associated to a continuous ultra-local symmety group are self-local observables satisfying a close asymptotic algebra.
}

\section{Semi-locality and twist fields}\label{secttwist}

Now that we have a theory of locality and extensivity, we have the general framework to explain the concept of twist field.

The first occurence of a twist fied -- although it was not identified as such -- was in the two-dimensional Ising model of classical statistical mechanics at equilibrium \cite{kadanoff1971determination}. It was introduced in order to understand its second order phase transition controlled by the temperature. Looking for a counterpart to the order parameter, the average magnetisation, that takes a nonzero value in the low-temperature, ordered phase, the ``disorder parameter'' was introduce, that takes a nonzero value in the high-temperature, disordered phase. These parameters are related by the high-temperature / low-temperature Kramers-Wannier duality of the Ising model. The most important particularity of the disorder parameter was that it looked {\em highly non-local}: it is not an observable supported on a few sites of the lattice, and instead it {\em has a tail that start at one point -- what we deem to be the position of the observable -- and ends up at the boundary of the system, or at infinity if the system is on infinite volume.} A similar phenomenon happens with the Jordan-Wigner transformation, that allows us to write certain spin chains in terms of free fermions: some of the original spin variables possess such a tail, when written in terms of the free fermions (and {\em vice versa}), see e.g.~\cite{itzykson2006quantum}. Crucially, correlations are essentially invariant under changes of the shape of the tail -- {\em the tail is allowed to ``wiggle''!}

This is the main aspect of {\em semi-locality} of an observable: there is a tail that starts at the point where the observable is, and extends all to way to infinity, whose shape does not affect the results.

In the following, for pedagogical reason I will get rid of the details of these original constructions, and discuss twist fields in their ``purest'' form, in the general setting of the previous Section. I will come back to these examples in the next Section.

In this Section, the main context is that of quantum many-body physics, and in particular the locality principle based on commutators as discussed in Subsec.~\ref{ssectdyn}, and the extensivity principles of Subsec.~\ref{ssectextdyn}. I consider the algebra of observables under multiplications, not just the Lie bracket; but I still assume that\footnote{In Subsec.~\ref{ssectdyn}, our arguments only fully made sense for the Lie algebra of observables, because only for this structure the adjoining of observables with linear tails could be ``shown'' not to give rise to divergencies. A more accurate analysis would be necessary.} $\int \dd x\,o(x)=0\Rightarrow o(x) = \p o'(x)$ for local $o'(x)$.

\subsection{Exchange relations}\label{ssectexch}

Recall the notion of a local observable Eq.~\eqref{local}, and the notion of a family of mutually local observables Eq.~\eqref{localset}. We now extend the notion of a family of mutually local observables, to that of a family $\mathfrak L$ of {\em mutually  semi-local} observables. The notion presented here is based on ``internal symmetries'', Eq.~\eqref{internal}, and is the most widely used notion of twist fields. But we will see below how it can be extended in Subsec.~\ref{ssectcon}.

\subsubsection*{Mutually semi-local observables}

Let $\mathfrak L_h\subset \mathfrak A$ be a subspace of local observables for some Hamiltonian density $h$. By our definition \eqref{local}, in general, observables do not commute with each other at large separations. But this is so general that it is not that useful. It is useful to have some commutation or exchange properties at large separations. Here, we want something weaker than the set of mutually local observables, Eq.~\eqref{localset}. We say that $\mathfrak L$ is a subspace of mutually semi-local observables, if the following holds.

First, although we don't assume that all our local observables are mutually local, still in application there usually is a subspace of mutually local observables, so it is convenient to separate it out:
\beq
	\mathfrak L_0\subset\mathfrak L\quad\mbox{with}\  \mathfrak L_0\ \mbox{mutually local observables, Eq.~\eqref{localset}.}
\eeq
Second, we consider the twist fields:
\beq
	\mbox{a subset}\ \mathfrak T\subset \mathfrak L\ \mbox{(not necessarily a subspace).}
\eeq
This is a set of observables which we will denote $\mathcal T\in \mathfrak T$, which {\em does not contain any of the mutualy local observables} $\mathfrak T \cap \mathfrak L_0 = \emptyset$. It is such that by taking translations $\mathcal T(x) = \iota_x \mathcal T$, and multiplications and linear combinations amongst elements of $\mathfrak L_0\cup \{\iota_x \mathfrak T:x\in\R\}$, we get the full space of local observables $\mathfrak L$. We write this as
\beq
	\mathfrak L =  \langle \mathfrak L_0 \cup\mathfrak T \rangle
\eeq
where $\langle S\rangle$ is the space generated by the set $S$, via translations, multiplications and linear combinations.

The most important property is as follows. For every $\mathcal T\in\mathfrak T$, there exists an associated, non-trivial homogeneous automorphism $\sigma_{\mathcal T}\neq \1$ of $\mathfrak L$, which we require to also be a bijection $\sigma_{\mathcal T}:\mathfrak T\to \mathfrak T$, such that
\beq\label{exch}
	\mathcal T(x) \mathcal T'(x') = \lt\{\ba{ll}
	\sigma_{\mathcal T} (\mathcal T'(x'))\, \mathcal T(x) & (x'\gg x)\\
	\mathcal T'(x')\, \sigma_{\mathcal T'}^{-1} (\mathcal T(x)) & (x'\ll x)
	\ea\rt. \quad(\mathcal T,\mathcal T'\in \mathfrak T)
\eeq
and
\beq\label{exch2}
	\mathcal T(x) o(x') = \lt\{\ba{ll}
	\sigma_{\mathcal T} (o(x'))\, \mathcal T(x) & (x'\gg x)\\
	o(x')\mathcal T(x) & (x'\ll x)
	\ea\rt.\quad (\mathcal T\in \mathfrak T,\ o\in\mathfrak L_0).
\eeq
The automorphism $\sigma_{\mathcal T}$ is the {\em twist} of $\mathcal T$, and $\mathfrak T$ is our set of {\em twist fields}. Relations \eqref{exch}, \eqref{exch2} are {\em exchange relations}, instead of commutation relations. The result of the exchange is determined again only at large separations, but now it depends on the direction, $x'\gg x$ or $x'\ll x$, and the result is not just the observables again, but transformed observables. Eq.~\eqref{exch} for two twist fields $\mathcal T,\mathcal T'\in\mathfrak T$, Eq.~\eqref{exch2} for a twist field $\mathcal T\in\mathfrak T$ and an observable in the mutually local subspace $o\in\mathfrak L_0$, and the mutual locality relation \eqref{localset}, specify the full set of exchange relations in $\mathfrak L$. The algebraic and linear structure of the set $\mathfrak T$ is discussed in Subsec.~\ref{ssectalg}.

Note that the exchange relations \eqref{exch}, \eqref{exch2} occur at {\em large separations}. This is important especially in quantum chains, as we will make clear in Subsec.~\ref{ssectalg}. In QFT, exchange relations are usually written for $x>x'$ and $x<x'$; this can be seen as arising from taking the scaling limit. I will discuss this briefly in Subsec.~\ref{ssectqft}.

In \eqref{exch}, if we take $\sigma_{\mathcal T'}=\1$ then we recover \eqref{exch2}, so mutually local fields would be twist fields with trivial twist; be we do not call these ``twist fields'', to avoid ambiguity.

As $\sigma_{\mathcal T}$ is an automorphism of $\mathfrak L$, we can show from \eqref{exch}, \eqref{exch2} that for every $o\in\mathfrak L$,
\beq
	\mathcal T(x) o(x') = \sigma_{\mathcal T}(o(x'))\mathcal T(x)\quad (x'\gg x)
	\quad (\mathcal T\in \mathfrak T,\ o\in\mathfrak L).
\eeq
Also, by homogeneity $\sigma_{\mathcal T}(\mathcal T'(x)) = (\sigma_{\mathcal T}(\mathcal T'))(x) = \iota_x \sigma_{\mathcal T}(\mathcal T') \in \iota_x\mathfrak T$ is again a twist field translated to $x$, so that, in Eq.~\eqref{exch}, the right-hand side is again a product of twist fields at different positions.

If the algebra of observables has a locality-preserving anti-linear adjoint involution $o\mapsto o^\dag$, and if the symmetry preserves the adjoint structure, $\sigma_{\mathcal T}(o)^\dag = \sigma_{\mathcal T}(o^\dag)$, then one can show from the exchange relations \eqref{exch}, \eqref{exch2}, that $\sigma_{\mathcal T^\dag} = \sigma_{\mathcal T}^{-1}$:
\beq
	\mathcal T^\dag\quad \mbox{is a twist field with twist $\sigma_{\mathcal T}^{-1}$}.
\eeq
More generally, it is usually possible, given $\mathcal T$, to define a natural ``conjugate'' twist field with twist $\sigma_{\mathcal T}^{-1}$. This is usually denoted
\beq\label{Tbar}
	\b{\mathcal T}\quad \mbox{(conjugate to $\mathcal T$, with twist $\sigma_{\mathcal T}^{-1}$)}.
\eeq

Eqs.~\eqref{exch} and \eqref{exch2} are for twist fields with their ``tail on the right'': the twist field affects the other observable only if the other observable's position is far enough on the right of the twist field's position. There is, of course, also a definition of twist fields with their tail on the left, see Subsec.~\ref{ssectcon}. We will see below that, contrary to $\p^{-1}o$, Eq.~\eqref{opdef}, and to height fields, Eq.~\eqref{heightfield}, the tail is {\em not linear, but exponential}.

One may want to relax the condition that $\sigma_{\mathcal T}$ be a transformation of the set $\mathfrak T$; however all examples I know satisfy this condition.

I give the main explicit constructions of twist fields in Subsec.~\ref{ssectultra} - \ref{ssectriemann}, as well as a slightly more abstract construction in App.~\ref{appgen}. These are based on the presence of additional structures: ultra-local symmetries, spatially factorised Hilbert spaces, the fundamental field in the path integral formulation, Hilbert-space generating observables. The constructions are related to each other, but cover different, yet overlapping, contexts. In these various contexts, I show how to construct a local field that satisfies the exchange relations \eqref{exch}, \eqref{exch2}.

But before this, let me deduce the crucial properties of the automorphisms $\sigma_{\mathcal T}$, which follow from the above definitions.

\subsubsection*{Properties of the automorphisms}

The exchange relation \eqref{exch} implies important properties of the automorphisms $\sigma_{\mathcal T}$.

First, as $\mathfrak L$ is an algebra of local observable, it contains the Hamiltonian density $h$, which commutes with local observables at large separations, hence $h\in\mathfrak L_0$. Choosing $o=h$ in \eqref{exch2}, because $\mathcal T\in\mathfrak L$ is a local observable, we conclude that we must have
\beq\label{th}
	\sigma_{\mathcal T}(h) = h.
\eeq
That is, all automorphisms $\sigma_{\mathcal T}$ must be {\em symmetries}, which preserve not only the full Hamiltonian \eqref{symmetry}, but {\em which preserve the Hamiltonian density}. Thus, {\em every twist field is associated to a symmetry under which the Hamiltonian density is invariant} -- an internal symmetry (Subsec.~\ref{ssectsymmetry}). This simplifies the structure of twist fields, and probably underpins many of its properties, such as the simple analytic structure of twist field form factors in QFT (Subsec.~\ref{ssectqft}). But see Subsec.~\ref{ssectcon} for a discussion of {\em conical twist fields}, associated to the rotation symmetry, that is neither homogeneous nor an internal symmetry, and whose form factor expansion (Subsec.~\ref{ssectqft}) are simply related to those of branch-point twist fields   (Subsec.~\ref{ssectentanglement}), as explained in \cite{castro2018conical}.
\st{
For $\mathcal T$ to be a local observable, its tail must be associated to an internal symmetry: a symmetry which preserves the Hamiltonian density.
}

Second, analysing the two possible ways of fully exchanging $\mathcal T(x)\mathcal T'(x')\mathcal T''(x'')$, under the condition $x''\gg x'\gg x$ and the condition $x''\gg x\gg x'$, we obtain the following conditions on the algebra of these automorphisms: $\sigma_{\mathcal T'}\circ \sigma_{\sigma_{\mathcal T'}^{-1}(\mathcal T)}
	=
	\sigma_{\sigma_{\mathcal T}(\mathcal T')}\circ \sigma_{\mathcal T}
	=
	\sigma_{\mathcal T}\circ  \sigma_{\mathcal T'}$.
This holds if and only if $\sigma_{\mathcal T}$ satisfies the following consistency relation under compositions:
\beq\label{relationsym}
	\sigma_{\mathcal T}\circ \sigma_{\mathcal T'}\circ \sigma_{\mathcal T}^{-1} = \sigma_{\sigma_{\mathcal T}(\mathcal T')}.
\eeq
We will come back to what this implies about the symmetry transformation of twist fields themselves in Subsec.~\ref{ssectalg}, after we have developed various explicit constructions of twist fields.

\subsection{Twist fields as half-line products from ultra-local symmetries}\label{ssectultra}

The simplest and still the most useful example of twist fields are those emerging from the presence of ultra-local symmetries in quantum systems where the Hilbert space can be factorised into a tensor product of local Hilbert spaces associated to each spatial point:
\beq\label{Htensor}
	\mathcal H = \bigotimes_x \mathcal H_x.
\eeq
In discrete space, such as in a quantum chain, Eq.~\eqref{Htensor} can be made precise (in the infinite-volume limit) using the notion of quasi-local $C^*$ algebras \cite{brattelioperator1,brattelioperator2}. In continuous space, this is a somewhat more formal description, but still conceptually very useful. We assume homogeneity: each site to be isomorphic, $\mathcal H_x\simeq \mathcal H_{x'}$ for all $x,x'$.

In order to illustrate ultra-local symmetries and the twist field construction, I start with an example.

\subsubsection*{Example: spin twist fields in quantum chains}

Take a spin chain which conserves the total spin $S^3 = \sum_{x\in\Z} \sigma_x^3$, such as
\beq\label{heisenberg}
	H = \sum_{x\in\Z} h(x),\quad h(x) = \vec\sigma_{x+1}\cdot\vec\sigma_x + h\sigma_x^3.
\eeq
Then the corresponding global transformation operator $\prod_{x\in\Z}e^{\ri \lambda\sigma^3_x}$ for $\lambda\in\R$, which generates spin rotation automorphisms $\sigma_\lambda$ about the spin-z axis by similarity transformations
\beq\label{sigmalambdachain}
	\sigma_\lambda(o) =  \prod_{x\in\Z}e^{\ri \lambda\sigma^3_x}\, o\, \prod_{x\in\Z}e^{-\ri \lambda\sigma^3_x},
\eeq
commutes with the energy density -- it is an internal symmetry (here, as usual, $h=h(0)$):
\beq
	\prod_{x\in\Z} e^{\ri \lambda\sigma^3_x}\, h\, \prod_{x\in\Z} e^{-\ri \lambda\sigma^3_x} = h.
\eeq
We define a twist field, associated to this symmetry, as
\beq\label{Tspin}
	\mathcal T_\lambda(x) = \prod_{x'\geq x} e^{\ri \lambda\sigma^3_{x'}}.
\eeq
The tail of the twist field is, explicitly, the product of $e^{\ri \lambda\sigma^3_{x'}}$ for $x'\gg x$. This is the example \eqref{Tspinintro}.

Taking $o(x)$ any local operator around site $x$, supported on a finite number of sites no matter how large, we see that \eqref{exch2} holds, with
\beq\label{exsgima}
	\sigma_{\mathcal T} = \sigma_\lambda.
\eeq
Indeed, if $o(x)$ is supported on, say, the interval $[x-c,x+c]$ for some fixed $c>0$, then
\beqa
	\mathcal T_\lambda(x) o(x')
	&=&
	\prod_{x''\geq x} e^{\ri \lambda\sigma^3_{x''}} o(x')
	\prod_{x''\geq x} e^{-\ri \lambda\sigma^3_{x''}}\ 
	\mathcal T_\lambda(x)\n
	&=&
	\lt\{\ba{ll}
	\prod_{x''\in[x-c,x+c]} e^{\ri \lambda\sigma^3_{x''}} o(x')
	\prod_{x''\in[x-c,x+c]} e^{-\ri \lambda\sigma^3_{x''}}\ 
	\mathcal T_\lambda(x)
	& (x'\geq x+c)\\
	o(x')\mathcal T_\lambda(x)
	& (x'< x-c)
	\ea\rt.\n
	&=&
	\lt\{\ba{ll}
	\prod_{x''\in\Z} e^{\ri \lambda\sigma^3_{x''}} o(x')
	\prod_{x''\in\Z} e^{-\ri \lambda\sigma^3_{x''}}\ 
	\mathcal T_\lambda(x)
	& (x'\geq x+c)\\
	o(x')\mathcal T_\lambda(x)
	& (x'< x-c).
	\ea\rt.
\eeqa
We see that $x'\gg x$ is implemented, for this observable, as $x'\geq x+c$, etc.

In this example, we also have
\beq\label{TTcomm}
	\mathcal T_\lambda(x) \mathcal T_{\lambda'}(x') = \mathcal T_{\lambda'}(x') \mathcal T_\lambda(x),
\eeq
and this is in agreement with \eqref{exch} and the definition \eqref{exsgima}, giving $\sigma_{\lambda}(\mathcal T_{\lambda'}(x')) = \mathcal T_{\lambda'}(x')$.

Of course, the quantity \eqref{Tspin}, and similar quantities defined below and in the following sections, does not quite make sense in the infinite quantum spin chain on $\Z$, because the tail is infinite. Usually one considers multi-point correlation functions, where tails cancel, making them finite, as will be clear in the applications, Section \ref{sectappli}. For instance, with the natural conjugate twist fields \eqref{Tbar},
\beq
	\mathcal T_\lambda(x) \b{\mathcal T}_\lambda(x')
	=
	\prod_{y\in[x,x'-1]\cap \Z} U(y).
\eeq

\subsubsection*{Ultra-local symmetries in quantum chains}

Except for \eqref{TTcomm}, the above is otherwise a typical example: if a global transformation preserves the Hamiltonian density $h$, and is represented via a similarity transformation with by a product over $x$ of local operators,
\beq\label{Uultra}
	U = \prod_x U(x),\quad 
	\sigma_{U} (o) = U o U^{-1},\quad
	\sigma_U( h) = h,
\eeq
for some on-site operator $U(x)\in\1\otimes \cdots\otimes \1 \otimes\Aut(\mathcal H_x)\otimes \1\otimes \cdots\otimes \1$ (where $\Aut(\mathcal H_x)$ is on site $x$) with homogeneity $\iota_x (U(x')) = U(x'+x)$, then a natural twist field is the half-line product
\beq\label{defTU}
	\mathcal T_U(x) = \prod_{x'\geq x} U(x')
\eeq
and the associated homogeneous algebra automorphism is
\beq
	\sigma_{\mathcal T_U}  = \sigma_{U}.
\eeq
An internal symmetry that can be represented as \eqref{Uultra} is referred to as {\em ultra-local}.  On $\mathcal H$ we have a natural adjoint involution $^\dag$. It is preserved by the symmetry if $U$ is unitary, and in this case $\mathcal T_U^\dag = \mathcal T_{U^{-1}}$. More generally, we denote
\beq
	\b{\mathcal T}_U = \mathcal T_{U^{-1}}.
\eeq

Note that in general, it may be necessary to {\em augment the Hilbert space} to $\mathcal H^{\rm aug} = \mathcal H\otimes \mathcal H^{\rm aux}$ in order to define $U(x)$ that implements a required internal symmetry $\sigma\in\Aut(\mathcal L)$.

Intuitively the ``ultra-local'' part of it is the fact that the tail of the twist field exactly factorises into a product of operators acting on each point of space. Another important point is that, as mentioned, it is an internal symmetry, i.e.~it preserves the Hamiltonian density $h$, not just the full Hamiltonian $H$ (a symmetry). It is often the case that spatial factorisation of a symmetry and invariance of $h$ go hand-in-hand. Indeed, if $\sigma$ is a symmetry, i.e.~an automorphism of the algebra of observables that preserve the total Hamiltonian, then $\int \dd x\,(\sigma(h(x))-h(x)) = 0$. By our result \eqref{oop}, $\sigma(h(x))-h(x) = \p_x o(x) = o(x+1)-o(x)$ for some local $o(x)$, where $\p_x$ is the discrete derivative. Because of the derivative, $\p_x o(x)$ is supported on a larger interval than $o(x)$. Yet, if $\sigma = \sigma_U$, because of the spatial factorisation, $\sigma(h(x))-h(x)$ is supported on the same region as $h(x)$. So we would need $o(x)$ to be supported on a region that is strictly within the support of $h(x)$, and there are less observables like this. In addition, many are ``trivial'': any $h(x)$ of the form $\p_x a(x)$ where $a(x)$ is not invariant under $\sigma$ would do, with $o(x) = \sigma(a(x))-a(x)$, but then the Hamiltonian in trivial, $\int h = 0$. I do not know of non-trivial examples: it appears as though spatial factorisation of a symmetry implies that it is internal.

In general, there may be many ultra-local symmetries, i.e.~many on-site operators $U(x)$'s giving internal symmetries. Clearly, $\sigma_U\circ\sigma_{U'} = \sigma_{UU'}$. Importantly, contrary to the specific example \eqref{sigmalambdachain}, the automorphisms do not necessarily act trivially on twist fields themselves. Indeed, from the definition \eqref{defTU} we have
\beq\label{UT}
	\sigma_{U} (\mathcal T_{U'}(x))
	= \mathcal T_{UU'U^{-1}}(x)
\eeq
hence
\beq\label{exchU}
	\mathcal T_U(x) \mathcal T_{U'}(x') = \lt\{\ba{ll}
	\mathcal T_{UU'U^{-1}}(x')\, \mathcal T_U(x) & (x'\gg x)\\
	\mathcal T_{U'}(x')\, \mathcal T_{{U'}^{-1}UU'}(x) & (x'\ll x)
	\ea\rt.
\eeq
which is immediate to verify from \eqref{defTU}.  In this case, we may organise the space of local observables $\mathfrak L = \langle \mathfrak L_0\cup \mathfrak T\rangle$ as
\beq\label{spaceultra}
	\mathfrak L_0 = \{\mbox{all finitely supported operators}\},
	\quad
	\mathfrak T = \{\mathcal T_U\,:\, U\mbox{\ giving ultra-local symmetry}\}.
\eeq
For instance, in the model \eqref{heisenberg} at $h=0$, we may restrict to unitary $U(x)$ and take all
\beq\label{SU2ex}
	U(x) = U_{\vec\lambda}(x) = e^{\ri\vec\lambda\cdot\vec\sigma_x},
	\quad\vec\lambda\in\R^3.
\eeq

\subsubsection*{Ultra-local symmetries for spatially factorised Hilbert spaces}

The construction \eqref{defTU} for an ultra-local symmetry $\sigma_U$, Eq.~\eqref{Uultra}, is not restricted to quantum chains: it works also in models with continuous spatial coordinates $x\in\R$, including QFT. If the Hilbert space has (formally) the structure of a tensor product over Hilbert spaces associated to each spatial point \eqref{Htensor}, then we still have a notion of ultra-local symmetry \eqref{Uultra}, and for every such symmetry, we may define a twist field by \eqref{defTU}.

It is easiest to understand this with an example. For instance, suppose we are given a QFT with a real bosonic field $\phi(x)$ and its canonical conjugate $\pi(x)$, $[\phi(x),\pi(x')] = \ri \delta(x-x')$, and suppose that it has $\Z_2$ symmetry group, with symmetry $\sigma$ given by $\sigma(\phi(x))=-\phi(x)$, $\sigma(\pi(x))=-\pi(x)$. For instance, we may have the Hamiltonian density $h(x) = \frc12 ((\p_x \phi(x))^2 + \pi(x)^2) + V(\phi(x)^2)$. This symmetry satisfies \eqref{Uultra} ($\sigma = \sigma_U$) with the following operator\footnote{As mentioned, I use $\pi$ for $3.14159265...$, and $\pi(x)$ for the field.}:
\beq\label{Uboson}
	U = \exp\Big[\pi \int \dd x\,\phi(x)\pi(x)\Big]
\eeq
so that it is indeed an ultra-local symmetry with, formally, $U(x) = e^{\pi \dd x\,\phi(x)\pi(x)}$. Then,
\beq\label{Tboson}
	\mathcal T_U(x)
	=\exp\Big[\pi \int_{x}^\infty \dd x'\,\phi(x')\pi(x')\Big]
\eeq
and all arguments above go through. This is the example \eqref{Tbosonintro}, and another example of this type is \eqref{TZintro}. Example \eqref{Tfermionintro} is also of this type.

\medskip

We note that the set of all ultra-local symmetries $\{\sigma_U\}$ forms a representation of the group $\{U\}$ on local observables $\mathfrak L$, and in this representation, twist fields $\mathcal T_U$ transform amongst each other via the inner automorphisms of the group, Eq.~\eqref{UT}. Thus we find that:
\st{
For every group $\{U\}$ giving rise to ultra-local symmetries $\sigma_U$ of a model with spatially factorised Hilbert space, there is a family of twist fields $\{\mathcal T_U\}$ with twist $\sigma_{\mathcal T_U} = \sigma_U$, products of on-site operators on the half-line Eq.~\eqref{defTU}, which transform by inner group automorphisms, Eq.~\eqref{UT}.
}

\subsection{The standard exponential form and continuous symmetry  groups}\label{ssectexp}

In the examples above, the twist fields \eqref{Tspin} and \eqref{defTU} are written as products over local operators. However, as is clear in the example \eqref{Tboson}, it can also be written in exponential form, e.g.
\beq\label{tlambdaspin}
	\mathcal T_\lambda(x) = \exp\Big[
	\ri \lambda\sum_{x\geq x'}\sigma^3_{x'}\Big].
\eeq
Also, here the parameter $\lambda\in\R$ can be chosen at will, so we have a one-parameter family of twist fields.

The exponential form is very useful, and the fact that there is a one-parameter family is related to the presence of a {\em continuous ultra-local symmetry group}, for which twist fields can usually be constructed in exponential form. Let me discuss these aspects.

\subsubsection*{Ultra-local symmetries from local observables}

In general, we may not have the Hilbert space structure of a tensor product over spatial points, Eq.~\eqref{Htensor}; for instance, the Hilbert space of a gauge theory in QFT does not have this structure. This encourages us to {\em define an ultra-local symmetry} more generally, as an internal symmetry that can be implemented as
\beq\label{Uultragen}
	\sigma_{\mathrm Q} = \exp\Big({\ad\int q}\Big),
	\quad \sigma_{\mathrm Q}(o) = \exp\Big(\int \dd x\,[q(x),\cdot]\Big)o,
	\quad \sigma_{\mathrm Q}(h(x)) = h(x)
\eeq
where $\mathrm Q = \ad \int q$, for some {\em local observable} $q(x)\in\mathfrak L_0$. Here, we recall the notion of a transformation induced by an extensive observable by adjoint action, Eq.~\eqref{transfoextensive}. That is, an ultra-local symmetry is a symmetry that is generated by an extensive observable,
\beq
	\sigma_{\mathrm Q} = e^{\mathrm Q},
\eeq
and that preserves the Hamiltonian density,
\beq\label{eQhsym}
	e^{\mathrm Q}(h) = h.
\eeq
The ``ultra-local'' nature of the symmetry is now the combined fact that it is generated by a local observable and that it is internal, i.e.~it preserves the Hamiltonian density, not just the full Hamiltonian $H = \int \dd x\,h(x)$. We recall that the inverse automorphism also is an ultra-local symmetry,
\beq
	\sigma_{\mathrm Q}^{-1} = \sigma_{\mathrm -Q}.
\eeq

With such a symmetry, we define
\beq\label{defTUgen}
	\mathcal T_q(x)
	=\exp\Big[\int_{x}^\infty \dd x'\,q(x')\Big].
\eeq
This clearly generalises \eqref{Tspin} (with $q(x) = \ri \lambda \sigma^3_{\lfloor x\rfloor}$), \eqref{defTU} (with $q(x) = \log U(x)$, choosing a branch) and \eqref{Tboson} (with $q(x) = \pi\phi(x)\pi(x)$). Eq.~\eqref{defTUgen} is the {\em standard exponential form of a twist field}. In this form, the tail of the twist field is the integral over $x'\gg x$ of a local observable $q(x')$ within the exponential. It is also natual to define the conjugate twist field as
\beq
	\b{\mathcal T}_q = \mathcal T_{-q},
\eeq
and this agrees with the adjoint $\mathcal T^\dag$ if $q^\dag = -q$.

There is one important subtlety: the twist field is defined via the local observable $q$. However, we know that $\mathrm Q$ depends not on $q$, but on $\int q$, the equivalence class \eqref{intoder} up to total derivatives. Therefore the twist field \eqref{defTUgen} {\em is not uniquely fixed by the symmetry $\sigma_Q$}: for any two local observables $q,q'$ in the same class, the twist fields $\mathcal T_q$ and $\mathcal T_{q'}$ are {\em different}, even though the symmetry $\sigma_{\mathrm Q}$ is the same. This is discussed in Subsec.~\ref{ssectalg}. For now, we simply mention that choosing two elements of the same equivalence class $q,q' \in \int q$, we have $q' = q + \p o$ for some local $o$, and therefore
\beq\label{Tambiguity}
	\mathcal T_q(x) = e^{o(x)} \mathcal T_{q'}(x). 
\eeq
That is, {\em the tail is unchanged}, with only the local observable $e^{o(x)}$ being multiplied at the position of the twist field.

Note that we do not ask, in general, for $e^{\lambda\mathrm Q}$ to be a symmetry for every $\lambda$; for instance, in \eqref{Uboson}, the quantity $\exp\lambda\int \dd x\,\phi(x)\pi(x)$ does not in general commute with the Hamiltonian (it is not a symmetry), let alone the Hamiltonian density. The transformation $e^{\lambda\mathrm Q}$ makes sense as a transformation on the space of local observables, but it is generically not a symmetry, except at $\lambda=1$.

We may consider many $q_1,q_2,\ldots$, with associated $\mathrm Q_1,\mathrm Q_2,\ldots$, that have the property \eqref{eQhsym}. These are all ultra-local symmetries. The Baker-Campbell-Hausdorff formula along with \eqref{adintlocal} guarantees that (at least formally) the set generated by $\{e^{\mathrm Q_n}\}$ under compositions $\circ$ forms a group of ultra-local symmetries.

Let us check that \eqref{defTUgen} satisfies the twist-field properties Eqs.~\eqref{exch}, \eqref{exch2}. We first check the latter:
\beqa
	\mathcal T_q(x) o(x')
	&=&
	e^{\int_x^\infty \dd x''\,
	[q(x''),\cdot]} o(x')\mathcal T_q(x)\n
	&=&
	\sum_{n=0}^\infty \frc1{n!}\Big[\int_x^\infty \dd x''\,
	q(x''),\cdot\Big]^n o(x')\; \mathcal T_q(x)\n
	&=&
	\sum_{n=0}^\infty \frc1{n!}
	\lt\{\ba{ll}
	\Big[\int_{-\infty}^\infty \dd x''\,
	q(x''),\cdot\Big]^n o(x') & (x'\gg x)\\
	\delta_{n,0} o(x') & (x'\ll x)
	\ea\rt.\times \mathcal T_q(x)\n
	&=&
	\lt\{\ba{ll}
	e^{\mathrm Q} (o (x'))\mathcal T_q(x) & (x'\gg x)\\
	o(x') \mathcal T_q(x)& (x'\ll x)
	\ea\rt.
	\label{TUo}
\eeqa
where we have use commutativity at large separations, Eq.~\eqref{localset}. This is indeed \eqref{exch2}. For \eqref{exch}, we consider two twist fields $\mathcal T_q(x)$, $\mathcal T_{q'}(x')$, and use the property of similarity transformations to write two equivalent expressions:
\beq
	\mathcal T_q(x) \mathcal T_{q'}(x')
	=
	\exp\Big[\int_{x'}^\infty\dd y' \,e^{\int_x^\infty \dd y\,
	[q(y),\cdot]} q'(y')\Big]\mathcal T_q(x)
	=
	\mathcal T_{q'}(x')\exp\Big[\int_{x}^\infty \dd y\, e^{-\int_{x'}^\infty \dd y'\,
	[q'(y'),\cdot]} q(y)\Big].
\eeq
We use the first way of writing for $x'\gg x$, and the second way for $x'\ll x$, and again commutativity at large separations, Eq.~\eqref{localset}, to find
\beq\label{TUTU}
	\mathcal T_q(x) \mathcal T_{q'}(x')=
	\lt\{\ba{ll}
	e^{\mathrm Q} (\mathcal T_{q'}(x'))\mathcal T_q(x) & (x'\gg x)\\
	\mathcal T_{q'}(x')e^{-\mathrm Q'}(\mathcal T_q(x))& (x'\ll x)
	\ea\rt.
\eeq
which is the exchange relation \eqref{exch}. We can also verify that the action of the automorphism on a twist field gives another twist field; this is a consequence of the group properties of ultra-local symmetries. Specifically,
\beq\label{twisttransfo}
	e^{\mathrm Q}(\mathcal T_{q'})
	=\mathcal T_{e^{\mathrm Q}(q')},
\eeq
and the right-hand side is a twist field in exponential form, because $q'' = e^{\mathrm Q}(q')$ generates an ultra-local symmetry: it is a local observable and, with $\mathrm Q'' = \ad\int q''$, we have
\beq
	e^{\mathrm Q''} = e^{\mathrm Q} \circ e^{\mathrm Q'}\circ
	e^{-\mathrm Q}\quad\Rightarrow\quad
	e^{\mathrm Q''}(h) = h.
\eeq

Note that here, any set of ultra-local symmetries $\{e^Q\}$ still forms a group, but because of the ambiguity expressed in \eqref{Tambiguity}, the associated twist fields are not in one-to-one correspondence with group elements: every group element is associated with a class of twist fields. By \eqref{twisttransfo}, these classes, though, still transform as per the group's inner automorphisms, and this is a general property of twist fields, see Eq.~\eqref{transfoclass} below. Hence we only conclude that:
\st{
For every ultra-local symmetry of a quantum theory $\sigma_{\mathrm Q} = e^{{\mathrm Q}}$, and every associated local density $q$  such that ${\mathrm Q}=\int q$, there is a twist field $\mathcal T_{q}(x)$ with twist $\sigma_{\mathcal T_q} = \sigma_{\mathrm Q}$, written as the exponential of the half-line charge, defined in \eqref{defTUgen}. These transform amongst each other as \eqref{twisttransfo}.
}

\subsubsection*{Continuous ultra-local symmetry groups}

Recall that we asked for the symmetry to be internal, i.e.~to preserve the Hamiltonian density. This was essential for the twist field to be local, Eq.~\eqref{th}, and was part of our general definition of ultra-local symmetry, Eqs.~\eqref{Uultragen}, \eqref{eQhsym}.

Now we may ask for a stronger condition: not just that the symmetry preserves the Hamiltonian density, but that its {\em generator annihilates the Hamiltonian density},
\beq\label{Qhzero}
	\mathrm Q(h) = 0.
\eeq
This implies that\footnote{Recall that the factor ``$\ri$'' here is by convention: if $q(x)$ is hermitian, then the result is a unitary transformation, and it  is often the case that continuous symmetry groups are unitarily represented. But as is clear from the example \eqref{Uboson}, unitarity is not necessary for the general construction (in that example, which is {\em not} part of a continuous symmetry group, we could have used for $q$ the {\em anti-Hermitian} local observable $-\frc{\ri}2(\phi(x)\pi(x) + \pi(x)\phi(x))$).}
\beq
	e^{\ri\lambda\mathrm Q}(h) = h\quad \forall\, \lambda
\eeq
and these form a continuous ultra-local symmetry group, Subsec.~\ref{ssectsymmetry}. More generally, suppose we have a continuous ultra-local symmetry group: a vector of extensive conserved quantities $\vec{\mathrm Q}$ with $\vec{\mathrm Q}(h) = 0$, with symmetries implemented as $\sigma_{\vec\lambda} = e^{\ri\vec\lambda\cdot\vec{\mathrm Q}}$, and with \eqref{groupalg} and \eqref{groupQq}. Then we may then extend \eqref{defTUgen} to the full symmetry group, and specify a twist field for every group element:
\beq\label{defTUgencont}
	\mathcal T_{\vec\lambda}(x)
	=\exp\Big[\ri\int_{x}^\infty \dd x'\,\vec\lambda\cdot\vec q(x')\Big].
\eeq
Because of the ambiguity in choosing the local density explained around Eq.~\eqref{Tambiguity}, this, as a function of the group element parametrised by $\vec\lambda$ (and determined by the Lie algebra element $\vec\lambda\cdot \vec{\mathrm Q}$), appears not to be well defined! However, the choice \eqref{groupQq} is expected to fix the conserved densities uniquely (given $Q_i$'s) -- although I do not know of a proof of this. This is how we define the family of twist fields associated to a continuous ultra-local symmetry goup. The examples \eqref{Tspinintro} and \eqref{Tfermionintro} are of this type. The conjugate twist field is simply
\beq
	\b{\mathcal T}_{\vec\lambda} = \mathcal T_{-\vec\lambda}.
\eeq
We may also write these twist fields in terms of the height fields \eqref{hf},
\beq\label{defTUgenconthf}
	\mathcal T_{\vec\lambda}(x)
	=\exp\Big[-\ri\vec\lambda\cdot\vec \varphi(x)\Big].
\eeq
Note that $\vec\lambda\cdot\vec\varphi$ is itself a height field associated to an ultra-local density, hence it is self-local. The example \eqref{Tvertexintro} is of this type, see also Subsec.~\ref{ssectjordan}.

The results \eqref{TUo} and \eqref{TUTU} still holds, and now the automorphism acts well on the set $\{\mathcal T_{\vec \lambda'}\}$, with the adjoint group action, as implemented on $\vec\lambda'$, given by the matrix $A_{\vec \lambda} = e^{B(\vec\lambda)},\, B_{kj} (\vec\lambda) = \sum_i\lambda_if_{ij}^k$ ,
\beq\label{TA}
	e^{\ri\vec\lambda \cdot\vec{\mathrm Q}}(\mathcal T_{\vec \lambda'})
	=
	\mathcal T_{A_{\vec\lambda} \vec\lambda'}.
\eeq
The adjoint group action is induced by the Lie algebra relation implemented by \eqref{groupQq},
\beq
	e^{\ri\vec\lambda \cdot\vec{\mathrm Q}}(\vec\lambda' \cdot\vec{q} )
	=
	e^{\ri\vec\lambda \cdot\vec{\mathrm Q}}(\vec{q} )\cdot \vec\lambda'
	=
	(A_{\vec \lambda}^{\rm T}\vec{q})\cdot \vec\lambda'
	=
	\vec{q}\cdot (A_{\vec \lambda} \vec\lambda').
\eeq
An example is the SU(2) group in the Heisenberg chain, Eq.~\eqref{SU2ex}, where
\beq
	\vec q(x) = \vec\sigma_x,
\eeq
which indeed satisfy \eqref{groupQq} for the $su(2)$ algebra.

\medskip

Hence we can say:
\st{
For every continuous ultra-local symmetry group $\sigma_{\vec\lambda}=e^{\ri\vec\lambda\cdot\vec{\mathrm Q}}$, there is a family of twist fields $\mathcal T_{\vec \lambda}$ with twist $\sigma_{\mathcal T_{\vec \lambda}} = \sigma_{\vec\lambda}$, defined in \eqref{defTUgencont}, which transform by  the adjoint group action \eqref{TA}.
}

\subsection{Twist fields as defects in the path integral formulation}\label{ssecttwistpath}

In the previous Subsections I discussed twist fields in the operator formalism of quantum theory, constructing explicit examples of operators satisfying \eqref{exch} and \eqref{exch2}. However, one of the most important aspects of twist fields is their representation in the path integral formulation of quantum theory, and the understanding within statistical field theory that emerges from this. This gives a more intuitive picture of what twist fields are. I now discuss these aspects, referring to App.~\ref{apppath} for my notation for the path integral formulation.

Recall that the path integral formulation is based on the fundamental, generically vector-valued field $\psi(x)$ and its eigenstates $|\Psi\ket$.  In this context, it is simplest to take the mutally local observables as those which act as \eqref{olocalpath} on $|\Psi\ket$, and therefore which are represented as \eqref{olocalpathrep}, as functionals of the field configuration:
\beq\label{L0path}
	\mathfrak L_0 = \{\mbox{observables acting as \eqref{olocalpath}, represented as \eqref{olocalpathrep}}\}.
\eeq
These commute at any non-zero distance, so this is a strong notion of locality. It is simple to generalise the following discussion to weaker notions. Clearly, the fundamental field is part of the space of mutually local observables, $\psi\in\mathfrak L_0$.

Paralleling previous sections, I  assume that there is an {\em ultra-local symmetry for the path-integral formulation}: here, the symmetry is associated to an operator $U_A$ on the Hilbert space that is implemented by multiplication by an invertible matrix $A$ (which may be 1 by 1) on the possibly-vector-valued field configurations:
\beq\label{ultrapath}
	U_A|\Psi\ket = |A\Psi\ket,\quad
	\sigma_A(o) = U_AoU_A^{-1}, \quad
	\sigma_A(h) = h.
\eeq
In particular, from \eqref{fundeq} this means that the symmetry acts internally on the fundamental field $\psi(x)$,
\beq\label{sigmapsi}
	\sigma_A(\psi(x)) = A^{-1}\psi(x),
\eeq
and it gives rise to a transformation of path-integral representation of observables as
\beq\label{sigmaopath}
	\sigma_A(o(x,t)) \stackrel{\text{path integral}}\to o[A^{-1}\Psi](x,t),\quad
	o\in \mathfrak L_0.
\eeq
By the construction of the path integral, the Lagrangian density is invariant under this transformation,
\beq\label{invaction}
	\mathcal L[A\Psi](x,t) = \mathcal L[\Psi](x,t).
\eeq
From \eqref{olocalpathrep} one also has the ``functional locality property''
\beq\label{oPsidep}
	\frc{\delta o[\Psi](x,t)}{\delta\Psi(x',t')} = 0 \quad \forall\;|x- x'|,\, |t-t'|>\ep
\eeq
for all $\ep>0$.

\subsubsection*{Twist fields as temporal defects}


Starting from this, I now provide an intrinsic way of representing twist fields in the path integral formulation. Note that \eqref{vacpi} does not allow for replacing $o_i$ with twist fields, because all observables lie in $\mathfrak L_0$ -- are mutually local. This was for simplicity, it is possible to include twist fields from the outset in developing the path integral formulation from operators. Instead, I will extend the formulation to include twist fields, as follows.

The idea is simple. Within the path integral, on the right-hand side of \eqref{vacpi}, the ordering of observables is unimportant, as these are no longer operators but functions of the field configuration $\Psi$. This correctly represents products of operators on the left-hand side, as long as they are time-ordered. Yet, exchange relations \eqref{exch}, \eqref{exch2} express relations between the product of operators {\em at the same time, but in two different orderings}. How do we express this  under the constraint of time ordering? We simply have to take product of observables at {\em infinitesimal, positive and negative time differences}: this is because there is continuity in time for product of operators:
\beq
	\lim_{t'\searrow t}\mathsf T\big[\mathcal T(x,t)o(x',t')\big]
	=o(x',t)\mathcal T(x,t),\quad
	\lim_{t'\nearrow t}\mathsf T\big[\mathcal T(x,t)o(x',t')\big]
	=\mathcal T(x,t)o(x',t).
\eeq
Thus, in space-time, a cut appears along the branch of the twist field implementing the relation between positive and negative time differences -- this is a {\em temporal defect} in space-time, emanating from the position of the twist field.

We define {\em the twist field $\mathcal T_A(x,t)$ associated to an ultra-local symmetry \eqref{ultrapath} for the path integral formulation} by how they modify the path integral:
\beq\label{Tpath}
	\frc{\bra \vac|\mathsf T\big[ \mathcal T_A(x,t) o_1(x_1,t_1)\cdots o_n(x_n,t_n)\big]|\vac\ket}{\bra \vac| \mathcal T_A(x,t)|\vac\ket}
	=
	\frc{\int_{\mathcal C_{x,t}^A} [\dd\Psi] \,e^{\ri S_{x,t}[\Psi]}
	o_1[\Psi](x_1,t_1)\cdots o_n[\Psi](x_n,t_n)}{
	\int_{\mathcal C_{x,t}^A} [\dd\Psi] \,e^{\ri S_{x,t}[\Psi]}}.
\eeq
Here $\mathcal C_{x,t}^A$ is the set of configurations on $\R^2\setminus ([x,\infty),t)$, with the following {\em jump condition} through the tail of the twist field\footnote{Recall, Footnote \ref{ft}, how in the usual path integral \eqref{vacpi}, the fields $\Psi(x,t)$ are continuous, but not necessarily their derivatives. This is why it is not necessary to impose conditions on derivatives in \eqref{twistcut}. If derivatives are continuous, then likewise, continuity of derivatives must be imposed in \eqref{twistcut}.}:
\beq\label{twistcut}
	\mathcal C_{x,t}^A : \Psi(x',t+ 0^+) = \lt\{
	\ba{ll}
	A \Psi(x',t - 0^+) & (x'> x)\\
	\Psi(x',t - 0^+) & (x'< x),
	\ea\rt.
\eeq
and the action avoids the tail of the twist field, where $\Psi(x',t')$ is ill-defined because of the jump condition \eqref{twistcut}:
\beq\label{Savoid}
	S_{x,t}[\Psi] = \int_{\R^2\setminus  ([x,\infty),t)} \dd x'\dd t'\,\mathcal L[\Psi](x',t')
\eeq
where we denote $([x,\infty),t) = \{(x',t):x'\in [x,\infty)\}$.  See Fig.~\ref{figcut}.
\begin{figure}
\bc\includegraphics[width=0.5\textwidth]{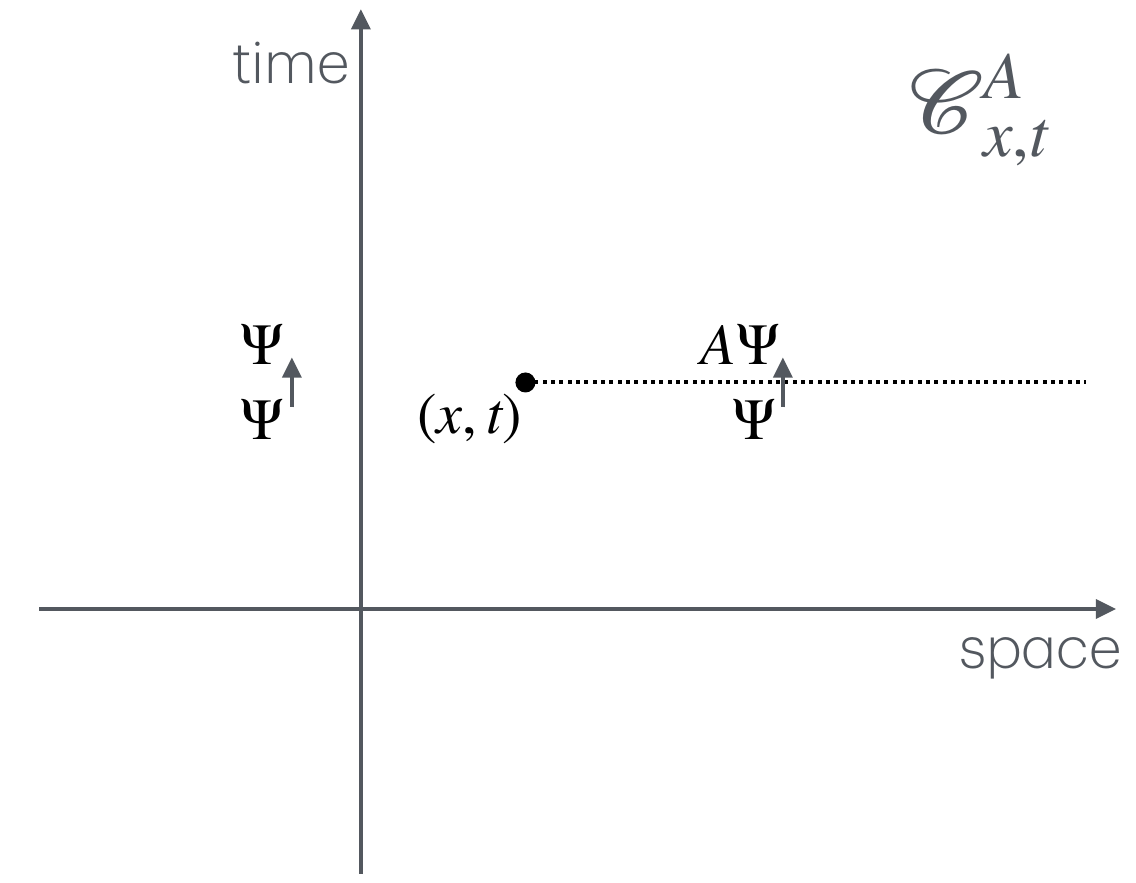}\ec
\caption{The cut condition \eqref{twistcut} in the path integral formulation of the twist field.}
\label{figcut}
\end{figure}

In fact, because in \eqref{twistcut} we also impose the condition of continuity on the spatial region at time $t$ away from the tail, we can also use the action $S_t[\Psi]=\int_{\R^2\setminus  (\R,t)} \dd x'\dd t'\,\mathcal L[\Psi](x',t')$ which avoids the {\em full time slice} $(\R,t)$ -- as the only effect of the infinitesimal-looking missing part of the spacetime integral on $((-\infty,x),t)$ is to impose continuity.

We define the conjugate twist field as
\beq
	\b{\mathcal T}_A = \mathcal T_{A^{-1}},
\eeq
and this is $\b{\mathcal T}_A = \mathcal T_A^\dag$ if and only if $U_A^\dag = U_{A^{-1}} = U_A^{-1}$.

Similarly, multiple insertions of twist fields $\mathcal T_A(x,t),\mathcal T_{A'}(x',t'),\ldots$ are implemented by multiple defect conditions $\mathcal C_{x,t}^A,\mathcal C_{x',t'}^{A'},\ldots$ simultaneously imposed on the path integral. Multi-point functions of twist fields are then related to {\em partition functions} with such connectivities. When these are at the same time, the discontinuity conditions get multiplied (see below), and  for instance,
\beq\label{twistpartition}
	\bra \vac|\mathcal T_A(x,t)  \b{\mathcal T}_{A}(x',t)|\vac\ket
	\propto
	\int_{\mathcal C_{x\to x',t}^A} [\dd\Psi] \,e^{\ri S_{x\to x',t}[\Psi]}
	=:Z(\mathcal C_{x\to x',t}^A)
\eeq
(where no time-ordering is required for this equal-time correlation) with
\beq\label{twistcut2}
	\mathcal C_{x\to x',t}^A : \Psi(x'',t+ 0^+) = \lt\{
	\ba{ll}
	A \Psi(x'',t - 0^+) & (x''\in( x,x'))\\
	\Psi(x'',t - 0^+) & \mbox{(otherwise)}
	\ea\rt.
\eeq
and $S_{x\to x',t}[\Psi]$ avoids the finite cut on $([x,x'],t)$; again, we can also use $S_t[\Psi]$ instead.

The tail of the twist field is explicit in \eqref{Tpath}, but contrary to Subsec.~\ref{ssectultra}, where it was a product of operators, and Subsec.~\ref{ssectexp}, where it was the exponential of an integral of a local observable, here it is a {\em temporal defect}, implemented as a cut in space-time, through which conditions on the discontinuity of the field configurations are imposed.

In \eqref{Tpath}, the normalisation of the twist field is factored out by dividing by its vacuum expectation value (VEV). The normalisation is, of course, irrelevant for the exchange relations \eqref{exch}, \eqref{exch2}. In relativistist QFT, the VEV can be fixed by the square-root of the large-distance saturation value of \eqref{twistpartition} under the requirement of the so-called ``conformal normalisation'' at short distances. I will discuss  VEV in Sect.~\ref{sectappli}. In any case, the physically meaninful part of \eqref{twistcut2} is how it depends on $x-x'$ (in this vacuum expectation value, it only depends on $x-x'$).

Let me show that \eqref{Tpath} gives rise to the exchange relations, along with the inner-automorphism transformation property
\beq\label{transTpath}
	\sigma_A(\mathcal T_{A'}) = \mathcal T_{AA'A^{-1}}.
\eeq
For this purpose, I use the fact that, as mentioned, in this formulation, equal-time products of operators are obtained by setting time differences to $\pm 0^+$, in such a way as to reproduce the order of operators in the product by the time ordering $\mathsf T$.

Choosing $t_1 = t-0^+$ in \eqref{Tpath} and placing all other operators appropriately according to their time order, we get
\beqa
	\frc{\bra\vac|\cdots \mathcal T_A(x,t)o_1(x_1,t) \cdots|\vac\ket}{
	\bra \vac| \mathcal T_A(x,t)|\vac\ket}
	&=&
	\frc{\int_{\mathcal C_{x,t}^A} [\dd\Psi] \,e^{\ri S_{x,t}[\Psi]}
	o_1[\Psi](x_1,t-0^+)\cdots }{
	\int_{\mathcal C_{x,t}^A} [\dd\Psi]\,e^{\ri S_{x,t}[\Psi]} }\n
	&=&
	\lt\{\ba{ll}
	\frc{\int_{\mathcal C_{x,t}^A} [\dd\Psi] \,e^{\ri S_{x,t}[\Psi]}
	o_1[A^{-1}\Psi](x_1,t+0^+)\cdots }{
	\int_{\mathcal C_{x,t}^A} [\dd\Psi] \,e^{\ri S_{x,t}[\Psi]}}
	& (x_1>x) \z
	\frc{\int_{\mathcal C_{x,t}^A} [\dd\Psi] \,e^{\ri S_{x,t}[\Psi]}
	o_1[\Psi](x_1,t+0^+)\cdots }{
	\int_{\mathcal C_{x,t}^A} [\dd\Psi]\,e^{\ri S_{x,t}[\Psi] }}
	& (x_1<x)
	\ea\rt.
	\n
	&=&
	\lt\{\ba{ll}
	\frc{\bra\vac|\cdots \sigma_A(o_1(x_1,t)) \mathcal T_A(x,t)\cdots|\vac\ket}{
	\bra \vac| \mathcal T_A(x,t)|\vac\ket}
	& (x_1>x) \z
	\frc{\bra\vac|\cdots o_1(x_1,t) \mathcal T_A(x,t)\cdots|\vac\ket}{
	\bra \vac| \mathcal T_A(x,t)|\vac\ket}
	& (x_1<x)
	\ea\rt.\n
\eeqa
where in the second step we use \eqref{twistcut} and \eqref{oPsidep}, in the last, \eqref{sigmaopath}. This shows \eqref{exch2}.

Now consider the product of twist fields $\mathcal T_A(x,t)$ and $\mathcal T_{A'}(x',t)$, at equal time. We can use the action $S_t[\Psi]$, where the part taken away does not depent on $x,x'$, and thus we only have to consider the defect conditions  \eqref{twistcut}. We combine them by putting them at slightly separates times, in two different ways:
\beq
	\mathcal T_A(x,t)\mathcal T_{A'}(x',t) \quad \to \quad \mathcal C_{x,t+0^+}^A\cup \mathcal C_{x',t}^{A'}: \Psi(y,t+ 0^+) = \lt\{
	\ba{ll}
	AA' \Psi(y,t - 0^+) & (y> x,x')\\
	A \Psi(y,t - 0^+) & (x'>y> x)\\
	A' \Psi(y,t - 0^+) & (x>y> x')\\
	\Psi(y,t - 0^+) & (x,x'<y)
	\ea\rt.
\eeq
and
\beq
	\mathcal T_{A'}(x',t)\mathcal T_{A'}(x',t) \quad \to \quad\mathcal C_{x,t}^A\cup \mathcal C_{x',t+0^+}^{A'}: \Psi(y,t + 0^+) = \lt\{
	\ba{ll}
	A'A \Psi(y,t - 0^+) & (y> x,x')\\
	A \Psi(y,t - 0^+) & (x'>y> x)\\
	A' \Psi(y,t - 0^+) & (x>y> x')\\
	\Psi(y,t - 0^+) & (x,x'<y).
	\ea\rt.
\eeq
We then observe that $\mathcal C_{x,t+0^+}^A\cup \mathcal C_{x',t}^{A'} = \mathcal C_{x,t}^A\cup \mathcal C_{x',t+0^+}^{AA'A^{-1}}$ if $x'>x$, as agreement is found in the three cases $y> x'>x$, $x'>y> x$ and $x,x'<y$, and likewise $\mathcal C_{x,t+0^+}^A\cup \mathcal C_{x',t}^{A'} = \mathcal C_{x,t}^{{A'}^{-1}AA'}\cup \mathcal C_{x',t+0^+}^{A'}$ if $x>x'$. This shows \eqref{exch} with \eqref{transTpath}.

\medskip
Therefore, we have found that:
\st{
For every group $\{A\}$ of ultra-local symmetries $\sigma_A$ for the path integral formulation, there is a family of twist fields $\mathcal T_A$ with twist $\sigma_{\mathcal T_A} = \sigma_A$ implemented as temporal defects, Eq.~\eqref{Tpath}, which transform by inner automorphisms of the group, Eq.~\eqref{transTpath}.
}

\subsection{Path integrals on resonant Riemann surfaces, monodromy and unwinding}\label{ssectriemann}

Let me take the same context as that of Subsec.~\ref{ssecttwistpath}.

A crucial representation of twist fields in the path integral formulations is in term of Riemann sheets. In order to see how this works, consider again the path-integral definition \eqref{Tpath}. Because of the invariance \eqref{invaction} of the Lagrangian density, we may define infinitely-many copies of the field configuration
\beq
	\ldots,\Psi_{-2},\Psi_{-1},\Psi_0,\Psi_1,\Psi_2,\ldots,\quad \Psi_0=\Psi
\eeq
under the ``resonant'' conditions
\beq\label{resonant}
	\Psi_{n}(x',t') =  A\Psi_{n+1}(x',t'),
\eeq
and replace $S[\Psi]\to \lim_{N\to\infty}\frc1{2N+1}\sum_{n=-N}^N S[\Psi_n]$ in the numerator and denominator, without changing the result. This is useful, because the resonant condition implies {\em twisted continuity conditions}: using $\Psi_{n+1}(x',t+0^+) = A\Psi_{n+1}(x',t-0^+)$ for $x'>x$, we get
\beq\label{riemannwinding}
	\Psi_{n+1}(x',t+0^+) = \lt\{\ba{ll}
	\Psi_{n}(x',t-0^+)& (x'> x)\\
	\Psi_{n+1}(x',t-0^+)& (x'<x).
	\ea\rt.
\eeq
We thus construct the field $\hat\Psi(\rz') = \Psi_n(x',t')$, which is continuous on the infinitely-many sheeted Riemann surface $\mathcal R_{x,t} = \cdots\times\R^2\times \R^2\times \R^2\times \cdots/\sim$, with the point $(x,t)$ identified, $(x,t,n)\sim (x,t,n+1)\,\forall\,n$, and with continuity as \eqref{riemannwinding}. Here $\rz' = (x',t',n)$ is a position on this Riemann surface. Because the field configuration is continuous on $\mathcal R_{x,t}$, we may forgo the restriction of avoiding the tail in the action \eqref{Savoid} if we interpret it as being integrated on the Riemann surface. Thus, normalisating by the ``volume'' of the Riemann surface ${\rm Vol}(\mathcal R_{x,t}) = 2N+1$ -- its number of sheets, taken to infinity -- we define the measure
\beq
	\int\sd^2 \rz' =  \lim_{N\to\infty} \frc1{{\rm Vol}(\mathcal R_{x,t})}\sum_{n=-N}^N\int\dd x'\dd t'
\eeq
and obtain
\beq
	S[\Psi] = \lim_{N\to\infty}\frc1{2N+1}\sum_{n=-N}^N S[\Psi_n]
	=
	S[\hat\Psi] := \int_{\mathcal R_{x,t}} \sd^2 \rz'\,L[\hat\Psi](z').
\eeq
The path integral is now for {\em field configurations on this resonant Riemann surface $\mathcal R_{x,t}^A$ }
\beq\label{Tpathriemann}
	\frc{\bra \vac|\mathsf T\big[ \mathcal T_A(x,t) o_1(x_1,t_1)\cdots o_n(x_n,t_n)\big]|\vac\ket}{\bra \vac| \mathcal T_A(x,t)|\vac\ket}
	=
	\frc{\int_{\mathcal R_{x,t}^A} [\dd\h\Psi] \,e^{\ri S[\h\Psi]}
	o_1[\Psi_0](x_1,t_1)\cdots o_n[\Psi_0](x_n,t_n)}{
	\int_{\mathcal R_{x,t}^A} [\dd\h\Psi] \,e^{\ri S[\h\Psi]}}
\eeq
where $\h\Psi$ has the continuity structure set by $\mathcal R_{x,t}$, but with the twist-dependent resonant condition \eqref{resonant} -- that is, the field configurations on different sheets are not independent. A similar construction can be made for multiple insertions of twist fields. For instance for $\mathcal T_A(x,t)\b{\mathcal T}_A(x',t)$ the jump condition \eqref{twistcut2} leads to the Riemann surface $\mathcal R_{x\to x',t}^A$ illustrated in Fig.~\ref{figresonant}.
\begin{figure}
\bc\includegraphics[width=0.5\textwidth]{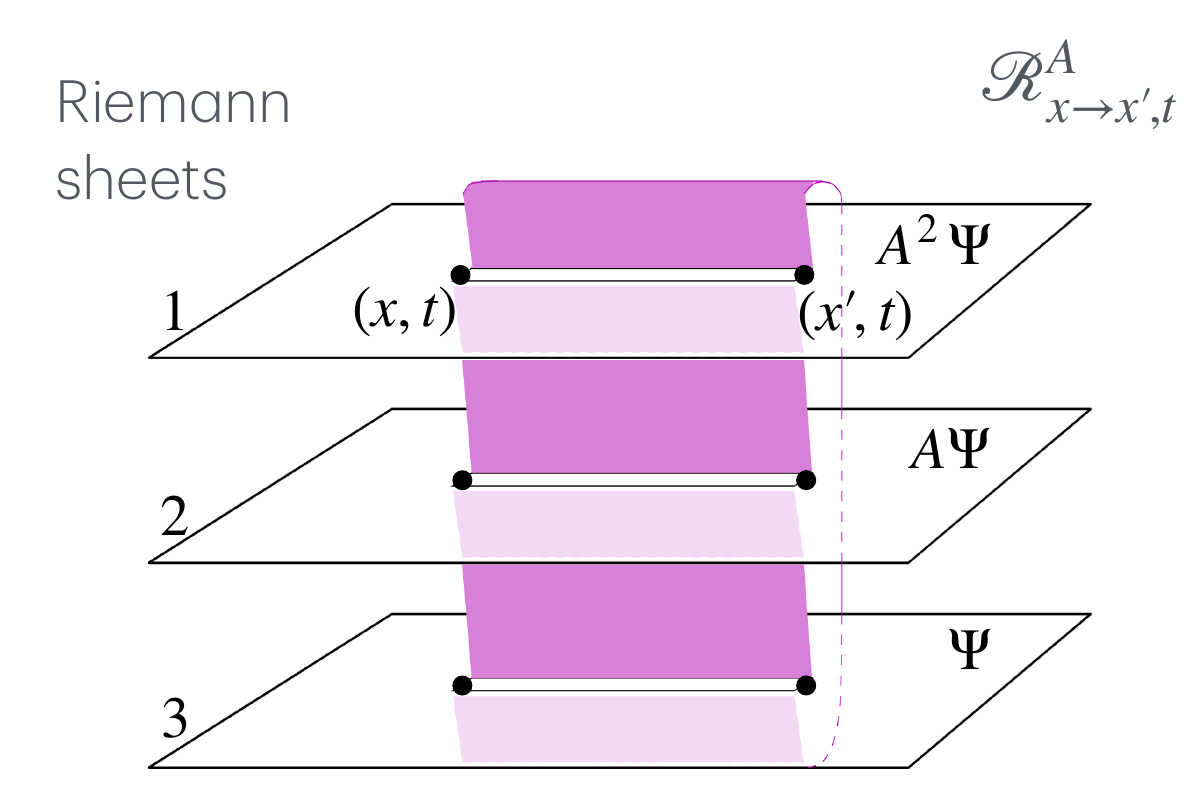}\ec
\caption{The field $\h\Psi$ lies on a resonant Riemann surface. This is the case of an idempotent symmetry group with $A^3=\1$, where $\h\Psi$ takes values $A^{3-n}\Psi$ on sheet $n$. Picture adapted from \cite{cardy2008form}.}
\label{figresonant}
\end{figure}

Note how in \eqref{Tpathriemann} the observables depend on the field configuration only on its original sheet $\Psi_0$. This geometric construction makes it clear, though, that performing a continuation of the corelation function in the space-time coordinates along any closed path that surrounds the point $x,t$, the value obtained {\em does not necessarily come back to its original value}. With $s\in[0,1]\mapsto \gamma(s)\in\R^2,\,\gamma(0) = \gamma(1) = (x_1,t_1)$ a path {\em winding counter-clockwise once around $(x,t)$}, for instance,
\beq\label{monodromy}
	\frc{\bra \vac|\mathsf T\big[ \mathcal T_A(x,t) o_1(x_1,t_1) \cdots\big]|\vac\ket}{\bra \vac| \mathcal T_A(x,t)|\vac\ket}\Big|_{(x_1,t_1):\stackrel{\gamma} \circlearrowleft} =
	\frc{\bra \vac|\mathsf T\big[ \mathcal T_A(x,t) \sigma_A(o_1(x_1,t_1)) \cdots \big]|\vac\ket}{\bra \vac| \mathcal T_A(x,t)|\vac\ket}
\eeq
where $o_1$ and $\cdots$ are local observbales in $\mathfrak L_0$.
Therefore, as a function of each coordinates $x_i,t_i$, the correlation function \eqref{Tpathriemann} lives on the Riemann surface $\R_{x,t}$, with {\em monodromy} \eqref{monodromy}. See Fig.~\ref{figmonodromy}.
\begin{figure}
\bc\includegraphics[width=0.6\textwidth]{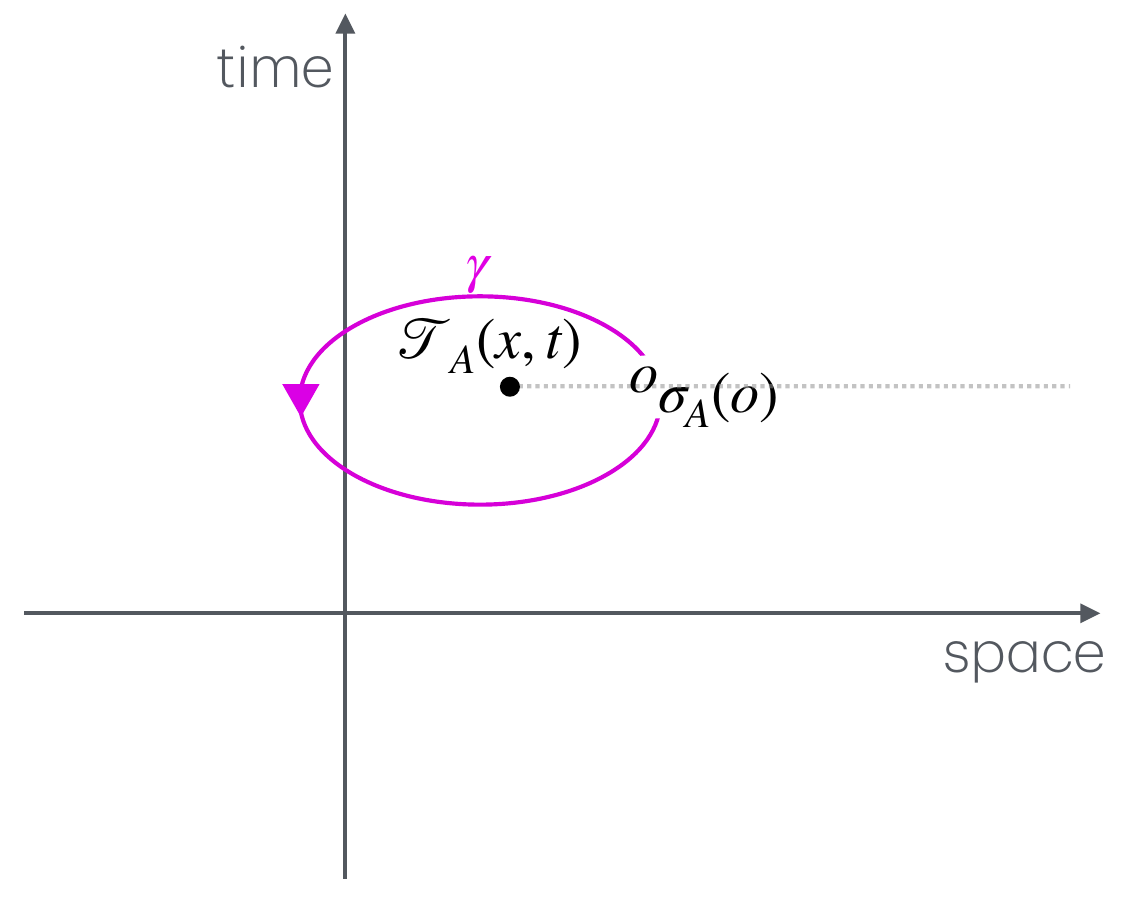}\ec
\caption{The monodromy property \eqref{monodromy}.}
\label{figmonodromy}
\end{figure}

The monodromy property \eqref{monodromy} is also valid in the Euclidean formulation, and at non-zero temperature, where the Riemann surface is composed of sheets that have the topology of cylinders, connected cyclically. It is in fact expected to be a completely general property: it should hold in most physical states, and, in models on continuous space-time, under the formulations of Subsec.~\ref{ssectultra}, \ref{ssectexp} and App.~\ref{appgen} as well.

If the symmetry is idempotent,
\beq
	A^N = \1\quad \mbox{for some $N>0$},
\eeq
then we can reduce the Riemann surface to an $N$-sheeted Riemann surface, and keep $n=0,1,\ldots,N$ with ${\rm Vol}(\mathcal R_{x,t}) = N$ finite.

\medskip
Hence we conclude that:
\st{
Every twist field $\mathcal T_A$ for the path integral formulation, can be expressed as a resonant path integral on an infinitely-many sheeted Riemann surface, Eq.~\eqref{Tpathriemann}. If $A^N=\1$, this can be restricted to an $N$-sheeted Riemann surface. In general, twist fields in continuous space-time give rise to the monodromy \eqref{monodromy}.
}

\subsubsection*{Relation to standard exponential form via unwinding gauge transformations}

The formulation \eqref{Tpathriemann} allows us to make the connection between the path-integral formulation of twist fields, and the standard exponential form \eqref{defTUgen}, where the twist is generated by an extensive observable.

In many applications to field theory, the exponent in the standard exponential form is made of local observables $q\in\mathfrak L_0$ as per \eqref{L0path}. Then, applying the techniques of App.~\ref{apppath}, we obtain, in principle, a twist field representation in the form
\beq\label{Tpathexp}
	\bra \vac|\mathsf T\big[ \mathcal T(x,t) o_1(x_1,t_1)\cdots o_n(x_n,t_n)\big]|\vac\ket
	=
	\int [\dd\Psi] \,e^{\ri S[\Psi]+\int_x^\infty \dd x\,q[\Psi](x',t)}o_1[\Psi](x_1,t_1)\cdots o_n[\Psi](x_n,t_n).
\eeq
This is, in fact, a slight abuse of notation, as the functional $q[\Psi](x',t)$ appearing in the exponent, is not necessarily the same as that which would appear by replacing, say, $o_1$ by $q$, because of ordering problems. But what this shows is that the twist field can be seen as {\em an extended source in the action, on the one-dimensional space-time region $([x,\infty),t)$}.

However, because of the ambiguity discussed around Eq.~\eqref{Tambiguity}, there is no guarantee that for $\mathcal T$ defined as \eqref{Tpathexp}, $\mathcal T_A$ defined as \eqref{Tpath}, and $e^{\ad\int q} = \sigma_A$, we would have $\mathcal T = \mathcal T_A$.

We can make a more direct relation in the case where {\em we have a continuous ultra-local symmetry group}, Subsec.~\ref{ssectsymmetry}. The result is as follows. Write
\beq\label{Alambda}
	A = A_\lambda = e^{\ri \lambda \mathfrak a}
\eeq
for some matrix $\mathfrak a$ (in the Lie algebra representation on the fundamental field).

In the path integral formulation \eqref{Tpathriemann} on the resonant Riemann surface $\mathcal R_{x,t}^{A_\lambda}$, perform a very particular change of variable
\beq\label{gauge}
	\t\Psi = e^{\ri\delta\lambda f(\rz')\mathfrak a}\h\Psi(\rz').
\eeq
Here $f(\rz)$ is an ``unwinding function'': it is smooth on $\mathcal R_{x,t}$, differs by $1$ on consecutive sheets,
\beq
	f(x',t',n+1) - f(x',t',n) = 1, \quad (x',t')\in\R^2,\,n\in\Z,
\eeq
is constant everywhere except on $\ep$-neighbourhoods of the half-lines
\beq\label{cutB}
	B_{x,t} = \{(x',t,n):x'>x,\,n\in\Z\}\subset\mathcal R_{x,t},
\eeq
and vanishes on the original sheet away from the $\ep$-neighbourhood of $B_{x,t}$,
\beq
	f(x',t',0)=0,\quad (x',t')\not\in [x,\infty)\times [-\ep,\ep].
\eeq
Then, it has smooth jumps through these $\ep$-neighbourhoods, where all of its winding is accumulated. An (almost) explicit expression for this function is \eqref{functionf}. As this is a linear transformation of the integration variable, the Jacobian does not depend on the variable itself, hence the path integral measure is only affected, at most, by an overall normalisation, which is cancelled in the ratio \eqref{Tpathriemann}. For every $\ep>0$, the unwinding function $f(z)$ takes away $\delta\lambda$ from the Riemann surface's winding. Thus $\t\Psi$ has the same structure as $\hat\Psi$, with continuity on $\mathcal R_{x,t}$ Eq.~\eqref{riemannwinding}, and resonance Eqs.~\eqref{resonant}, \eqref{Alambda}, but with the replacement $\lambda\to\lambda-\delta\lambda$. 

The transformation \eqref{gauge} is a {\em gauge transformation} associated to the symmetry, where the transformation parameter is made space-time dependent. However, its non-trivial space-time dependence is only on the $\ep$-neighbourhoods \eqref{cutB}. As the transformation is trivial on the original sheet and assuming observables $o_i[\Psi_0](x_i,t_i)$ are at times away from $[t-\ep,t+\ep]$, they are not affected in \eqref{Tpathriemann}. Further, the action is invariant, except for contributions from these neighbourhoods. By Noether's theorem, in the limit where $\ep\to0$, these contributions are the conserved density concentrated on the half-lines \eqref{cutB},
\beq
	\lim_{\ep\to0} (\ri S[\h\Psi]
	-\ri S[\t\Psi]) =
	\ri \delta\lambda \int_{\mathcal R_{x,t}} \sd^2 \rz'\,\delta(\rz'\in B_{x,t})\,q[\t\Psi](\rz') + \mathcal O(\delta\lambda^2)
\eeq
where $q$ is the local density associated to the Lie algebra element $\mathfrak a$ (see Subsec.~\ref{ssectexp}), and
\beq
	\int_{\mathcal R_{x,t}} \sd^2 \rz'\,\delta(\rz'\in B_{x,t}) g(\rz') = 
	\lim_{N\to\infty}\frc1{2N+1}\sum_{n=-N}^N \int_x^\infty \dd x'\,g(x',t,n).
\eeq
By Eq.~\eqref{groupQq},  $q$ is invariant under the transformation it generates, so we have
\beq\label{qintilde}
	q[e^{\ri\lambda'\mathfrak a}\t\Psi](x',t',n)
	=
	q[\t\Psi](x',t',n).
\eeq
Hence, the half-lines may be moved towards positive or negative infinitesimal times and each sheet contributes the same, so we may write
\beq\label{shiftSa}
	\lim_{\ep\to0}(\ri S[\h\Psi]
	-
	\ri S[\t\Psi])
	= \ri \delta\lambda \int_x^\infty \dd x' q[\Psi_0](x',t+0^+) + \mathcal O(\delta\lambda^2).
\eeq
Thus we arrive at
\beq\label{twistpathexpsmall}
	\frc{\bra \vac|\mathsf T\big[ \mathcal T_{A_{\lambda}}(x,t) o_1(x_1,t_1)\cdots \big]|\vac\ket}{\bra \vac| \mathcal T_{A_{\lambda}}(x,t)|\vac\ket}
	=
	\frc{\bra \vac|\mathsf T\big[ \mathcal T_{A_{\lambda-\delta\lambda}}(x,t) e^{\ri\delta\lambda \int_x^\infty q(x',t+0^+)} o_1(x_1,t_1)\cdots \big]|\vac\ket}{\bra \vac| \mathcal T_{A_{\lambda-\delta\lambda}}(x,t)|\vac\ket}.
\eeq

Repeating the process, we see that the twist field, as defined in \eqref{Tpath} or equivalently \eqref{Tpathriemann}, is identified with {\em an infinite stack of infinitesimal standard exponential forms, infinitesimally time-separated},
\beq\label{twistpathexp}
	\mathcal T_{A_\lambda}(x,t) =
	\lim_{\delta\lambda\to 0} \prod_{k=1}^{\lambda/\delta\lambda} e^{\ri \delta \lambda \int_x^\infty\dd x'\, q(x',t+k\delta\lambda^2)}
	= \lim_{\delta\lambda\to 0} \prod_{k=1}^{\lambda/\delta\lambda} e^{-\ri \delta \lambda \varphi(x,t+k\delta\lambda^2)}
\eeq
where $k\delta\lambda^2$ implements the infinitesimal time displacements, and we may replace it by $k h(\delta\lambda)$ for any $h(\delta\lambda)$ that decays as $\delta\lambda\to0$ faster than $\delta\lambda$. Here we used the height-field expression \eqref{defTUgenconthf}. In terms of the time-ordered exponential,
\beq\label{TAstacktimeordered}
	\mathcal T_{A_\lambda}(x,t) =
	\lim_{s\to 0}\mathsf T\exp \Big[\ri \lambda \int_{t}^{t+s} \frc{\dd t'}s\,\int_x^\infty \dd x'\,q(x',t')\Big].
\eeq
 If $q[\Psi](x',t')$ is invariant not only under \eqref{Alambda}, but under gauge transformations, such as \eqref{gauge}, then in \eqref{shiftSa} the higher orders in $\delta\lambda$ vanish, and therefore the result is simply 
\beq\label{twistpathexpsimple}
	\mathcal T_{A_\lambda}(x,t) =
	e^{\ri \lambda \int_x^\infty q(x',t)}
	=
	e^{-\ri \lambda \varphi(x,t)}
	\quad\mbox{(if $q[\Psi](x',t')$ is gauge invariant).}
\eeq
This is the case, for instance, for the number operator in Fermionic or Bosonic systems, $q[\Psi](x',t') = |\Psi(x',t')|^2$ with the symmetry $A_\lambda = e^{-\ri \lambda}$.

The logic of the above derivation is quite subtle. The transformation \eqref{gauge} keeps the continuous structure on the Riemann surface $\mathcal R_{x,t}$, because $f$ is smooth on it, hence \eqref{riemannwinding} still holds for $\t\Psi(x,t,n)$. However, by unwinding, it modifies the resonance condition \eqref{resonant}. The modification of the action is due to the $\ep$-width region along the cut. Then we take the limit $\ep\to0$ on this modification of the action. Of course, the change of variable would look singular in this limit: the limit on $f$ is a function that is no longer continuous through the cuts, thus not smooth on the Riemann surface. However, the limit on the result of the change of variable is not singular. Indeed, for every $\ep$, not only the measure is invariant, but the change of variable keeps the continuity on the Riemann surface -- we integrate over all continuous field configurations with resonance \eqref{resonant} modified to $\lambda\to\lambda-\delta\lambda$. So both the space of functions over which we integrate, and the measure, are independent of $\ep$.

I illustrate the techniques in App.~\ref{appunwinding} for the complex relativistic Boson. 

\medskip
This, I believe, is a new result:
\st{
Every twist field $\mathcal T_A$ for the path integral formulation, can be expressed as an infinite product of closely time-separated infinitesimal twist fields in exponential form, Eq.~\eqref{twistpathexp}, equivalently \eqref{TAstacktimeordered}. 
}

%

\subsection{Topological in(co)variance}\label{ssecttopo}

In our main constructions of twist fields, that based on ultra-local symmetries in Subsec.~\ref{ssectultra} and \ref{ssectexp}, and that based on the path integral in Subsec.~\ref{ssecttwistpath} and \ref{ssectriemann}, the tail of the twist field was explicit: either an exponetial of a local observable integrated on $\{(x',t):x'>x\}$ (which in Subsec.~\ref{ssectultra} is written as a product of operators on the half-line $x'>x$), or a defect in space-time lying on $\{(x',t):x'>x\}$, through which discontinuity conditions are imposed (which in Subsec.~\ref{ssectriemann} became a branch cut associated to a Riemann surface). In this Subsection, I show that {\em the shape of the tail may be changed without (significantly) changing the result}: there is topological invariance, or covariance. I show this first for ultra-local symmetries that form a continuous Lie group, where it is simple to understand. Then I show it for ultra-local symmetries that are not necessarily part of a continuous Lie group, and for the path integral formulation.

\subsubsection*{Continuous ultra-local symmetry group}

Consider the setup of a continuous ultra-local symmetry group. As recalled in Subsec.~\ref{ssectsymmetry}, every conserved density is associated with a conserved current and continuity relation \eqref{conteq} and explicit current \eqref{jiexplicit}. Then, we may use Stokes' theorem to write, using the fully anti-symmetric symbol $\ep_{\mu\nu}$,
\beq
	\int_x^\infty \dd y\,q_i(y,t) = -\int_{\gamma_{x,t}}
	\dd s^\mu\ep_{\mu\nu}\, j_i^\nu(s)
\eeq
where $s\in\R^2$ and $\dd s^\mu$ is the vector parallel to the integration direction, with $x^0 = t,\,x^1 = x$, and we can take any differentiable path
\beq\label{gammaxt}
	\gamma_{x,t}: (x,t)\to(\infty,t).
\eeq
Therefore, the twist field \eqref{defTUgencont} can be written as
\beq\label{defTUgencontpath}
	\mathcal T_{\vec\lambda}(x,t)
	=\exp\Big[-\ri\vec\lambda\cdot\int_{\gamma_{x,t}} \dd s^\mu\ep_{\mu\nu}\,\vec j^\nu(s)\Big].
\eeq
This makes it clear that {\em the shape of the tail $\gamma_{x,t}$ is not important} -- it means that the twist field only depends on $x,t$. It was, of course, crucial that the twist field be associated to a {\em symmetry} for its invariance under deformations of its tail.

The deformation of the shape of the tail is meaningful because the currents $j^\mu(x,t)$ are local observables. Hence, with a more precise, model-dependent analysis, we can in principle determine the decay of their correlations with other observables. By moving the tail, we can then reduce its influence on other observables, or at least use local-physics arguments to determine universal effects of the twist field. This makes it clear that the twist field is truly ``supported at $x,t$'', the point from which the tail emanates. Such arguments will be discussed in Subsec.~\ref{ssectld}. In fact, the symmetry is internal, and a stronger expression of locality holds -- the expression \eqref{defTUgenconthf}, along with the fact that the height field $\vec\lambda\cdot\vec\varphi$ is self-local.

The end-point $(\infty,t)$ may be modified to $(\infty, t')$ for any $t'$ (as we will see below), and, in some situations like in the vacuum, even to different asymptotic rays of velocities $v\neq 0$, e.g.~$(x',x'/v)|_{x'\to\infty}$. But if this can be done or not without changing the result of what is being calculated, depends on what is being calculated. Frameworks to address this are discussed in Subsec.~\ref{ssectld}.

\subsubsection*{Ultra-local symmetry}

Continuous ultra-local symmetry groups are special cases of ultra-local symmetries. In the more general situation, the twist field still is an exponential of an integral on the half-line, Eq.~\eqref{defTUgen}, but neither \eqref{Qih0}, nor the continuity equation \eqref{conteq}, hold. Can we still change the shape of the tail?

It turns out that yes we can. Using again the fact that the tail is associated with a symmetry, we first show that
\beq
	\mathcal T_q(x,t)\mathcal T_{-q}(x)   \in \mathfrak L_0.
\eeq
That is, the product of the $t$-evolved twist field and its opposite at time 0, for any fixed time $t$ is, in fact, {\em a mutually local observable}, and no longer a twist field! We show this by verifying that it satisfies \eqref{localset} with $\mathfrak L_0$. Consider $x'\gg x$. Then by \eqref{exch2}, the automorphism property \eqref{autoglobal} for the time evolution, and the symmetry property \eqref{symmetry},
\beqa
	\mathcal T_q(x,t)\mathcal T_{-q}(x) o(x')
	&=& 
	\mathcal T_q(x,t)\sigma_{-\mathrm Q}(o(x')) \mathcal T_{-q}(x) \n
	&=& 
	e^{\ri \mathrm H t} (\mathcal T_q(x))\sigma_{-\mathrm Q}(o(x')) \mathcal T_{-q}(x) \n
	&=& 
	e^{\ri \mathrm H t} \Big(\mathcal T_q(x)(e^{-\ri \mathrm H t}\circ \sigma_{-\mathrm Q})(o(x')) \Big)\mathcal T_{-q}(x) \n
	&=& 
	e^{\ri \mathrm H t} \Big((\sigma_{\mathrm Q}\circ e^{-\ri \mathrm H t}\circ \sigma_{-\mathrm Q})(o(x'))\mathcal T_q(x) \Big)\mathcal T_{-q}(x) \n
	&=& 
	e^{\ri \mathrm H t} \Big(e^{-\ri \mathrm H t}(o(x'))\mathcal T_q(x) \Big)\mathcal T_{-q}(x) \n
	&=& 
	o(x')\mathcal T_q(x,t) \mathcal T_{-q}(x).
	\label{calculocalt}
\eeqa
A similar calculation can be done for $x'\ll x$, with the same result.

Note that
\beq
	\mathcal T_q(x,\ep)\mathcal T_{-q}(x)
	= e^{\ep\int_{x}^\infty \dd x'\,\dot q(x') + \mathcal O(\ep^2)}.
\eeq
Thus we define a ``linearly tailed observable'' $j$, that is not local, by
\beq\label{defjultralocal}
	j(x) = \int_{x}^\infty \dd x'\,\dot q(x'),
\eeq
which is such that $e^{\ep j(x,t)}$, for infinitesimal $\ep$, is local and ``supported'' around $(x,t)$. Then we obtain, say for $t'>t$,
\beq
	\mathcal T_q(x,t') = \lim_{\ep\to0} e^{\ep j(x,t'-\ep)}\cdots e^{\ep j(x,t+\ep)}e^{\ep j(x,t)}
	\mathcal T_q(x,t)
	=
	\Pexp\Big[\int_t^{t'}\dd s\,j(x,s)\Big]\mathcal T_q(x,t)
\eeq
where the time-ordered exponential {\em is local}
\beq
	\Pexp\Big[\int_t^{t'}\dd s\,j(x,s)\Big]
	\in \mathfrak L_0\quad \forall\;t,t'.
\eeq
Denoting
\beq
	\mathcal T_q(x,t)\mathcal T_{-q}(x',t) = \mathcal T_q(x\to x',t)
\eeq
which is also in $\mathfrak L_0$, as can be easily checked, we may represent any path in space-time from $(x,t)$, formed by a succession of horizontal (spatial) and vertical (temporal) intervals $[x_i,x_{i+1}]\times \{t_i\}$, $\{x_{i+1}\}\times [t_i,t_{i+1}]$, $i=1,2,\ldots,N-1$:
\beq\label{tailultra}
	\mathcal T_q(x,t)
	=
	\prod_{i=1}^{N-1}\Big(
	\mathcal T_{q}(x_i\to x_{i+1},t_i)
	\Pexp\Big[\int_{t_{i+1}}^{t_i}\dd s\,j(x,s)\Big]\Big)
	\mathcal T_q(x_N,t_N)
\eeq
where the factors in the product are ordered from $i=1$ (left) to $i=N-1$ (right). Taking intervals to be infinitesimal, and $N\to\infty$, this is the sense in which the shape of the tail can be modified to any path $\gamma_{x,t}$ as Eq.~\eqref{gammaxt}.

We see that, thanks to the fact that locality is preserved by finite time translations, the end-time can be chosen as $t'\neq t$. Effectively, the local current observable that we ``should'' put at $x=\infty$, integrated over $[t,t']$, decorrelates  from all other local observables, so can be neglected (it does so both from the algebraic viewpoint and the state viewpoint).

Formula \eqref{tailultra} has a similar meaning as \eqref{defTUgencontpath}, even though it is more involved as there is no local current. Nevertheless, the precise correlation properies of $e^{\ep j(x,t)}$, supported around $(x,t)$, and $\mathcal T_q(x\to x',t)$, with a tail lying on $[x,x']$ at time $t$, can in principle be deduced from the above construction in specific models, and thus deforming the tail is useful and shows again that the twist field is more truly supported at $x,t$.

\subsubsection*{Path integral formulation}

The final aspect of topological invariance that I discuss is that which occurs in the path integral formulation. In the case of a symmetry that is not part of a continuous group, this is arguably the clearest way of understanding twist-field topological invariance. There is a subtelty: above, we saw how invariance occured at the level of the operator itself. In the path integral formulation, instead, we concentrate on averages of time-ordered observable. Because the deformations of the tail in space-time modify the time positions where it is located, we have to account for this. The effect is not an invariance, but a {\em covariance}, where local observables are affected whenever {\em the tail, as it is deformed, passes through them}. I take the temporal defect formulation \eqref{Tpath}, where a cut in space-time is introduced on the tail, but of course the same arguments can be given for the Riemann-surface formulation \eqref{Tpathriemann}, where the tail is where the branching structure is imposed. I also consider only vacuum expectation values for simplicity.

The result is as follows. For any simple path $\gamma_{x,t} : (x,t)\to(\infty,\infty)$, that is, a simple path starting from $(x,t)$ and asymptotically going to infinity on $\R^2$ (and we have to assume that it has at most a finite winding number around $(x,t)$), we have
\beq\label{pathgammagen}
	\frc{\bra \vac|\mathsf T\big[ \mathcal T_A(x,t) \prod_i o_i(x_i,t_i)\big]|\vac\ket}{\bra \vac| \mathcal T_A(x,t)|\vac\ket}
	=
	\frc{\int_{\mathcal C_{\gamma_{x,t}}^A} [\dd\Psi] \,e^{\ri S_{\gamma_{x,t}}[\Psi]}
	\prod_i o_i[A^{\chi_{x,t}(x_i,t_i)}\Psi](x_i,t_i)
	}{
	\int_{\mathcal C_{\gamma_{x,t}}^A} [\dd\Psi] \,e^{\ri S_{\gamma_{x,t}}[\Psi]}}.
\eeq
Here $\chi_{x,t}(x',t')$ is the sum of $1$ if a continuous path deformation $u\in [0,1]\mapsto \gamma_{x,t}^{u}$ from the horizontal path $\gamma_{x,t}^0 = \eta_{x,t}:(x,t)\to (\infty,t)$ to $\gamma_{x,t}^1 = \gamma_{x,t}$ crosses $(x',t')$ ``counter-clockwise''; $-1$ if the path deformation crosses $(x',t')$ ``clockwise''; and 0 otherwise. Crossing counter-clockwise means that $(x',t')$ is to the left (with respect to the path direction) of $\gamma_{x,t}^{u-\delta u}$ and to the right of $\gamma_{x,t}^{u+\delta u}$ for some $u$ and infinitesimal $\delta u>0$. There can be multiple values of $u$ where such crossings happens, and $\chi_{x,t}(x',t')$ is the sum of the described numbers; this is a topological quantity, independent of the choice of path deformation. Further,
\beq\label{Savoidgammagen}
	S_{\gamma_{x,t}}[\Psi] = \int_{\R^2\setminus  \gamma_{x,t}} \dd x'\dd t'\,\mathcal L[\Psi](x',t'),\quad\mathcal C_{\gamma_{x,t}}^A : 
	\Psi(\gamma(s)^+) = A\Psi(\gamma(s)^-)
\eeq
where $\gamma(s)^+$ ($\gamma(s)^-$), for $s\in[0,\infty)$, are the points lying just to the left of (to the right of) $\gamma$. See Fig.~\ref{figtopologicalpi}
\begin{figure}
\bc\includegraphics[width=0.5\textwidth]{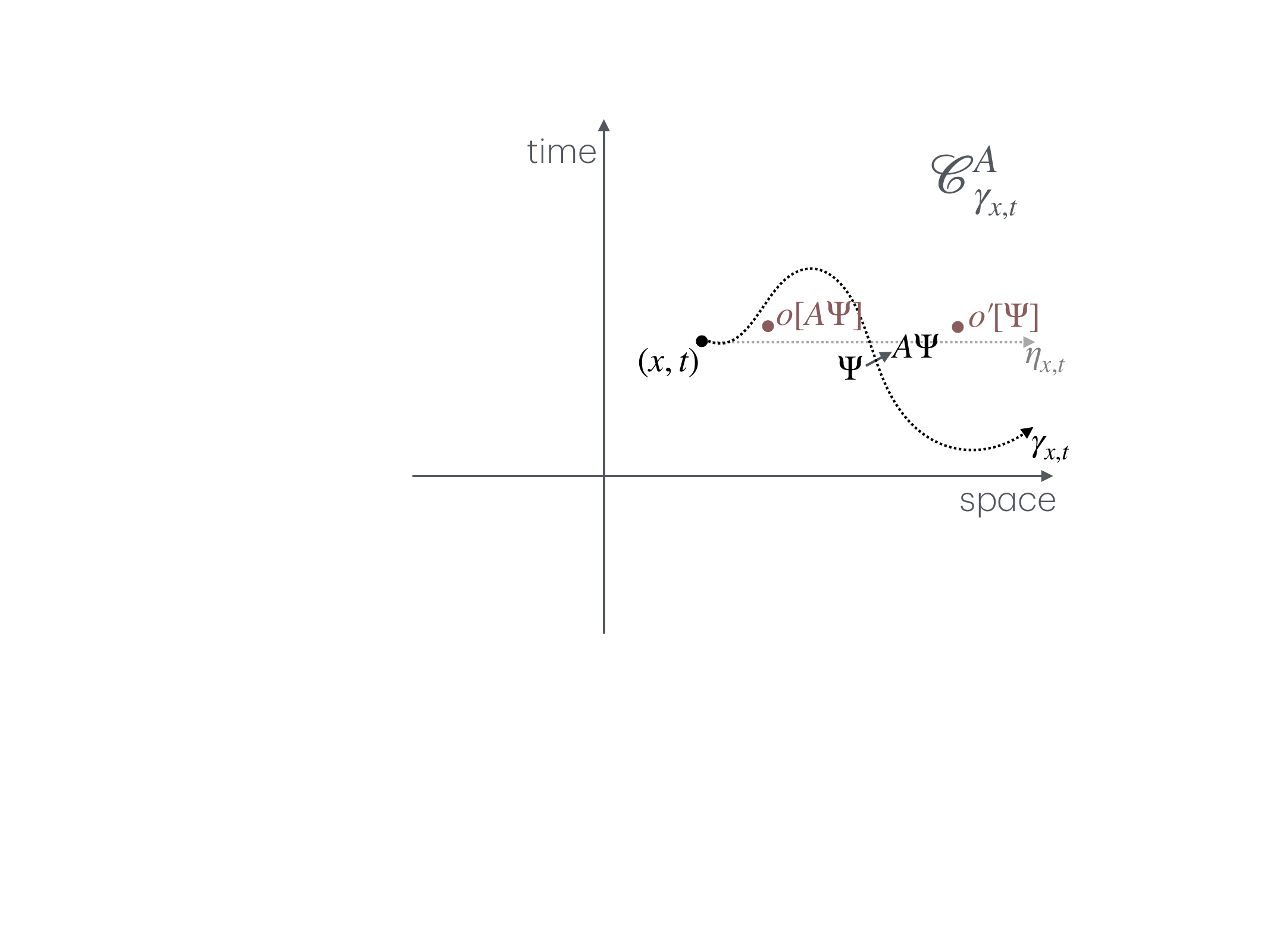}\ec
\caption{The jump condition \eqref{Savoidgammagen} for a deformation of the tail of the twist field, along with local observable insertions $o(x_1,t_1)$ and $o'(x_2,t_2)$.}
\label{figtopologicalpi}
\end{figure}

Of course, by winding there are many continuous path deformations from $\eta_{x,t}$ to $\gamma_{x,t}$, but the result is the same no matter the winding, because of invariance of the action, $S_{\gamma_{x,t}}(A\Psi)=S_{\gamma_{x,t}}(\Psi)$, and of the path integral measure and function space, under $\Psi\mapsto A\Psi$. The result \eqref{pathgammagen} shows that, as we continuously deform the contour to a different shape, the local fields that are ``swept'' by this deformation are affected by the transformation $A$, or $A^{-1}$, depending in which direction (counter-clockwise, or clockwise) the contour is passing through. Interestingly, here there is no condition for the contour to end on $(\infty,t')$ for some finite $t'$ -- this is because we are evaluating the correlation function in the vacuum. For correlation functions involving many twist fields, similar results hold. For instance,
\beq\label{twistpartitiondeformed}
	\bra \vac|\mathsf T\big[ \mathcal T_A(x,t)  \b{\mathcal T}_{A}(x',t')\big]|\vac\ket
	\propto
	\int_{\mathcal C_{(x,t)\to (x',t')}^A} [\dd\Psi] \,e^{\ri S_{(x,t)\to(x',t')}[\Psi]}
\eeq
for any path $(x,t)\to(x',t')$ and $S_\gamma,\,C_\gamma$ defined as in \eqref{Savoidgammagen}.

Let us see how this is obtained.

In the path integral, introduce a new cut on a closed (continuous, non-self-intersecting) space-time contour $\gamma: s\in[0,1]\mapsto\gamma(s)\in\R^2,\,\gamma(0)=\gamma(1)$; below I will also use $\gamma$ to mean the contour as a subset, $\gamma([0,1])\subset \R^2$. Assume that $\gamma$ does not intersect the tail $([x,\infty),t)$. Through the new cut, impose continuous defect conditions -- thus this is a ``phantom cut'', which does not affect the result. That is, re-write the right-hand side of \eqref{Tpath} as
\beq\label{piSA}
	\frc{\int_{\mathcal C_{x,t;\gamma}^A} [\dd\Psi] \,e^{\ri S_{x,t;\gamma}[\Psi]}
	\prod_i o_i[\Psi](x_i,t_i)}{
	\int_{\mathcal C_{x,t;\gamma}^A} [\dd\Psi] \,e^{\ri S_{x,t;\gamma}[\Psi]}}.
\eeq
where the full defect condition is
\beq\label{twistcutgamma}
	\mathcal C_{x,t;\gamma}^A : \begin{aligned}
	&
	\Psi(x',t+ 0^+) = \lt\{
	\ba{ll}
	A \Psi(x',t - 0^+) & (x'\geq x)\\
	\Psi(x',t - 0^+) & (x'< x),
	\ea\rt.
	\\ &
	\Psi(\gamma(s)^+) = \Psi(\gamma(s)^-) \quad (s\in[0,1])
	\end{aligned}
\eeq
with $\gamma(s)^+\in \R^2\setminus D$  ($\gamma(s)^-\in D$) are the points infinisemally closed to the contour just outside (inside) of the region it encloses, $D = $ bounded component of $\R^2\setminus \gamma$. In \eqref{piSA}
\beq\label{Savoidgamma}
	S_{x,t;\gamma}[\Psi] = \int_{\R^2\setminus  (([x,\infty),t)\,\cup\, \gamma)} \dd x'\dd t'\,\mathcal L[\Psi](x',t')=S_{x,t}^{\R^2\setminus D}[\Psi]
	+
	S^{D}[\Psi],
\eeq
and in the last equality, we used locality of the Lagrangian density, Eq.~\eqref{lagrangiandensity}, where the superscript means that we restrict the integral on the domain indicated. 
Invariance of the Lagrangian density \eqref{invaction} gives
\beq
	S_{x,t;\gamma}[\Psi] = S_{x,t}^{\R^2\setminus D}[\Psi]
	+
	S^{D}[A^{-1}\Psi].
\eeq
By factorisation of the path-integrale measure we may change the path-integral variable $\Psi|_D\to A\Psi|_D$ on $D$, and we obtain
\beq\label{pathgamma}
	\frc{\int_{\t{\mathcal C}_{x,t;\gamma}^A} [\dd\Psi] \,e^{\ri S_{x,t;\gamma}[\Psi]}
	\prod_i o_i[A^{\chi_D(x_i,t_i)}\Psi](x_i,t_i)
	}{
	\int_{\t{\mathcal C}_{x,t;\gamma}^A} [\dd\Psi] \,e^{\ri S_{x,t;\gamma}[\Psi]}}
\eeq
where $\chi_D(x,t) = 1$ if $(x,t)\in D$ and 0 otherwise, with
\beq\label{twistcutgamma}
	\t{\mathcal C}_{x,t;\gamma}^A : \begin{aligned}
	&
	\Psi(x',t+ 0^+) = \lt\{
	\ba{ll}
	A \Psi(x',t - 0^+) & (x'\geq x)\\
	\Psi(x',t - 0^+) & (x'< x),
	\ea\rt.
	\\ &
	\Psi(\gamma(s)^+) = A\Psi(\gamma(s)^-).
	\end{aligned}
\eeq

Now choose $\gamma = \t\gamma_{\rm uhp} \cup \t\gamma_{\rm straight}$ to be on the upper half of space-time, $\gamma\in\R\times (0,\infty)$, and to have a straight portion $\t\gamma_{\rm straight} = \{(x',t+\ep):x'\in[y_1,y_2]\}$ lying infinitesimally close to $([y_1,y_2],t)$ for some $x<y_1<y_2$. Then along the straight part of $\gamma$, the cut conditions cancel each other, and the contribution to the action of the infinitesimal region in-between this straight part and the tail, is infinitesimal. With
\beq
	\kappa_{x,t} = \{(x',t):x'\in (x,y_1]\}\cup \gamma_{\rm uhp}\cup
	\{(x',t):x'\in[y_2,\infty)\}
\eeq
in the limit $\ep\to 0$ we obtain
\beq\label{pathgamma2}
	\frc{\int_{\mathcal C_{\kappa_{x,t}}^A} [\dd\Psi] \,e^{\ri S_{\kappa_{x,t}}[\Psi]}
	\prod_i o_i[A^{\chi_D(x_i,t_i)}\Psi](x_i,t_i)
	}{
	\int_{\mathcal C_{\kappa_{x,t}}^A} [\dd\Psi] \,e^{\ri S_{\kappa_{x,t}}[\Psi]}}
\eeq
with
\beq\label{Savoidgamma2}
	S_{\kappa_{x,t}}[\Psi] = \int_{\R^2\setminus  \kappa_{x,t}} \dd x'\dd t'\,\mathcal L[\Psi](x',t'),\quad\mathcal C_{\kappa_{x,t}^A} : 
	\Psi(\kappa(s)^+) = A\Psi(\kappa(s)^-)
\eeq
and, by convention, the domain enclosed by $\kappa$ contains the lower half-plane.

The result \eqref{pathgamma2} with \eqref{Savoidgamma2} shows topological co-variance of the twist field in the path integral formulation: importantly, the local observables that are within the new region ``added'' to the lower half-plance, are affected by the symmetry. That is, as we continuously deform the contour towards the upper half-plane, the local fields swept by this deformation are affected by the transformation $A$. 

This shows how to deform a portion of the tail toward the upper half-plane. Clearly, a similar calculation can be obtained for a deformation towards the lower half-plane, where the transformation of the local fields swept by this deformation is $A^{-1}$ instead of $A$. The general result, deforming different parts of the contour in either directions, and extending these to unbounded portions of the contour (which is doable because only a finite number of observables may be affected), is \eqref{pathgammagen}.

\subsubsection*{Exponential form in the path integral and contact terms}

The main difference between the operator formulation and the path-intergal formulation of topological invariance is the fact that in the latter, but not the former, the other local fields inserted within the expectation value are affected by the symmetry transformation if the contour deformation passes through them -- this is topological {\em covariance}. This difference is easily understandable from the time-ordering requirement in the path-integral formulation. Another way to undertand it is to attempt to reproduce the calcuations done in the operator realm, but directly in the path-integral formulation. That is, given the relation to the standard exponential form, Eq.~\eqref{twistpathexp} derived in Subsec.~\ref{ssectriemann}, can't we simply use this and deform the contour of the integral in the standard exponential form {\em directly in the path integral formulation}, thus obtaining, for instance in the case of a continuous symmetry group, the equivalent of Eq.~\eqref{defTUgencontpath}?

The answer is that we can perform this calculation, but we obtain, again, Eq.~\eqref{pathgammagen}, where local observables passed by during contour deformation are affected by the symmetry transformation. This is because of the path-integral {\em contact terms}.

Take the example of a continuous symmetry group. In the path-integral formulation, equations of motion, which imply the continuity equation \eqref{conteq}, are obtained by the total-derivative identity:
\beq
	\int [\dd\Psi]\frc{\delta}{\delta \Psi(x,t)} e^{\ri S[\Psi]} = 0\quad \Rightarrow\quad \int [\dd\Psi]e^{\ri S[\Psi]} \p_u j_i^\mu[\Psi](x,t) = 0.
\eeq
When other observables are inserted at positions $x_i,t_i$, the identity $\p_u j_i^\mu[\Psi](x,t) = 0$ still holds, but only if $x,t\neq x_i,t_i$: indeed, at $(x,t)=(x_i,t_i)$, the functional derivative $\delta/\delta\Psi(x,t)$ receives additional contributions
\beq
	\int [\dd\Psi]\Big(\frc{\delta}{\delta \Psi(x,t)} e^{\ri S[\Psi]}\Big)\prod_i o_i[\Psi](x_i,t_i) =
	-\int [\dd\Psi]e^{\ri S[\Psi]}\Big(\frc{\delta}{\delta \Psi(x,t)} \prod_i o_i[\Psi](x_i,t_i)\Big).
\eeq
As a consequence, there are {\em contact terms}, which turn out to be exactly given by the transformations of the observables under the symmetry action,
\beqa
	\lefteqn{\int [\dd\Psi]e^{\ri S[\Psi]} \p_u j_i^\mu[\Psi](x,t) \prod_i o_i[\Psi](x_i,t_i)} && \n &=& -
	\int [\dd\Psi]e^{\ri S[\Psi]} \sum_j \delta(x-x_j)\delta(t-t_j) o_j[\mathfrak a_i\Psi](x_j,t_j) \prod_{i\neq j} o_i[\Psi](x_i,t_i)
\eeqa
where we write, as in \eqref{Alambda}, $A_{\vec \lambda} = e^{\ri \vec\lambda\cdot \vec{\mathfrak a}}$. Using Stokes' theorem
\beq
	\int_D \dd x\dd t\,\p_uj^\mu_i(x,t) = \int_{\p D} \dd s^\mu\ep_{\mu\nu}\, j_i^\nu(s)
\eeq
the result \eqref{pathgammagen} then follows from the path-integral representation of the exponential form \eqref{twistpathexp}.

%
%
%
%
%
%
%
%
%

\subsection{Twist families and algebra of twist fields}\label{ssectalg}

We saw in the discussion around Eq.~\eqref{Tambiguity} that, for a given symmetry generator $\mathrm Q = \ad\int q$, there are many possible definitions of twist fields, $\mathcal T_q$, because there are many densities $q$ giving the same symmetry transformation. But the difference is in a multiplication by a local observable at $x$.

This is a general principle: given a twist field $\mathcal T$, it is a simple matter to see that the new twist fields $\mathcal T'$, defined by
\beq\label{tprimet}
	\mathcal T'(x) = o(x)\mathcal T(x)
\eeq
for some local $o\in\mathfrak L_0$ in the subspace of mutually local observables, satisfies the same exchange relations \eqref{exch}, \eqref{exch2} as does $\mathcal T$, that is $\sigma_{\mathcal T'} = \sigma_{\mathcal T}$. Therefore, given a symmetry $\sigma$ for which there exists at least one twist field $\mathcal T$, then there are many twist fields with the same twist. We call this the {\em twist family} associated to $\sigma$: denoting $[\mathcal T] = \{o\mathcal T:o\in\mathfrak L_0\}$, we have
\beq
	\mathfrak T_\sigma = [\mathcal T]\subset \mathfrak L,\quad\mbox{for}\ 
	\sigma = \sigma_{\mathcal T}.
\eeq
This is now a {\em linear subspace of the space of local observables $\mathfrak L$}. Thus, there is a set of symmetries $\Sigma$ such that our set of twist fields separates into disjoint linear subspaces, one subspace for each symmetry in $\Sigma$,
\beq
	\mathfrak T = \bigcup_{\sigma\in \Sigma} \mathfrak T_\sigma.
\eeq

In the example of Subsec.~\ref{ssectultra}, Eq.~\eqref{spaceultra}, the set of symmetries can be taken as $\Sigma = \{\sigma_U:U\mbox{\ giving ultra-local symmetry}\}$, and we may extend $\mathfrak T$ to
\beq
	\mathfrak T = \{o\mathcal T_U\,:\, o\mbox{\ mutually local},\ U\mbox{\ giving ultra-local symmetry}\}.
\eeq
The resulting space of local observables $\mathfrak L = \langle \mathfrak L_0\cup \mathfrak T\rangle$ generated by these, is the same as that generated by \eqref{spaceultra}, but now we have extracted {\em all twist fields within $\mathfrak L$}.

Therefore, although our main definition was {\em associating a symmetry to every twist field}, the above, along with the discussion in Subsecs.~\ref{ssectultra}, \ref{ssectexp} and App.~\ref{appgen}, suggests that it is often better to consider {\em all twist fields associated to a given symmetry}.

Now recall the consistency condition \eqref{relationsym} on the twists. This implies the following relation for the transformation properties of twist fields:
\beq\label{transfoclass}
	\sigma(\mathcal T'(x))\in \mathfrak T_{\sigma\circ\sigma'\circ\sigma^{-1}},\quad \mathcal T' \in \mathfrak T_{\sigma'}.
\eeq
This generalises the specific transformations \eqref{UT}, \eqref{TA} and \eqref{gtransfo} for particular constructions of twist fields, and is the formal description of the more general transformation property \eqref{twisttransfo}.

Finally, in addition to the structure of linear subspace obtained by multiplying by local observables, twist fields also have an algebraic structure. Indeed, it is simple to check that if $\mathcal T\in\mathfrak T_{\sigma}$ and $\mathcal T'\in\mathfrak T_{\sigma'}$, then their product, translated to any positions $x,x'$, lies in the space obtained by the composition of the symmetry transformations,
\beq\label{productTT}
	\mathcal T(x)\mathcal T'(x') \in \mathfrak T_{\sigma\circ\sigma'},\quad
	\mathcal T\in\mathfrak T_{\sigma},\ \mathcal T'\in\mathfrak T_{\sigma'}.
\eeq
For instance, in the real of quantum chains, take $\mathcal T_U$ defined in \eqref{defTU}, and assume that $x'>x$. Then
\beq\label{TUprod}
	\mathcal T_U(x) \mathcal T_{U'}(x')
	=
	\prod_{x''=x}^{x'-1} U(x'')\,\mathcal T_{UU'}(x')
	\in\mathfrak T_{\sigma_U\circ\sigma_{U'}}
\eeq
where the local observable in front of the ``fundamental'' twist fields $\mathcal T_{UU'}(x')$ is $o(x,x') = \prod_{x''=x}^{x'-1} U(x'')\in\mathfrak L_0$, and is indeed supported on a finite interval, $[x,x'-1]$. If $x=x'$, we directly recover a fundamental twist field.

Note that it is important here that the exchange relations \eqref{exch}, \eqref{exch2} occur at {\em large separations}. Once we accept that we can multiply by local observables, which are supported on finite intervals, then only large separations keep their meaning. In QFT, local observables are often taken to be supported on single points, and exchange relations can be written for $x>x'$ and $x<x'$. As mentioned, I discuss this in Subsec.~\ref{ssectqft}.

\subsection{Extensions: lattices, classical Hamiltonians, conical singularities}\label{ssectcon}

The constructions of Subsec.~\ref{ssectultra} - \ref{ssectriemann}, and of App.~\ref{appgen}, are based on the exchange relations of Subsec.~\ref{ssectexch}, and formulated within the context of continuous-time quantum many-body systems. This is just a part of the general theory of twist fields. It is not hard to extend to other setups: Statistical (Euclidean) field theory and classical lattice systems, classical Hamiltonian systems, discrete time evolution and quantum circuits, and space-time and other external symmetries. All these extensions have their applications and interest, some of which I will touch upon in the next Section. Not all constructions above have a natural equivalent in some of these extensions, but the notion of twist families, Subsec.~\ref{ssectalg}, always apply.

Let me explain how some of the constructions above are modified to account for these extensions. I start, however, with the natural alternative definition of twist fields, with tails going towards the left instead of the right.

\subsubsection*{Twist fields with left-going tails}

We may modify the definition \eqref{exch}, \eqref{exch2}, at the basis of all our constructions, to have a tail going towards the left ($x'<x$) instead of the right ($x'>x$). Given a (right-tailed) twist field $\mathcal T$, there is a left-tailed twist field, which we denote $\mathcal T^-(x)\in\mathfrak T^-$. They satisfy
\beq\label{exchleft}
	\mathcal T^-(x) {\mathcal T^-}'(x') = \lt\{\ba{ll}
	\sigma_{\mathcal T}^{-1} ({\mathcal T^-}'(x'))\, \mathcal T^-(x) & (x'\ll x)\\
	\mathcal T'(x')\, \sigma_{\mathcal T'} (\mathcal T^-(x)) & (x'\gg x)
	\ea\rt. \quad(\mathcal T^-,{\mathcal T^-}'\in \mathfrak T^-)
\eeq
and
\beq\label{exch2left}
	\mathcal T^-(x) o(x') = \lt\{\ba{ll}
	\sigma_{\mathcal T}^{-1} (o(x'))\, \mathcal T^-(x) & (x'\ll x)\\
	o(x')\mathcal T^-(x) & (x'\gg x)
	\ea\rt.\quad (\mathcal T^-\in \mathfrak T^-,\ o\in\mathfrak L_0).
\eeq
Note how the asymptotic relations between $x',x$, and the automorphism, have been inverted.

If there exists a global invertible symmetry operator $U_{\mathcal T}$ such that
\beq
	U_{\mathcal T} o(x)U_{\mathcal T}^{-1} = \sigma_{\mathcal T}(o(x))
\eeq
for all local observables and all $x$, then the main relation is
\beq
	U_{\mathcal T}^{-1}\mathcal T(x) \in \mathfrak T^-.
\eeq
That is, $U_{\mathcal T}$ takes away the symmetry change along the right-going tail, and puts it along a new left-going tail. Constructions in Subsec.~\ref{ssectultra} - \ref{ssecttwistpath} have explicit $U_{\mathcal T}$, and the construction of App.~\ref{appgen} may be repeated for left-tailed twist fields. A similar general theory of twist families, Subsec.~\ref{ssectalg}, holds for left-tailed twist fields. We leave as an exercise computing the exchange relations mixing right-tailed and left-tailed twist fields.

In Subsec.~\ref{ssecttwistpath} and \ref{ssectriemann}, the result for left-tailed twist fields is a branch cut towards the left,
\beq\label{twistcuteft}
	\mathcal C_{x,t}^{A, -} : \Psi(x',t- 0^+) = \lt\{
	\ba{ll}
	A \Psi(x',t + 0^+) & (x'< x)\\
	\Psi(x',t+  0^+) & (x'> x),
	\ea\rt.
\eeq
but the monodromy \eqref{monodromy} is the same: for a single-winding counter-clockwise path $\gamma$ around $x,t$,
\beq\label{monodromyleft}
	\frc{\bra \vac|\mathsf T\big[ \mathcal T_A^-(x,t) o_1(x_1,t_1) \cdots\big]|\vac\ket}{\bra \vac| \mathcal T_A^-(x,t)|\vac\ket}\Big|_{(x_1,t_1):\stackrel{\gamma} \circlearrowleft} =
	\frc{\bra \vac|\mathsf T\big[ \mathcal T_A^-(x,t) \sigma_A(o_1(x_1,t_1)) \cdots\big]|\vac\ket}{\bra \vac| \mathcal T_A^-(x,t)|\vac\ket}
\eeq
where $o_1$ and $\cdots$ are local observables in $\mathfrak L_0$.
Topological covariance of Subsec.~\ref{ssecttopo} holds in the same way, and in the path-integral formulation of the vacuum state, because topological covariance admits rotation deformations of the path, we get
\beqa
	\frc{\bra \vac|\mathsf T\big[ \mathcal T_A(x,t) \prod_i o_i(x_i,t_i)\big]|\vac\ket}{\bra \vac| \mathcal T_A(x,t)|\vac\ket}
	&=&
	\frc{\bra \vac|\mathsf T\big[ \mathcal T_A^-(x,t) \prod_{i:t_i>t} \sigma_A(o_i(x_i,t_i)) \prod_{i:t_i<t}o_i(x_i,t_i)\big]|\vac\ket}{\bra \vac| \mathcal T_A^-(x,t)|\vac\ket}\n
	&=&
	\frc{\bra \vac|\mathsf T\big[ \mathcal T_A^-(x,t) \prod_{i:t_i>t} o_i(x_i,t_i) \prod_{i:t_i<t}\sigma_A^{-1}(o_i(x_i,t_i))\big]|\vac\ket}{\bra \vac| \mathcal T_A^-(x,t)|\vac\ket}\n
	\label{leftrightequiv}
\eeqa
which, from $U_{A}|\vac\ket = |\vac\ket$ and choosing the normalisation appropriately, indicates that
\beq\label{Ualeftright}
	U_{A}^{-1}\mathcal T_A(x) = \mathcal T_A^-(x).
\eeq
This is in fact a general property, and we can define such left-tailed twist fields in the context of ultra-local symmetries more generally, Subsec.~\ref{ssectultra}, \ref{ssectexp}, e.g.
\beq\label{Uleftright}
	U^{-1}\mathcal T_U(x) = \mathcal T_U^-(x).
\eeq

\subsubsection*{Statistical field theory}

It is in the context of classical lattice systems that twist fields have their origin. In the continuum limit, at or near criticality, these are described by Euclidean field theory. Naturally, the clearest way of generalising the constructions above to Euclidean field theory is that based on the path integral. So I start with the Euclidean version of the constructions of Subsec.~\ref{ssecttwistpath} and \ref{ssectriemann}.

The equivalent formulation is for the infinite plane (as this corresponds to the vacuum state in the quantum formulation), Eq.~\eqref{inftypi}, or more precisely \eqref{statpi} with $\mathcal S = \R^2$. Considering positions $ z = (x,\tau), z_i = (x_i,\tau_i)\in\R^2$, we simply set
\beq\label{Tpathstat}
	\frc{
	\bra \mathcal T_A( z) o_1( z_1)\cdots o_n( z_n)\ket}{\bra  \mathcal T_A( z)\ket}
	=
	\frc{\int_{\mathcal C_{ z}^A} [\dd\Psi] \,e^{-S_{{\rm E},\, z}[\Psi]}
	o_1[\Psi]( z_1)\cdots o_n[\Psi]( z_n)}{
	\int_{\mathcal C_{ z}^A} [\dd\Psi] \,e^{-S_{{\rm E},\, z}[\Psi]}}
\eeq
where $\mathcal C_{ z}^A$ is \eqref{twistcut} with $t$ replaced by $
\tau$, and $S_{{\rm E},\, z}$ is the Euclidean equivalent of \eqref{Savoid},
\beq\label{SavoidE}
	S_{{\rm E},\, z}[\Psi] = \int_{\R^2\setminus  ([x,\infty],\tau)} \dd x'\dd \tau'\,\mathcal L_{\rm E}[\Psi](x',\tau').
\eeq
Because the Euclidean Lagrangian density is invariant under the transformation $A$, the Euclidean equivalent of topological covariance, \eqref{pathgamma2}, \eqref{Savoidgamma2}, follows, as well as the Euclidean Riemann surface formulation, with the monodromy  \eqref{monodromy}
\beq\label{monodromyeucl}
	\frc{\bra \mathcal T_A(z) o_1(z_1) \cdots \ket}{\bra \mathcal T_A(z)\ket}\Big|_{z_1:\stackrel{\gamma} \circlearrowleft} =
	\frc{\bra  \mathcal T_A(z) \sigma_A(o_1(z_1))\cdots \ket}{\bra \mathcal T_A(x,t)\ket}
\eeq
for $\gamma$ surrounding $z$ once counterclockwise. One may also use the exponential form \eqref{twistpathexp} in the Eulicean setting (or its implified version \eqref{twistpathexpsimple}), which can be derived similarly by Noether's theorem.

The quantum finite-temperature state $\rho$ in \eqref{statesquantum} is represented as a cylinder in the Euclidean field theory, and we may take more general domains $\mathcal S$, Eq.~\eqref{statpi}. On the cylinder, {\em it is not possible to deform the cut to bring it from one side to the other}. That is, \eqref{pathgammagen}, \eqref{Savoidgammagen} are now restricted to paths $\gamma_{x,\tau}: (x,\tau) \to (\infty,\tau')$. This is because otherwise, the deformation would involve infinite winding around the cylinder, and the of sweeping these infinte windings through local insertions would be ill-defined. That is, the equivalent of \eqref{leftrightequiv} in finite-temperature states does not hold. In fact, \eqref{Ualeftright} still holds, but we see that $U_A$ has the effect of {\em modifying the density matrix}, $\rho \to U_A  \rho$.

On a finite domain $\mathcal S$, the cut must terminate at the boundary $\p \mathcal S$. The point where it terminates corresponds to the insertion of a {\em boundary twist field}, and, unless the boundary conditions are very special, it cannot be moved by topological invariance.

\subsubsection*{Classical lattice systems}

The Euclidean field theory formulation can be seen as the continuum limit of lattice models, and under this identification, we obtain a natural formulation of twist fields in classical lattice statistical mechanics. In fact, the transfer-matrix formulation of classical statistical lattice systems can be used to produce a theory based on operators and exchange relations, similar to that we developed in the quantum case.

For simplicity, take the example of the square lattice, with nearest-neighbour interactions. Then the result is simply \eqref{Tpathstat} where now $z,\,z_i\in\Z^2$. The integral $\int_{\mathcal C_z^A} [\dd\Psi]$ is the integration over lattice configurations, and if we have discrete degrees of freedom on each site, such as in the Ising model, it is a sum instead of an integral, $\int_{\mathcal C_z^A} [\dd\Psi] \to \sum_{\Psi,\,\mathcal C_z^A}$. Most importantly, the tail is now represented as {\em a separation of each lattice site along\footnote{We denote $[[x,\infty)) = [x,\infty)\cap \Z$.} $([[x,\infty)),\tau)$ into two adjacent sites, one connecting above, the other below, with the discontinuity condition relating them}. That is, the lattice becomes the graph
\beq\label{latticewithcut}
	\mathbb L =  \Z^2\ \mbox{where the tail}\ ([[x,\infty)),\tau)\ \mbox{is replaced by}\ 
	([[x,\infty)),\tau)^{\rm upper} \times ([[x,\infty)),\tau)^{\rm lower}
\eeq
and  there is no edge connecting the sites of $([[x,\infty)),\tau)^{\rm upper}$ with those of $([[x,\infty)),\tau)^{\rm lower}$. Accordingly the action
\beq\label{SavoidElattice}
	S_{{\rm E},\, z}[\Psi] = \sum_{(x',\tau')\in \mathbb L}\,\mathcal L_{\rm E}[\Psi](x',\tau')
\eeq
has no interaction term that connects $([[x,\infty)),\tau)^{\rm upper}$ with $([[x,\infty)),\tau)^{\rm lower}$. It is the discontinuity condition that connects these:
\beq\label{twistcutlattice}
	\mathcal C_{x,t}^A : \Psi(x',\tau)^{\rm upper} = 
	A\Psi(x',\tau)^{\rm lower} \quad (x'\geq x).
\eeq
This formulation guarantees that topological covariance still holds in the lattice formulation, as well as the Riemann surface formulation. See Fig.~\ref{figlatticecut}
\begin{figure}
\bc\includegraphics[width=0.5\textwidth]{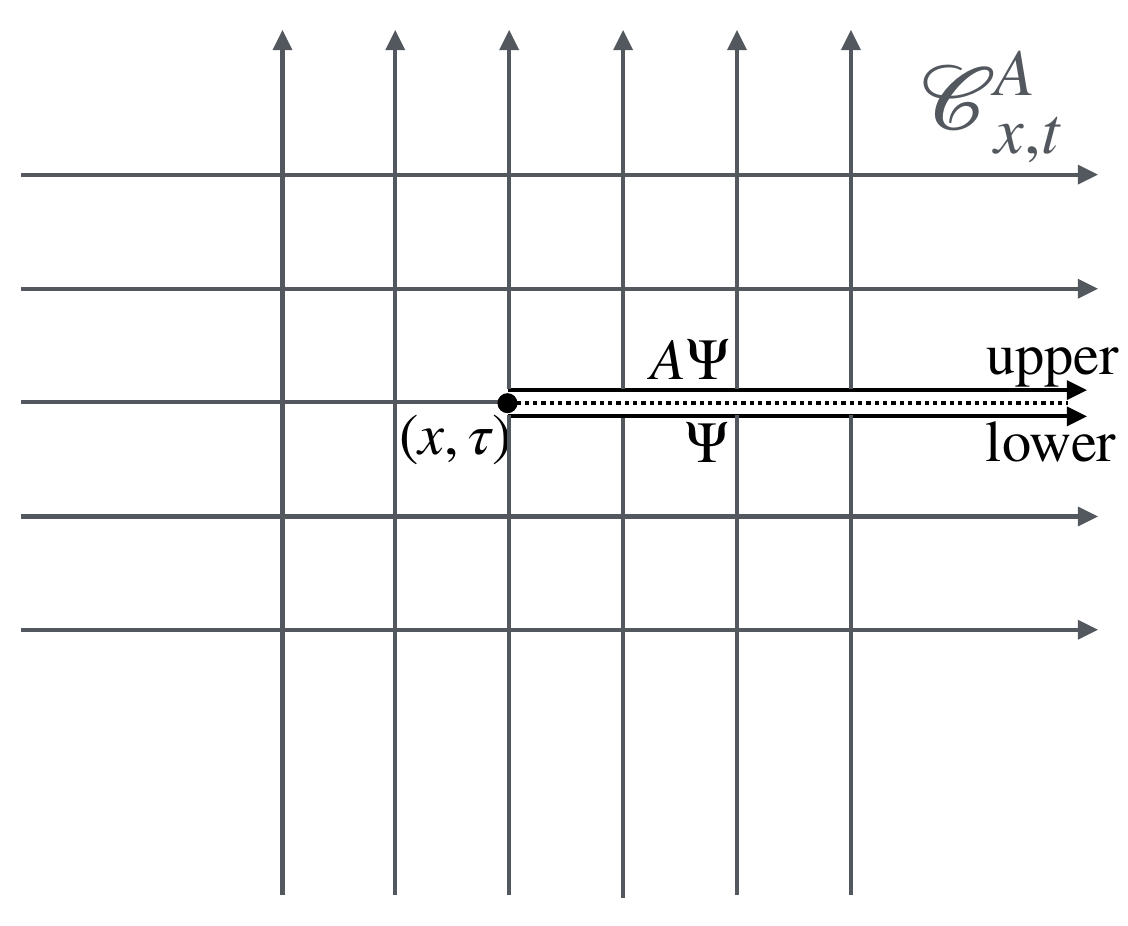}\ec
\caption{The jump condition \eqref{twistcutlattice} on the graph $\mathbb L$, Eq.~\eqref{latticewithcut}: the vertices are the line crossings, and the edges are the segments connecting them. The jump condition is implemented through a doubling of vertices along the tail of the twist field.}
\label{figlatticecut}
\end{figure}

It turns out that this is equivalent to {\em a change of the interaction terms on the edges just above (or just below) the tail $([[x,\infty)),\tau)$}. For instance, if interactions along the vertical (horizontal) edges are given by the matrix $B$ ($C$) acting on the internal degrees of freedom,
\beq\label{modelBC}
	S_{\rm E} = \sum_{(x,\tau)\in\Z^2}\big(\Psi(x,\tau)^{\rm T}B\Psi(x,\tau+1)+\Psi(x,\tau)^{\rm T}C\Psi(x+1,\tau)\big),
\eeq
with $A^{\rm T}BA = B,\,A^{\rm T}CA= C$ (so that $A$ represents an internal symmetry), then the part of the modified action along $x',\tau$, $x'\geq x$ corresponding to the vertical edges is
\beqa
	S_{{\rm E},\,z}^{\rm tail,\,vertical} &=& \sum_{x'\geq x}\Big(\Psi(x',\tau-1)^{\rm T}B\Psi(x',\tau)^{\rm lower} \Big)+
	(\Psi(x',\tau)^{\rm upper})^{\rm T}B\Psi(x',\tau+1)\n
	&=&
	\sum_{x'\geq x}\Big(
	\Psi(x',\tau-1)B\Psi(x',\tau)^{\rm lower} +
	(\Psi(x',\tau)^{\rm lower})^{\rm T}A^{\rm T}B\Psi(x',\tau+1)\Big).
	\label{edgechange}
\eeqa
Discarding the now irrelevant $([[x,\infty)),\tau)^{\rm upper}$, we obtain the model on $\Z^2$ with the interaction along the vertical edges $\big((x',\tau),(x',\tau+1)\big),\,x'\geq x$ modified to $A^{\rm T}B=BA^{-1}$ instead of $B$. See Fig.~\ref{figedges}.
\begin{figure}
\bc\includegraphics[width=0.5\textwidth]{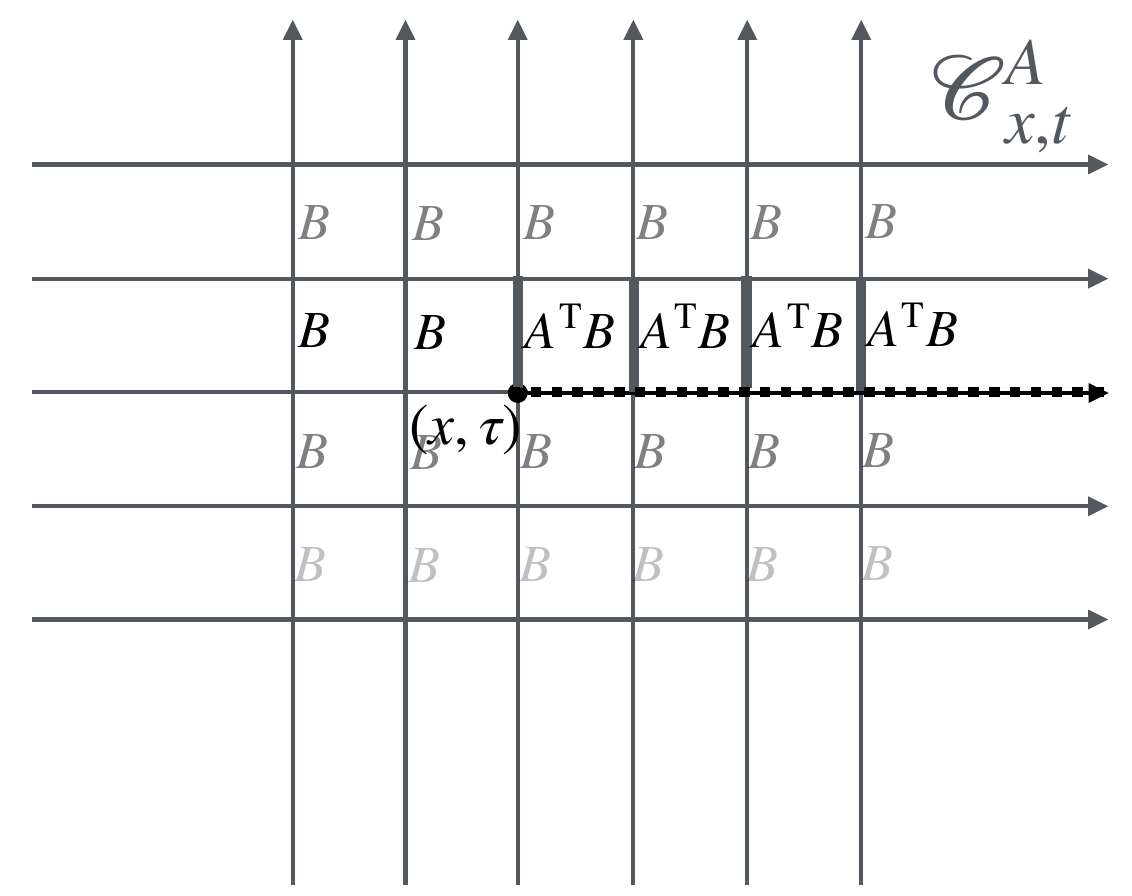}\ec
\caption{For the model \eqref{modelBC}, the jump condition \eqref{twistcutlattice} on the lattice $\mathbb L$ gives rise to a change of edge interaction on the original lattice $\Z^2$ along the tail of the twist field, Eq.~\eqref{edgechange}.}
\label{figedges}
\end{figure}

The exampe \eqref{spintwistintro} is the case $B=\1$, with $A=R\in O(d)$ and $\Psi(x,\tau) = \sigma_{(x,\tau)}\in\R^d,\,|\sigma_{\vec x}|=1$.

Such a change of interactions along a string of edges has a natural interpretation in the operator formulation of lattice systems based on transfer matrices. It is, in fact, the original disorder parameter in the Ising model, characterising the disordered phase \cite{kadanoff1971determination}: in the example above, $\Psi(x',\tau') \to \sigma_{x',\tau'}\in \{\pm 1\}$ is the local spin,  $B=C = -\beta J$ where $\beta$ is the inverse temperature and $J$ the coupling, and $A = -1$ represents the $\Z_2$ symmetry (the case $d=1$ of the example \eqref{spintwistintro}). This is perhaps the closest connection between twist fields and the ``dislocations'' I talked about in Section \ref{sectmodels} in order to heuristically introduce twist fields -- the disorder parameter measures ``spin dislocations''.

\subsubsection*{Classical Hamiltonian systems}

In classical Hamiltonian systems, exchange relations \eqref{exch}, \eqref{exch2} do not immediately make sense, as the order of observables does not matter. Nevertheless, the standard exponential form of Subsec.~\ref{ssectexp} can be directly written, and then, a natural replacement of the exchange relation \eqref{exch2} holds: with $\mathcal T(x) = \mathcal T_q(x)$ from Eq.~\eqref{defTUgen},
\beq\label{exch2cl}
	\exp\Big[\int_{x}^\infty \dd y\,\{q(y),\cdot\}\Big] o(x') = \lt\{\ba{ll}
	\sigma_{Q} (o(x')) & (x'\gg x)\\
	o(x') & (x'\ll x)
	\ea\rt.\quad (\mathcal T\in \mathfrak T,\ o\in\mathfrak L_0).
\eeq
In this replacement, the algebra is invovled. But for evaluating averages of the twist field $\mathcal T_q(x)$ itself, the algebra does not play as fundamental role. Instead, it is locality based on the state, Subsec.~\ref{ssectstate} that is perhaps the most relevant notion, and that should be taken as the basis of the standard exponential form. I will come back to this in Subsec.~\ref{ssectld}.

There are path-integral formulations based on a transfer-matrix formalism in classical Hamiltonian systems, so there should be formulations paralleling Subsec.~\ref{ssecttwistpath} and \ref{ssectriemann} (which, as far as I am aware, have not been developed yet). Topological invariance, Subsec.~\ref{ssecttopo}, also applies, from local conservation laws.

\subsubsection*{Discrete time}

Although our formalism was developed in continuous time, there is no difficulty in adapting it to discrete time evolution, such as in quantum circuits \cite{fisher2023random,piroli2024approximating}. Locality, as defined in Subsec.~\ref{ssectdyn}, Eq.~\eqref{local}, is now with respect to a {\em pivot} observable $h$, which does not need to be a Hamiltonian density, but which defines the topological notion on which we base locality. Exchange relations \eqref{exch}, \eqref{exch2} makes sense, as they do in any operator formalism, with this notion of locality. We only ask, then, that the discrete time evolution be implemented by a {\em local transformation},
\beq
	\tau_t = e^{\ri \mathrm F t},\quad t\in\Z
\eeq
for some ``Floquet'' extensive observable $\mathrm F = \ad \int f$ and $f\in\mathfrak L_0$ a mutually local observable. A symmetry satisfies, instead of \eqref{symmetry},
\beq
	\sigma\circ \tau_t = \tau_t\circ\sigma \quad\forall \; t\in\Z.
\eeq
Ultra-local symmetries in factorised Hilbert spaces (Subsec.~\ref{ssectultra}) and more generally from local observables (Subsec.~\ref{ssectexp}), and their associated twist fields, can immediately be defined, and similarly the construction of App.~\ref{appgen} applies.

Topological invariance based on ultra-local symmetries also holds, Subsec.~\ref{ssecttopo}, where 
\beq
	\mathcal T_q(x,t)\mathcal T_{-q}(x,t+1)
	= \mathcal T_q(x,t\to t+1)
\eeq
and we have
\beq\label{tailultradiscrete}
	\mathcal T_q(x,t)
	=
	\prod_{i=1}^{N-1}\Big(
	\mathcal T_{q}(x_i\to x_{i+1},t_i)
	\prod_{t=t_i}^{t_{i+1}-1}\mathcal T_q(x,t\to t+1)\Big)
	\mathcal T_q(x_N,t_N)
\eeq
in place of \eqref{tailultra}. Locality of $\mathcal T_q(x,t\to t+1)$, more precisely the fact it is supported around $x,t$, results from the calculation \eqref{calculocalt} where $\mathrm H$ is replaced by $\mathrm F$.

\subsubsection*{Beyond internal symmetries: conical fields}

In our construction, we emphasise the importance that the symmetry associated to a twist field be {\em internal}: it preserves the Hamiltonian density $h$, or more generally the pivot observable, Eq.~\eqref{internal}. As explained in Subsec.~\ref{ssectexch}, this guarantees that twist fields $\mathcal T(x)$ are {\em local observable}, in accordance to our main definition \eqref{local}. However, the explicit twist field constructions in Subsec.~\ref{ssectultra} and \ref{ssectexp}, and App.~\ref{appgen}, can be easily extended to seem {\em any local symmetry}, and topological invariance, Subsec.~\ref{ssecttopo} still holds.

Yet, the path integral construction of Subsec.~\ref{ssecttwistpath}, \ref{ssectriemann} is drastically modified if the symmetry is not internal, because if it does not act on the internal space of the field configuration, the discontinuity condition is more involved. In fact, this is symptomatic of a more fundamental aspect of twist fields that are not associated to internal symmetries, which for instance modifies the way to construct their matrix elements in QFT. In order to illustrate this, I consider the path integral formulation for a certain type of twist fields associated to an ``external'' (i.e.~not internal) symmetry: that of {\em space-time Lorentz transformations}, or rather, as I will take the Euclidean field theory setup, {\em Euclidean rotations}. As we will see, such space-time-symmetry twist fields {\em alter the structure of space-time}. This was studied in \cite{castro2018conical}.

Consider a rotation-invariant Euclidean field theory, and the space-time  symmetry for counter-clockwise rotations by $\theta\in[0,2\pi)$, acting on scalar observables as
\beq
	\sigma_\theta (o(x,\tau)) = o(\cos\theta \,x + \sin\theta \,\tau,
	\sin\theta \,x - \cos\theta\,\tau).
\eeq
This does not preserve $h(x,\tau)$, and {\em neither does it commute with the translation symmetry}. It is implemented by the boost operator (that is $U_A\to e^{-\theta B}$ in \eqref{ultrapath}),
\beq
	B = \int \dd x\,xh(x),\quad \sigma_\theta(o(x,\tau)) = e^{-\theta B}o(x,\tau)e^{\theta B},\quad \sigma_\theta (h(x,\tau)) \neq h(x,\tau)
\eeq
where $h(x)$ is the Hamiltonian density. Assuming that the fundamental field is a scalar, this introduces the rotation differential operator $R$ on field configurations,
\beq
	\sigma_\theta(\psi(x,\tau)) \stackrel{\text{path integral}}\to
	e^{-\theta R}\Psi(x,\tau) =\Psi
	(\cos\theta \,x + \sin\theta \,\tau,
	\sin\theta \,x - \cos\theta\,\tau).
\eeq
The twist field for this symmetry is then naturally defined in the path-integral formalism as
\beq\label{Tpathconical}
	\frc{\bra \vac|\mathsf T\big[ \mathcal T_\theta(0) o_1(z_1)\cdots o_n(z_n)\big]|\vac\ket}{\bra \vac| \mathcal T_\theta(0)|\vac\ket}
	=
	\frc{\int_{\mathcal C_{0}^\theta} [\dd\Psi] \,e^{- S_{{\rm E},\,0}[\Psi]}
	o_1[\Psi](z_1)\cdots o_n[\Psi](z_n)}{
	\int_{\mathcal C_{0}^\theta} [\dd\Psi] \,e^{- S_{{\rm E},\,0}[\Psi]}}
\eeq
with
\beq\label{twistcutconical}
	\mathcal C_{x,t}^\theta : \Psi(x, 0^+) = \lt\{
	\ba{ll}
	e^{\theta R}\Psi(x, - 0^+) & (x> 0)\\
	\Psi(x, - 0^+) & (x< 0).
	\ea\rt.
\eeq
One can check that this satisfies the exchange relations \eqref{exch2} with other scalar observables $o(x,\tau)$.

Relation \eqref{twistcutconical} connects the field configuration above the tail to the {\em clockwise} rotation of that below it. Thus, it renders the section of the field configuration lying on the wedge of angles in $[2\pi-\theta,2\pi)$ {\em irrelevant} to the evaluation of the path integral -- it cuts it out. There remains a path integral on a reduced space, where this wedge has been taken out:
\beq\label{Tpathconicalreduced}
	\frc{\bra \vac|\mathsf T\big[ \mathcal T_\theta(0) o_1(z_1)\cdots o_n(z_n)\big]|\vac\ket}{\bra \vac| \mathcal T_\theta(0)|\vac\ket}
	=
	\frc{\int_{\mathcal C_{0}^{{\rm conical},\,\theta}} [\dd\Psi] \,e^{- S_{{\rm E},\,0}^\theta[\Psi]}
	o_1[\Psi](z_1)\cdots o_n[\Psi](z_n)}{
	\int_{\mathcal C_{0}^{{\rm conical},\,\theta}} [\dd\Psi] \,e^{- S_{{\rm E},\,0}^\theta[\Psi]}}
\eeq
with
\beq\label{twistcutconical}
	\mathcal C_{x,t}^{{\rm conical},\,\theta} : \Psi(r, 0) = 
	\Psi(r \cos\theta, -r\sin\theta) \quad (r> 0)
\eeq
and
\beq\label{SavoidEconical}
	S_{{\rm E},\, 0}^{\rm \theta}[\Psi] = \int_{\R^2\setminus
	\{(r\cos\alpha,r\sin\alpha): r\geq 0,\,\alpha\in [2\pi-\theta,2\pi)\}} \dd x'\dd \tau'\,\mathcal L_{\rm E}[\Psi](x',\tau').
\eeq
Thus, this is a path integral where {\em the angle around the point $(0,0)$ is reduced $2\pi-\theta<2\pi$}. The twist field has introduced a {\em conical singularity}, an infinite positive curvature point, and therefore changed the structure of space-time. This is we refer to as a {\em conical fied}. A similar construction can be done for $\theta=-\alpha<0$, in which case the conical twist field introduces a negative infinite curvature, where the angle has been increased to $2\pi+\alpha>2\pi$. See Fig.~\ref{figconical}.
\begin{figure}
\bc\includegraphics[width=0.3\textwidth]{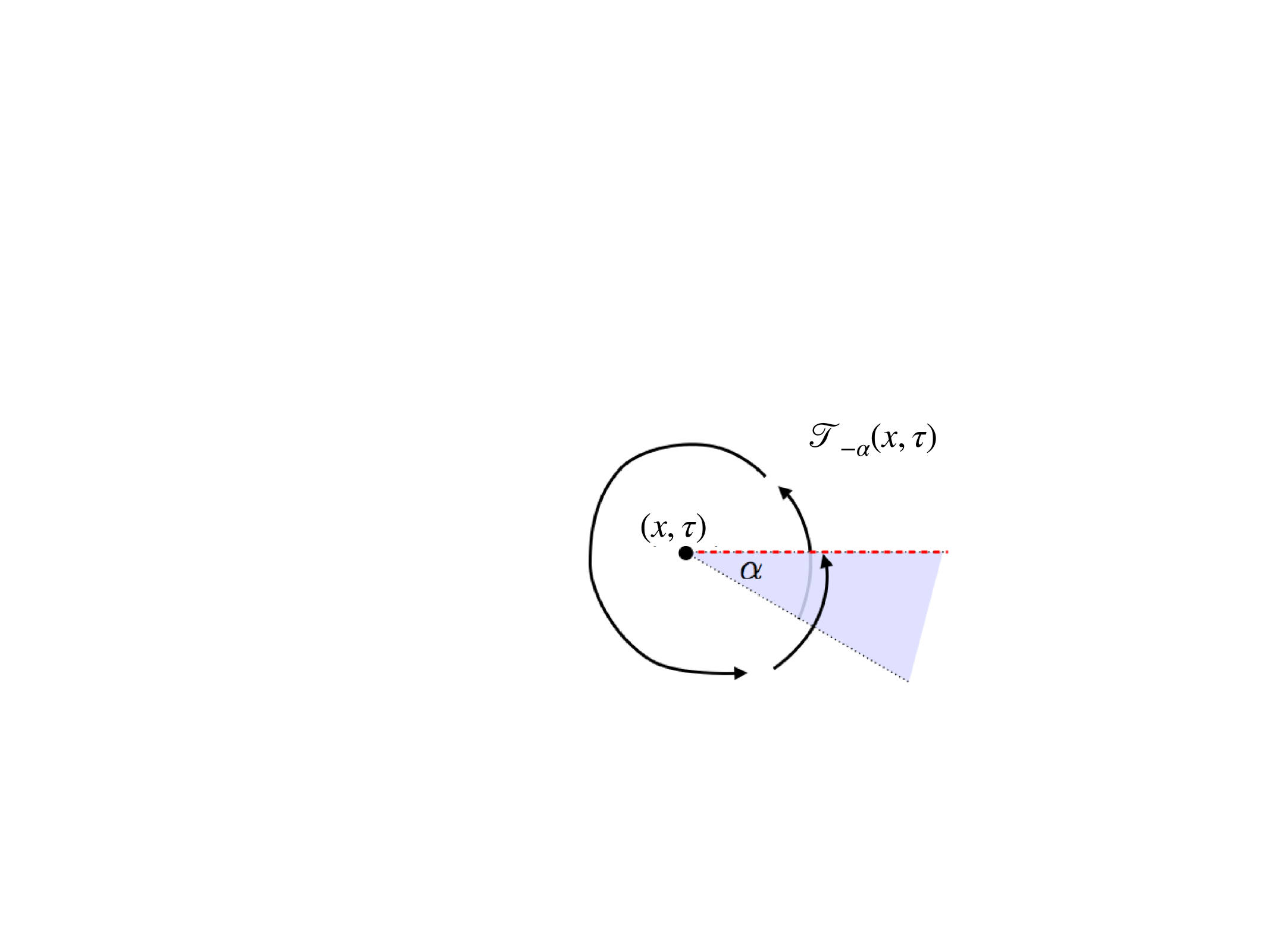}\ec
\caption{The conical twist field at negative angles $\theta=-\alpha$ introduces an infintie negative curvature singularity, increasing the angle around it. Picture adapted from \cite{castro2018conical}.}
\label{figconical}
\end{figure}

It is also possible to see this with the resonant Riemann surface construction of Subsec.~\ref{ssectriemann}. In the monodromy \eqref{monodromy}, the right-hand side involves a counter-clockwise rotation by $\theta$, and therefore a full $2\pi$-angle counter-clockwise continuation of the observable brings it an angle $\theta$ too far, in agreement with a conical singularity of angle $2\pi-\theta$.

\section{Some applications}\label{sectappli}

\subsection{Jordan-Wigner transformation, bosonisation, height fields}\label{ssectjordan}

One of the most basic and earliest applications where twist fields occur is Jordan-Wigner transformations. The main idea is that a Jordan-Wigner ``string'', which allows one to write spin operators in terms of Fermion operators and vice versa, is in fact a product of two twist fields, one at the start, the other at the end, of the string. This is closely related to bosonisation, which relates Fermions to Bosons instead of spins. Bosonisation makes clear the fact that un-compactified Bosons are in fact height fields, and this, then, connects to the notion of twist fields from height fields more generally, with applications in classical systems. I briefly discuss all these subjects here.

\subsubsection*{Jordan-Wigner transformation}

A simple example is the Ising quantum chain,
\begin{equation}
            H = -\frac{J}{2} \sum_{x = 1}^L \sigma_x^1 \sigma_{x+1}^1 + h \sigma_x^3,
            \label{eq:IsingHamiltonian}
\end{equation}
where $L$ is the lattice length, $h$ is the strength of the magnetic field and $\sigma_x^\alpha$ are the Pauli matrices acting on site $x$, with periodic boundary conditions $\sigma_{L+1}^\alpha := \sigma_1^\alpha$. This Hamiltonian preserves the total spin $S^3 = \sum_{x=1}^L \sigma_x^3$, which generates spin rotations via $U_\lambda = e^{\ri \lambda S^3}$, hence the associated twist field \eqref{Tspin} can be constructed.

In the Jordan–Wigner transformation, we introduce Fermionic creation and annihilation operators $c_x^\dag$ and $c_x$ in order to represent Pauli matrices as
\begin{equation}
            \sigma_x^3 =1- 2c_x^\dagger c_x,\qquad \sigma_x^1 = \prod_{x' = 1}^{x-1} \left ( 1-2c_{x'}^\dagger c_{x'}\right )\left(c_x + c_x^\dagger\right).
        \label{eq:ferminintro}
\end{equation}
This is useful, as in terms of these Fermionic operators, the Hamiltonian becomes almost quadratic -- more precisely,
\begin{equation}
            H = P_{\rm e} H_{\rm e} + P_{\rm o} H_{\rm o},
\end{equation}
where
\begin{equation}
            P_{\rm e/o} = \frac{1\pm e^{i \pi N}}{2}
\end{equation}
are projectors on the subspaces of the Hilbert space with even and odd fermion number $N = \sum_{x=1}^L c_x^\dag c_x = (1+S^3)/2$, and $H_{\rm e}$ and $H_{\rm o}$ are quadratic in $c_x$'s, $c_x^\dag$'s and preserve fermion number, with, respectively, periodic and anti-periodic boundary conditions.

\subsubsection*{Spin twist fields}

One can check that the inverse transformation is 
\begin{equation}
            c_x = \prod_{x' = 1}^{x-1}\left(\sigma_{x'}^3\right) \sigma_x^+, \quad c_x^\dag = \sigma_x^- \prod_{x' = 1}^{x-1}\left (\sigma_{x'}^3\right) ,
\end{equation}
where $\sigma_x^\pm = \frac{1}{2}(\sigma_x^1 \pm i \sigma_x^2)$. It is also simple to verify that
\beq
	e^{\frc12 \ri\pi \sigma_x^3} = \ri \sigma^3_x,
\eeq
and therefore {\em fermionic operators are products of twist fields associated to the spin rotation symmetry}: with $\mathcal T_\lambda^{S^3}(x)$ constructed as in Eq.~\eqref{Tspin},
\beq
	c_x = \ri^{x-1} \mathcal T_{-\pi/2}^{S^3}(1)\mathcal T_{\pi/2}^{S^3}(x)\sigma_j^+,\quad
	c_x^\dag = \ri^{1-x} \sigma_j^-\mathcal T_{-\pi/2}^{S^3}(x)\mathcal T_{\pi/2}^{S^3}(1).
\eeq
Note how the product of a twist field and its inverse (that associated with the inverse symmetry) guarantees that the tail is supported on a finite interval, and therefore allows us to define the result on a finite volume $L$. Note also how the Jordan-Wigner transformation involves a twist field at the {\em arbitrary} point $x=1$ -- the chain is periodic, hence any other points could have been taken. This only affects where the Fermionic anti-periodic boundary conditions are imposed in $H_{\rm o}$.

\subsubsection*{$U(1)$ and $\Z_2$ twist fields of the Fermionic theory}

Most importantly for computational purposes, however, is the fact that in the Fermionic representations, the spin observables $\sigma_x^\pm$ {\em also are twist fields, but now with respect to the generator of the $\Z_2$ subgroup of the $U(1)$ symmetry of the fermionic theory}. The $U(1)$ symmetry of the Fermionic model is
\beq
	U_\lambda = e^{\ri\lambda N},\quad \sigma_\lambda(o) = U_\lambda o U_{-\lambda},
\eeq
and $\mathcal T_\lambda^{N}(x) = e^{-\ri x\lambda/2}\mathcal T_{-\lambda/2}^{S^3}(x)$ are the $U(1)$ twist fields associated to the symmetries $\sigma_\lambda$. We note that
\begin{equation}
            e^{i \pi N} = \prod_{x = 1}^L (1-2c_x^\dagger c_x)
\end{equation}
and therefore, 
\beq
	\sigma_x^+ = \mathcal T_{\pi}^{N}(1)
	\mathcal T_{\pi}^{N}(x) c_x,
	 \quad
	 \sigma_x^- = c_x^\dag\mathcal T_{\pi}^{N}(x)
	 \mathcal T_{\pi}^{N}(1) ,
\eeq
where we use $\mathcal T_{\pi}^{N}(x) = \mathcal T_{-\pi}^{N}(x)$ are the $\Z_2$ twist fields, for the symmetry $\sigma_\pi$ generating the $\Z_2$ group.

Observe that $\mathcal T_{\pi}^{N}(x) c_x$ still is a twist field. This appears to be of the form \eqref{tprimet}, where we multiply by a local observable, the Fermion observable $c_x$. However, this is only partially true: Fermions {\em anti-commute}, and as a consequence, with respect to Fermionic local observables, $\mathcal T_{\pi}^{N}(x) c_x$ has a tail on the left,
\beq\label{Tminusfermion}
	\mathcal T_\pi^-(x):=\mathcal T_{\pi}^{N}(x) c_x \in \mathfrak T^-_{\sigma_\pi}.
\eeq
It is different from ${\mathcal T_\pi^N}^-\in \mathfrak T^-_{\sigma_\pi}$ built as \eqref{Uleftright} by multiplying by $U_\lambda$. In particular, these two left-tailed twist fields transform differently under the $\Z_2$ symmetry $\sigma_\pi$: the latter is even, the former is odd,
\beq
	\sigma_\pi({\mathcal T_\pi^N}^-(x)) = {\mathcal T_\pi^N}^-(x),\quad
	\sigma_\pi(\mathcal T_\pi^-(x)) = -\mathcal T_\pi^-(x).
\eeq
We see that twist fields in the same family may have different transformation properties under the symmetry to which they are associated. In particular, the exchange relations \eqref{exch} are affected by this.

With these $\Z_2$-odd, left-tailed, $\Z_2$ twist fields, we write
\beq
	\sigma_x^+ = \mathcal T_{\pi}^{N}(1)
	\mathcal T_{\pi}^-(x),
	 \quad
	 \sigma_x^- = \big(\mathcal T_{\pi}^-(x)\big)^\dag
	 \mathcal T_{\pi}^{N}(1) ,
\eeq
and, for instance, the vaccum two-point function simplifies to
\beq\label{twopttwist}
	\bra\vac| \sigma_x^- \sigma_{x'}^+|\vac\ket
	=
	\bra\vac | \big(\mathcal T_{\pi}^-(x)\big)^\dag\mathcal T_{\pi}^-(x')
	|\vac\ket
	=
	\bra\vac | \big(\mathcal T_{\pi}(x)\big)^\dag\mathcal T_{\pi}(x')
	|\vac\ket,
\eeq
which we have re-written for convenience in terms of $\Z_2$-odd, right-tailed twist fields $\mathcal T_{\pi}(x) = e^{\ri\pi N}\mathcal T_{\pi}^-(x)$. In the ferromagnetic phase of the Ising model, $|h|<J$, the vacuum in infinite volume breaks the $\Z_2$ symmetry, and $\sigma_x$ aquires a non-zero vacuum expectation value (VEV). In terms of twist fields, this means that
\beq\label{vev}
	\lim_{|x-x'|\to\infty}\bra\vac | \big(\mathcal T_{\pi}(x)\big)^\dag\mathcal T_{\pi}(x')
	|\vac\ket
	=: |\bra\vac|\mathcal T_\pi|\vac\ket|^2\neq 0,
\eeq
which {\em defines} the (absolute value of the) twist-field VEV $|\bra\vac|\mathcal T_\pi|\vac\ket|$.

Expression \eqref{twopttwist} is useful, as there are strong methods to evaluate twist fields two-point functions in free-Fermion models, especially near critical points, some of which I briefly overview in Subsec.~\ref{ssectfree}.

\subsubsection*{Bosonisation and height fields}

A related transformation is that of bosonisation \cite{sachdevbook}. Take for instance a QFT model with a complex Fermion $\psi(x),\,\psi^\dag(x)$. As a part of the bosonisation transformation, one writes the density of the Fermion number conserved quantity $Q = \int \dd x\,\psi^\dag(x)\psi(x)$, as a total derivative of a bosonic field, $\psi^\dag(x)\psi(x) = \p_x \varphi(x)$. The Fermion number is the generator of the $U(1)$ group of phase multiplications, $e^{\ri \lambda Q}\psi(x)e^{-\ri \lambda Q} = e^{-\ri\lambda} \psi(x)$. In our construction \eqref{defTUgenconthf}, this means that {\em the twist field associated to the $U(1)$ continuous ultra-local symmetry group is an exponential of the Bosonic field},
\beq\label{vertex}
	\mathcal T_\lambda^{\rm bos} = e^{\ri\lambda \int_x^\infty\dd x'\,\psi^\dag(x')\psi(x')}
	=
	e^{-\ri \lambda\varphi(x)}.
\eeq

A much-studied example of bosonisation is the relation between the Thirring model of an interacting relativistic Dirac Fermions $\psi$ (with internal spin structure), and the sine-Gordon model of an interacting relativistic Boson $\phi$ \cite{korepin1997quantum,sachdevbook}. In this case, the $U(1)$ height field is identified as $\varphi =\phi$.

We may see directly in the sine-Gordon model that $\phi$ is a height field with the properties explained in Subsec.~\ref{ssectsymmetry}, without making the connection \eqref{vertex} with the Fermion number density operator $\psi^\dag(x)\psi(x)$ of the Thirring model. That is, we only have to argue that $\lim_{\ell\to\infty} (\phi(\ell/2)-\phi(-\ell/2))$ is extensive. Indeed, in the sine-Gordon model, there is a non-compact $\Z$ symmetry of discrete shifts of the field $\phi\mapsto \phi+2\pi n /\beta,\,n\in\Z$ because of periodicity of the potential (Eq.~\eqref{HQFT} with $V(\phi^2) = g\cos(\beta\phi)$). Therefore, the sine-Gordon field is not compactified by the interaction, and its differences can take large values. Extensivity in the algebraic sense of \eqref{extensivehf} is seen by admitting, within the algebra of observables of the sine-Gordon model, the space of observables generated by the Dirac Fermion $\psi$ and its conjugate, expressed via bosonisation in terms of sine-Gordon fields. Then, one chooses in \eqref{extensivehf} $o = \psi$ in order to get a non-zero result. These natural Fermionic local observables have non-zero overlap with the sector of asymptotic states with single particles that correspond, semi-classically, to sine-Gordon topological solitons. Topological solitons exist thanks to the non-confining property of the potential -- there is thus logical consistency in our arguments. Extensivity in the sense of the state, Eqs.~\eqref{cond1}, \eqref{cond2}, is perhaps more immediate in the path-integral formulation: as the interaction potential is not confining, one essentially has a Brownian motion in space, and expects
\beq\label{phiellext}
	\bra (\phi(\ell/2)-\phi(-\ell/2))^2\ket \propto \ell,
\eeq
and therefore $Q_\ell = \phi(\ell/2)-\phi(-\ell/2)$ is, in the limit $\ell\to\infty$, an extensive quantity $\mathsf Q$. The scaling \eqref{phiellext} can be verified numerically in the Euclidean path integral of a finite-temperature state, for instance. See the discussion in \cite{del2023exact}.

The fields $e^{-\ri \lambda\varphi(x)}$ are the natural scaling fields in the sine-Gordon model, mapping to ``vertex operators'' at the CFT fixed point (Subsec.~\ref{ssectqft}); they are called vertex operators also in the sine-Gordon model. The viewpoint of such scaling fields, or vertex operators, as $U(1)$ twist field is particularly powerful for evaluating correlation functions: either in terms of integrable PDEs at values of the sine-Gordon coupling constants where the Thirring model specialises to a free relativistic Fermion, Subsec.~\ref{ssectfree} \cite{bernard1994differential}, or at all values of couplings using hydrodynamic methods, Subsec.~\ref{ssectld} \cite{del2023exact}.

But the notion of height fields goes beyond. Take the infinite Toda chain, Eq.~\eqref{Hclassical} with $f(a) = e^{-a}$. The differences $q_{x+1}-q_x$ tend to be positive, hence the total stretch between positions $\ell$ and $-\ell$,
\beq
	Q_\ell = q_{\ell}-q_{-\ell},
\eeq
tends to take values that grow with $\ell$. In fact, $Q_\ell$ is extensive in the sense of states, Eqs.~\eqref{cond1}, \eqref{cond2}, within thermal states or other generalised Gibbs ensembles of the integable Toda chain. Therefore,
\beq\label{heightfieldtoda}
	\varphi(x) = q_x
\eeq
is a height field for the density $q(x) = q_{x+1}-q_x$ of the total stretch \eqref{totalstretch}, and our general theory of twist fields allows us to evaluate the generating function of stretch cumulants,
\beq
	\mathcal T_\lambda^{\rm Toda}(x) =  e^{\lambda \varphi(x)},\quad
	\bra \mathcal T_\lambda^{\rm Toda}(x) \b{\mathcal T}_\lambda^{\rm Toda}(0)\ket =  \bra e^{\lambda (q_x - q_0)}\ket.
\eeq
Techniques of Subsec.~\ref{ssectld} apply.

\subsection{QFT: renormalisation, form factors, twisted modules}\label{ssectqft}

The exchange relations \eqref{ssectexch}, and the constructions of Subsec.~\ref{ssectultra} - \ref{ssecttwistpath} for twist fields, can be applied to relativistic quantum field theory (QFT)\footnote{The constuction of App.~\ref{appgen} can also in principle, although I do not know if Hilbert-space generating local observables of the type needed exist.} . There are, however, a number of subtleties.

\subsubsection*{Renormalisation and primary twist fields}

These come from the fact that in QFT, the number of degrees of freedom on any finite section of space, or space-time, is infinite. In the viewpoint of QFT as an emergent theory near critical points, this comes from the fact that finite lengths in QFT are infinite lengths in any underlying microscopic model (i.e.~UV regularisation of the QFT model). This infinite amount of degrees of freedom means that observables in QFT correspond to {\em renormalised observables in the underlying microscopic model}. For instance, if the approach to a critical point is parametrised by a correlation length $\xi>0$, then a QFT observable may be defined via the {\em scaling limit}\footnote{As we are discussing relativistic QFT, the dynamical critical exponent is 1.}
\beq\label{multiplicativeRG}
	o(x,t) = \lim_{\xi\to\infty}\xi^d\,o^{\rm micro}(x/\xi,t/\xi)
\eeq
for some $d\geq 0$. This is in the sense that for a quantum critical point of the ground state $|{\rm gs}\ket_g$ of a microscopic model, with correlation length $\xi(g)$ diverging at the critical parameter value $\xi(g)\stackrel{g\to g_c}\to\infty$, one would set, for instance,
\beq
	\bra\vac| o_1(x_1,t_1)o_2(x_2,t_2)|\vac\ket_{\rm QFT}
	= \lim_{g\to g_c}
	\xi(g)^{d_1+d_2}\,{}_g\bra{\rm gs}|o_1^{\rm micro}(\tfrac {x_1}{\xi(g)},\tfrac{t_1}{\xi(g)})
	o^{\rm micro}_2(\tfrac{x_2}{\xi(g)},\tfrac{t_2}{\xi(g)})|{\rm gs}\ket_g,
\eeq
while for a finite-temperature critical point of a classical statistical model in a state $\bra\cdots\ket_T$ with correlation length $\xi(T)$ diverging at the critical temperature $\xi(T)\stackrel{T\to T_c}\to\infty$,
\beq
	\bra o_1(z_1)o_2(z_2)\ket_{\text{Euclidean QFT on the plane}}
	=\lim_{T\to T_c}
	\xi(T)^{d_1+d_2}\bra o_1^{\rm micro}(\tfrac{x_1}{\xi(T)}\tfrac{\tau_1}{\xi(T)})
	o^{\rm micro}_2(\tfrac{x_2}{\xi(T)},\tfrac{\tau_2}{\xi(T)})\ket_{T}
\eeq
where $z_i = (x_i,\tau_i)$.

Renormalisation is an intricate process, and the relation between QFT observables and microscopic observables may be more complicated than \eqref{multiplicativeRG}. Under scale transformations, the QFT model is not invariant, but instead its parameters change, thus giving rise to a flow, the renormalisation group (RG) flow, in the space of QFT's. Both at large length scales (small temperatures, large distances), and at small length scales, the QFT tends to a conformal field theory (CFT), a fixed point of the RG flow, where conformal invariance, including scale invariance, emerges. In this sense, there is a mapping that takes observables $o(x,t)$ in QFT, to observables $o^{\rm UV, IR}(x,t)$ in the corresponding CFT, either at the UV (small scales) or IR (large scales) fixed point. The IR fixed point is often trivial, composed of a single observable, the identity $\1$. This happens if all correlation functions in the microscopic model vanish at large distances -- that is, all correlation lengths are finite before the QFT limit is taken. But, in UV-complete models, the UV fixed point is non-trivial. The mapping $o\to o^{\rm UV}$ is not injective: there are ``ghost fields'', which map to 0. But these ghost fields usually are spanned by fields of the form $\xi^{-a} o$ where $o$ is not a ghost field, and $\xi^{-a}$ is a striclty negative power of a correlation length $\xi$ (all correlation lengths diverge at the UV fixed point). Then, observables $o$ in QFT that are not ghost fields can be classified as per the {\em transformation properties under the conformal symmetry} of the observables $o^{\rm UV}$. It is usually possible to organise these in terms of {\em primary observables}, and their descendants. Primary observables are such that $o^{\rm UV}$ transforms into itself under scale transformations and rotations\footnote{There are cases where this is not possible, and logarithmic observables appear, where a number of observables transform into themselves in a non-diagonalisable way under scale transformations.}: they have a well-defined scaling dimension and a well-defined spin. They are usually obtained from the microscopic model via a multiplicative renormalisation \eqref{multiplicativeRG}, where the scaling dimension would be $d$.

Because of the scaling limit, e.g.~\eqref{multiplicativeRG}, microscopic distances are infinite for every finite QFT distances. Then, it is conventional to define twist fields in QFT by equal-time exchange relations \eqref{exch}, \eqref{exch2} that hold {\em for every non-zero distances}:
\beq\label{exchQFT}
	\mathcal T(x) \mathcal T'(x') = \lt\{\ba{ll}
	\sigma_{\mathcal T} (\mathcal T'(x'))\, \mathcal T(x) & (x'> x)\\
	\mathcal T'(x')\, \sigma_{\mathcal T'}^{-1} (\mathcal T(x)) & (x'< x)
	\ea\rt. \quad(\mathcal T,\mathcal T'\in \mathfrak T)
\eeq
and
\beq\label{exch2QFT}
	\mathcal T(x) o(x') = \lt\{\ba{ll}
	\sigma_{\mathcal T} (o(x'))\, \mathcal T(x) & (x'> x)\\
	o(x')\mathcal T(x) & (x'< x)
	\ea\rt.\quad (\mathcal T\in \mathfrak T,\ o\in\mathfrak L_0).
\eeq

Further, it is usually possible to define {\em primary twist fields} as those that additionally have a well-defined scaling dimension, which is the smallest possible. In many examples, one observes that fixing its transformation properties under all other symmetries of the model, a primary twist field is {\em unique}, and often, those constructed as \eqref{defTU}, \eqref{defTUgencont} and \eqref{Tpath} give rise, in the scaling limit under multiplicative renormalisation \eqref{multiplicativeRG}, to such unique primary twist fields. In particular, a twist field $\mathcal T$ that is spinless and that has a well-defined scaling dimension, along with its conjugage $\b{\mathcal T}$, has two-point function that behaves as
\beq\label{shortdist}
	\bra\vac|\mathcal T(x,t)\b{\mathcal T}(0,0)|\vac\ket \sim
	|x^2-t^2|^{-d}\qquad (x^2-t^2\to0),
\eeq
where $d$ is the scaling dimension of $\mathcal T$ and $\b{\mathcal T}$, and where the factor 1 in front is the {\em conformal normalisation}. This, in fact, holds also in finite-temperature states, and any normalisable state of the QFT, and has an equivalent in Euclidean QFT,
\beq\label{shortdisteucl}
	\bra\mathcal T(x,\tau)\b{\mathcal T}(0,0)\ket \sim
	|x^2+\tau^2|^{-d}\qquad (x^2+\tau^2\to0).
\eeq

\subsubsection*{Descendant and composite twist fields}

Because of the renormalisation \eqref{multiplicativeRG}, the product of two fields at the same point is generically not well-defined in QFT. Instead, one has the notion of operator product expansion (OPE). Then, instead of defining twist families as in \eqref{tprimet}, one uses the OPE to obtain fields in a given twist family: if $\mathcal T\in \mathfrak T_\sigma$ and $o\in\mathfrak L_0$, then
\beq\label{oTOPE}
	o(x,t) \mathcal T(0,0) \sim \sum_a C_{o,\mathcal T}^a\lt(\frc{x-t}{x+t}\rt)^{\frc{s_a-s-s_o}2}|x^2-t^2|^{\frc{d_a-d-d_o}2} \mathcal T_a(0,0),\quad
	\mathcal T_a\in \mathfrak T_\sigma,
\eeq
where $d_a,\,d,\,d_o$ are, respectively, the scaling dimensions of $\mathcal T_a,\,\mathcal T,\,o$, and $s_a,\,s,\,s_o$ are their spin. The twist fields $\mathcal T_a$ are called {\em descendant twist fields}, and the leading one is usually denoted
\beq\label{oTdesc}
	(o\mathcal T)(0,0) = \mathcal T_{a_{\rm min}}(0,0)
\eeq
where $a_{\rm min}$ is the value of $a$ that minimises $d_a$ (if it is unique).

The monodromy property restricts the OPE coefficients $C_{o,\mathcal T}^a$. Indeed, passing to Euclidean space $t=-\ri \tau$, with $z=x+\ri \tau$, 
\beq
	o(z) \mathcal T(0) \sim \sum_aC_{o\mathcal T}^a e^{\ri\arg(z)(s_a-s-s_o)}|z|^{d_a-d-d_o} \mathcal T_a(0,0)
\eeq
and with \eqref{monodromyeucl}, we must have
\beqa\label{monodroconstraints}
	&& \sum_aC_{o,\mathcal T}^a e^{2\pi\ri (s_a-s-s_o)} e^{\ri\arg(z)(s_a-s-s_o)}|z|^{d_a-d-d_o} \mathcal T_a(0,0)
	\n
	&=&
	\sum_aC_{\sigma(o),\mathcal T}^a e^{\ri\arg(z)(s_a-s-s_o)}|z|^{d_a-d-d_o} \mathcal T_a(0,0).
\eeqa

One may also consider the OPE between two twist fields, associated to {\em potentially different symmetries}, setting $o = \mathcal T'$ in \eqref{oTOPE}. The result is an expansion in terms of twist fields in the familiy associated to the product of symmetry transformations, as per \eqref{productTT}. In QFT, these are usually referred to as ``composite twist fields'' \cite{Horvath_2022}\footnote{Nomenclature varies. Sometimes ``composite twist fields'' are also used for what I called ``descendant twist fields'' in Subsec.~\ref{ssectalg}, referring to twist fields in the same twist families, obtained via the OPE \eqref{oTOPE} in QFT. I prefer reserving ``composite'' to the composition of two twist fields, as ``composing'' a twist field with a local observable is not a composition of twists. In the context of symmetry resolved entanglement, it is truly composite twist fields that occur, composing branch-point twist fields with a copy-internal symmetry twist field.}. The simplest case of a spin-less twist field and its conjugate is not a composite twist field, but the identity
\beq
	\mathcal T(x,t)\b{\mathcal T}(0,0) \sim |x^2-t^2|^{-d} \bf 1,
\eeq
see \eqref{shortdist}.

\subsubsection*{Form factors and vacuum correlation functions}

Once twist fields are defined in QFT, the question arises as to the evaluation of correlation functions involving them.

In QFT, the Hilbert space is spanned by asymptotic states \eqref{qftstates}, and matrix elements of operators in asymptotic states can be obtained the LSZ reduction formula, which relates correlation functions to matrix elements. That is, one constructs wave packets of small extent $\ep$ in rapidity space,
\beq\label{asymptoper}
    A_a^{\rm (in,out)}(\theta;\ep) = \ri 
    \lim_{t\to\mp\infty}
    \int_{\theta-\ep}^{\theta+\ep}\frc{\dd\alpha}{2\ep}
        \int dx\,\big(f_{\alpha,a}(x,t) \p_t \psi_a(x,t) - \p_t f_{\alpha,a}(x,t)
        \psi_a(x,t)\big)
\eeq
with
\beq\label{wavepacket}
    f_{\alpha,a}(x,t) = \exp\lt[\ri m_a\big(t\cosh(\alpha) - x\sinh(\alpha)\big)\rt]
\eeq
where $\psi_a$ is a fundamental local field associated to the particle of type $a$ (App.~\ref{apppath}); here it is assumed to be Bosonic with spin 0 for simplicity. For the free relativistic Boson, the expression inside the limit on $t$ is in fact $t$-independent, and the limit $\ep\to0$ can be  taken immediately, giving the usual annihilation operator. More generally, $\ep$ guarantees that the wave packet is of finite extent on the asymptotic region $t\to\pm\infty$. Asymptotic states are obtained by taking the limit of infinite wave-packet spatial extent, after the infinite-time limit. For instance with $\theta_1>\cdots>\theta_n$,
\beq\label{asympstatesA}
	|\theta_1,\ldots,\theta_n\ket_{a_1,\ldots, a_n} = \lim_{\ep\to 0}A^{\rm (in)}_{a_1}(\theta_1;\ep)^\dag\cdots
	A_{a_n}^{\rm (in)}(\theta_n;\ep)^\dag|\vac\ket.
\eeq
The vacuum is defined as $\lim_{\ep\to 0}A^{\rm (in)}_{a_1}(\theta_1;\ep)\cdots A_{a_n}^{\rm (in)}(\theta_n;\ep)|\vac\ket=0$.

Matrix elements of local fields $o=o(0,0)$ between the vacuum and asymptotic states are called form factors. For instance,
\beq
	F_{a_1,a_2}^o(\theta_1,\theta_2)  = \bra \vac | o|\theta_1,\theta_2\ket_{a_1,a_2}.
\eeq
Functions such as $F_{a_1,a_2}(\theta_1,\theta_2) $ satisfy a number of analytic properties, part of the analytic $S$-matrix theory of QFT \cite{eden2002analytic}, which has been very successful in integrable QFT \cite{smirnov1992form} and has seen much recent developments beyond integrability \cite{poland2019conformal,guerrieri2021rigorous,mizera2024physics}. Some of these analytic properties follow from the observation that after Wick rotation, rapidity shifts become {\em Euclidean rotations}. Indeed, in \eqref{asympstatesA} setting $t=-\ri\tau$ and $\theta = \ri s$, we have
\beq
	f_{\ri s,a}^*(x,-\ri \tau) = \exp\lt[-m_a \mato{cc} \cos s & \sin s\matf
	\mato{c} t\\ x\matf\rt].
\eeq
Defining asymptotic states, via an analytic continuation in time, as wave packets lying in Euclidean space, this means that the pure imaginary shift $\theta \to \theta+\ri s$ is associated to a clockwise rotation by an angle $s$ of the wave packet in Euclidean space. In this way, an $s=\pi$ shift transforms an in-wave packet (with $\psi_a(x,t\to-\infty)$) into the out-wave packet, but travelling in the wrong direction; the use of charge conjugation symmetry then makes it and out-wave packet for the anti-particle (with $\psi_{\b a}(x,t\to+\infty)$). Further, an $s=2\pi-0^+$ shift exchanges the positions of the wave packets, putting them in the out-state configuration (where the higher-rapidity wave packet is positioned, in space, to the right of the lower-rapidity wave packet). 

The consequences of this for twist fields are most easily seen using the path integral formulation \eqref{Tpath}, and in particular the cut \eqref{twistcut} and the monodromy relation \eqref{monodromyeucl}. The symmetry acts as \eqref{sigmapsi}, explicitly
\beq
	\sigma_A(\psi_a) = \sum_b (A^{-1})_{ab}\psi_b
\eeq
and because it commutes with space-time translations, it must be that $A$ (and thus $A^{-1}$) is block diagonal, with $A_{ab} = 0$ if $m_a\neq m_b$. Two main effects are seen. (1) Because of the cut \eqref{twistcut} along the tail $([0,\infty],0)$, the $+\ri\pi$ shift is associated to a possible singularity, coming from different overlaps between out- and in-wave packets in the asymptotic regions $x\to\pm\infty$:
\beqa\label{kinematicpole}
	F_{a_1,a_2}^{\mathcal T_A}(\theta_1+\ri\pi,\theta_2) &\sim&
	\mbox{analytic part of} \ 
	\frc1{2\pi }\int \dd x\,e^{\ri x m_{a_2}(\sinh(\theta_2)-\sinh(\theta_1))}
	\times\\ &&\;\times\;
	\Big(
	\delta_{\b a_1,a_2}\Theta(x<0)+
	\sum_{b_2}(A^{-1})_{a_2,b_2}\delta_{\b a_1,b_2}\Theta(x>0)
	\Big)\quad (\theta_1\to\theta_2).\no
\eeqa
Note that we have chosen to take particle 2 through the cut in order to see the overlap with anti-particle 1. The analytic part of the integral is a pole, its Cauchy principal value part from $\int_0^\infty \dd x\,e^{\ri p x} = \prin\frc{\ri}p + \pi \delta(p)$, which is called {\em kinematic pole}. (2) The $+2\pi \ri$ shift gives an out-state, but with the particle whose rapidity has shifted being affected by the symmetry transformation, because of the monodromy \eqref{monodromyeucl}:
\beq\label{periodicityff}
	F_{a_1,a_2}^{\mathcal T_A}(\theta_1+2\pi \ri,\theta_2)
	= \sum_{b_1} A_{a_1,b_1}F_{a_2,b_1}^{\mathcal T}(\theta_2,\theta_1).
\eeq
Note that because this is a clockwise rotation, it is the symmetry $\sigma_{A}^{-1}$, associated to multiplication by the matrix $A$, Eq.~\eqref{sigmapsi}, that is involved. Other analytic properties are unchanged by semilocality.

Such properties of twist fields two-particle form factors in QFT that is {\em not necessarily integrable} were first discussed in \cite{doyon2009bipartite}, in the context of the branch-point twist field (Subsec.~\ref{ssectentanglement}); see the discussion there for more explanations. I believe, however, Eqs.~\eqref{kinematicpole} and \eqref{periodicityff} were never written at this level of generality.

In integrable systems, such relations hold for many-particle states as well. It is important, in this case, that there be a well-defined notion of asymptotic state for {\em any ordering of rapidities}; this is the case in integrable models because of the full scattering factorises into two-body scattering events. The theory was originally developed with the concept of ``semi-locality index'', where the symmetry acts by multiplication by a pure phase instead of a matrix $A$; I believe the earliest study of form factors of such twist fields was in \cite{karowski1978exact}, see the book \cite{smirnov1992form}. However, the notion is more general and applies to any internal symmetry of the QFT model as reviewed above, something that was first realised and exploited in \cite{cardy2008form} using the cyclic permutation symmetry of replica models in order to define branch-point twist fields and evaluate entanglement entropy, Subsec.~\ref{ssectentanglement}.

The set of form factor equations form what is usually called a {\em Riemann-Hilbert problem}: a problem of determining a function based on its analytic properties. It turns out that it is possible to solve this Riemann Hilbert problem, including \eqref{kinematicpole} and \eqref{periodicityff},  in integrable systems, giving unique solutions under some ``minimality'' assumption. The minimality assumption corresponds to asking for the twist field to be primary (see the discussion above). Having solved these equations, one may then {\em evaluate correlation functions of twist fields} by using a decomposition of the identity. This, along with relativistic invariance and using the fact that $\mathcal T_A$ is spinless, gives the following convergent series in the space-like region $x^2>t^2$:
\beq\label{ffseries}
	\bra\vac|\mathcal T_A^\dag(x,t)\mathcal T_{A}(0,0)|\vac\ket
	=
	\sum_{n=0}^\infty\frc1{n!}
	\sum_{a_1,\ldots,a_n}
	\int \dd^n\theta\,
	|\bra \vac |\mathcal T_A|\theta_1,\ldots,\theta_n\ket_{a_1,\ldots,a_n}|^2
	e^{-\sum_j m_{a_j}\cosh(\theta_j) \sqrt{x^2-t^2}}.
\eeq
This is the main use of form factors. See \cite{smirnov1992form,mussardo2010statistical}.

There is also a partial theory for form factors in finite-entropy-density states such as thermal states \cite{doyon2007finite,cortes2019thermodynamic}, however this still needs to be developed.

\subsubsection*{Radial and angular quantisation, twisted modules, vacuum expectation values}

In QFT, after passing to the path integral formalism, one may re-quantise in various forms, by choosing different foliations of space-time corresponding to different choices of a space and time direction. In order to make space and time essentially equivalent, one usually concentrates on the Euclidean formulation for the vacuum state (that is, on the plane). The above, where form factors are constructed, is the ``quantisation on the line'': the spatial sheets are the lines $(\R,t)$ for all times $t\in\R$, and a spanning set of states is parametrised by field configurations on the line. But one may also do {\em radial quantisation}, where the spatial sheets are the circles $\{(r\cos\theta,r\sin\theta):\theta\in[0,2\pi)\}$ for all ``times'' $r>0$. The canonical coordinates are in fact $(t_{\rm rad}, x_{\rm rad}) = (\log r,\theta)$, and we see that the spatial direction is compact. In this constuction, the set of ``asymptotic state'' is the {\em set of local observables}. This is what is referred to as the {\em state-field correspondence} in CFT \cite{francesco2012conformal}. In that context, it is possible to make a conformal transformation of the plane to the cylinder, that maps this quantisation scheme to the standard quantisation scheme for a system in finite volume, and the ``states'' in state-field correspondence are those on finite volumes. However, the notion of radial quantisation is more general.

In radial quantisation, the set of local observables (as asymptotic states) is then seen as a module for the algebra of local observables (as operators acting on field configurations on the circle). Then the notion of twist family from Subsec.~\ref{ssectalg} gives rise to {\em twisted modules} (see e.g.~the textbook \cite{lepowsky2012introduction}, and the more recent work \cite{bakalov2016twisted}). Translated to our language, a twisted module for an algebra of mutually local observables $\mathfrak L_0$ is a twist family, that is, a space $\mathfrak T_\sigma$ for the twist $\sigma$, where the module is obtained by inserting a twist fields $\mathcal T(0,0)$ within correlation functions, or equivalently modifying the partition function according to \eqref{twistcut}. As a consequence, a twisted module simply corresponds to a {\em quasi-periodic condition} on the circle -- in the spatial direction--, with quasi-periodicity determined by the symmetry $\sigma$ in accordance to the monodromy \eqref{monodromyeucl}. In CFT, normally one restricts twisted modules to be modules for the algebra of local observables generated by chiral conserved densities (energy-momentum tensor, etc.). Because the symmetry is ultra-local, the energy-momentum tensor is periodic, and therefore its expansion near $(0,0)$, controlled by the CFT, preserves its regular structure, with only pole singularities; thus the algebraic structure is not modified. One constructs the {\em orbifold theory} by concentrating on this algebra and adjoining to the space of modules the twisted modules for this symmetry. But in many-body physics, it is not necessary to make such restrictions -- observables that are not invariant under the symmetry simply acquire branch cuts.

Another useful quantisation scheme is {\em angular quantisation}. There, the spatial direction is the radial half-line, $(t_{\rm ang},x_{\rm ang}) = (\theta,\log r)$. Because time is compact, the operator expression for the theory on the Euclidean plane is a trace, with the density matrix $e^{-2\pi K}$ where $K$ is the rotation generator, as an operator acting on half-line field configurations. The beauty of this quantisation scheme is that, according to the standard exponential form \eqref{defTUgen}, {\em twist fields are exponentials of total ``charges''},
\beq
	\mathcal T_q(0,0) = e^{Q_{\rm ang}},
\eeq
where ``total'' means that the ``charge'' $Q_{\rm ang} = \int_0^\infty \dd x\,q(x)$ acts on the full spatial direction of the quantisation scheme. Crucially, according to topological invariance in the vacuum, Subsec.~\ref{ssecttopo}, the tail may be rotated. In the quantum language, this means that $e^{Q_{\rm ang}}$ commutes with $K$,
\beq
	[e^{Q_{\rm ang}},K]=0.
\eeq
That is, $e^{Q_{\rm ang}}$ is truly a symmetry operator in this quantisation scheme. This is at the basis of S.~Lukyanov's ``free-field'' construction of form factors for quantisation on the line \cite{lukyanov1995free,brazhnikov1998angular}, and of Al.B.~Zamolodchikov's trick to evaluate VEV such as those in \eqref{vev} (unpublished work, as far as I know it was first published by S.~Lukyanov and A.B.~Zamolodchikov in \cite[App B]{lukyanov1997exact}),
\beq
	\bra\vac|\mathcal T|\vac\ket = \Tr\Big( e^{-2\pi K}e^{Q_{\rm ang}}\Big)
\eeq
where $e^{-2\pi K},\,e^{Q_{\rm ang}}$ can be diagonalised simultaneoulsy. See also \cite{khoroshkin1999angular,doyon2003two}. Note that the universal QFT meaning of this quantity is as the first, constant saturation term in \eqref{ffseries} under the conformal normalisation \eqref{shortdist} (or \eqref{shortdisteucl}), and gives the universal part of \eqref{vev}. This is discussed in \cite{doyon2003two,cardy2008form}

\subsection{Free particles: tau functions of integrable PDEs}\label{ssectfree}

Above we saw how QFT techniques, such as form factors, conformal mapping and angular quantisation, allow us to evaluate vacuum correlation functions involving twist fields (and partially, e.g.~in CFT, correlation functions at finite temperatures). It turns out that in models of free Fermionic particles, i.e.~quadratic Fermionic Hamiltonians, there are other techniques, even more powerful, that allow us to evaluate correlation functions not only in the vacuum, but also in finite-entropy-density states. The literature on this subject is vast, and relatively complex; here I just give some of the main lines of arguments

Consider a state $\bra\cdots\ket$, which is either the vacuum state $\bra\vac|\cdots|\vac\ket$, a finite-entropy-density state $\bra\cdots\ket$, with density matrix $\rho = \frc1Ze^{-W}$ for some extensive observable $W$,
\beq
	\bra \cdots\ket = \frc1Z\Tr \big(e^{-W} \cdots\big),\quad Z = \Tr\big(e^{-W}\big),\quad W = \int \dd x\,w(x),
\eeq
or, in the Euclidean formulation, a path integral on the plane, or the cylinder possibly with quasi-periodic conditions, etc. Assume space-time translation invariance. We are given some (complex) free Fermion fields $\psi(x,t)$, $\psi^\dag(x,t)$, which may also have internal structure (spin, vector, etc.); the considerations below can also be applied to Majorana (``real'') Fermions, such as those occuring in the scaling limit of the Ising model, from the Jordan-Wigner transformation (Subsec.~\ref{ssectjordan}). Assume that we have a twist field $\mathcal T$ with twist acting as
\beq
	\sigma_{\mathcal T}(\psi) = A^{-1}\psi
\eeq
for some matrix $A$ (this agrees with the path integral formulation, Eq.~\eqref{sigmapsi}).

The main statement is that correlation functions of twist fields can be expressed in terms of solutions to integrable PDEs. The first time this was observed was in the seminal work of Wu, McCoy, Tracy and Barough \cite{wu1976spin}, expressing spin-spin two-point correlation functions in the statistical Ising model on the plane in terms of Painlev\'e transcendants, and shortly after by Perk in \cite{perk1980equations}, expressing correlation functions in the quantum XY model at finite temperature in terms of integrable PDEs. The modern way of formulating the result is, technically, in terms of tau functions, as first introduced in the pioneering work of Sato, Miwa and Jimbo \cite{sato1978holonomic1,sato1979holonomic2,sato1979holonomic3}, see the modern formulation in \cite{gamayun2012conformal}. Form factors (Subsec.~\ref{ssectqft}), conformal perturbation theory, and large-deviation and hydrodynamic techniques (Subsec.~\ref{ssectld}), then give asymptoptic results for tau functions of Painlev\'e equations and integrable PDEs, and angular quantisation techniques for vacuum expectation values (Subsec.~\ref{ssectqft}) provide solutions to connection problems \cite{doyon2003two}.

\st{Correlation functions of twist fields in free Fermion models are tau functions of integrable PDEs.}

\subsubsection*{Isomonodromy}

The first observation is based on the monodromy \eqref{monodromy}. Because the Fermion fields are free, they satisfy a linear equation of motion, $\mathsf D\psi=0$ with differential operator $\mathsf D$; for instance for a massive Dirac Fermion
\beq
	\mathsf D\psi = \Big(\sum_\mu \gamma^\mu\p_\mu -m\1 \Big)\psi = 0
\eeq
where $\gamma^\nu$ are Dirac's gamma matrices. Therefore, the function
\beq\label{fxt}
	f_{x,t}(y,s):=\bra \mathcal T(x,t)\psi^\dag(y_0,s_0)\psi(y,s)\ket
\eeq
satisfies
\beq\label{Dirac}
	\mathsf D f_{x,t}=0,
\eeq
along with the monodromy property
\beq\label{Diracwinds}
	f_{x,t}(y,s)|_{(y,s):\stackrel{\gamma} \circlearrowleft}
	= A^{-1} f_{x,t}(y,s)
\eeq
where $\gamma$ winds counter-clockwise around $(x,t)$ once. Thus, we are looking for solutions to \eqref{Dirac}, generically with a branch-point singularity at $(x,t)$ and a pole singularity at $(y_0,s_0)$, which satisfies the monodromy \eqref{Diracwinds}. The problem of studying the deformation of the solution under a change of $x,t$ is an {\em isomonodromic deformation problem}: we change the position of the branch-point singularity, without changing its monodromy. Isomonodromic deformations are usually studied in the context of linear systems with Fuschian singularities, which give rise to integrable PDEs and Painlev\'e equations \cite{harnad2002isomonodromic,filipuk2012isomonodromic}. The connection to $U(1)$ twist field (the case where $A$ is a pure complex phase) may be established via the theory of holonomic quantum fields developed in \cite{sato1978holonomic1,sato1979holonomic2,sato1979holonomic3}, or, using the path-integral formulation \eqref{twistpartition} (Subsec.~\ref{ssecttwistpath}) and textbook Fermionic Gaussian integral formulae \cite{zinn2021quantum}, via determinants of Dirac operators on function spaces with monodromy, as first established in \cite{palmer1990determinants}. In this way one identifies
\beq
	\tau(x,t) = \bra \mathcal T(x,t)\b{\mathcal T}(0,0)\ket
\eeq
with {\em tau functions associated to integrable PDEs} \cite{harnad2021tau}, giving a strong underpinning for the early exact results for spin-spin correlations in the Ising model \cite{wu1976spin}. It is my understanding that the general case, with matrix $A$, still requires more study.

\subsubsection*{Form factors and Fredholm determinants}

By taking appropriate Fourier transforms on the variables $y,s,y_0,s_0$ in \eqref{fxt} and related functions with various Fermion fields inserted, evaluated in the vacuum, we can construct form factors of the twist field $\mathcal T$. The form factor properties described in Subsec.~\ref{ssectqft} then follow, again, from the monodromy property. Here the simplification is that the relation between the asymptotic-state annihilating and creating operators $A(\theta)$, $A^\dag(\theta)$ and local fields $\psi(x),\,\psi^\dag(x)$ is linear -- a property of free-particle models. This also holds in finite-entropy-density states as explained in \cite{doyon2005finite}, and was used in \cite{doyon2005finite,doyon2007finite,chen2014form} to define and study form factors in such states.

In free Fermion models, conserved densities for internal symmetries are quadratic forms $q(x) = \psi^\dag(x) B\psi(x)$, for some matric $B$ (related to the matrix $A$ above). The state is also quadratic, and therefore, in the standard exponential form \eqref{defTUgen}, and also in its stacked form \eqref{twistpathexp} arising in the path integral formulation, the insertion of twist field leads to a state that satisfies Wick theorem. That is,
\beq
	\frc{\bra \mathcal T_1(x_1,t_1)\cdots \mathcal T_n(x_n,t_n)
	\psi^{(\dag)_1}(y_1,s_1)\cdots \psi^{(\dag)_n}(y_m,s_m)\ket}{
	\bra \mathcal T_1(x_1,t_1)\cdots \mathcal T_n(x_n,t_n)\ket}
\eeq
can be evaluted by Wick's theorem on $(y_i,s_i)$. As a consequence, with $n=1$, we find that {\em form factors of twist fields can be represented as determinants}, and using this, the form factor series expansion \eqref{ffseries} for twist-field two-point functions can usually be re-summed as a {\em Fredholm determinant of an integral operator} \cite{babelon1992form,bernard1994differential,doyon2007finite}. Such Fredholm determinants are, again, tau functions of integrable PDEs, satisfying nonlinear differential equations. Fredholm determinant methods were used in 
\cite{babelon1992form,its1993temperature,bernard1994differential} (see the book \cite{korepin1997quantum}), in particular for correlation functions of spin in the Ising model and of the vertex operator \eqref{vertex} in the sine-Gordon model.

\subsubsection*{Doubling trick and Hirota bilinear form}

Tau functions of integrable PDEs are known to satify what is referred to as {\em Hirota bilinear equations} (see e.g.~\cite{harnad2021tau}). These are the fundamental forms of ``interacting'' integrable PDEs, much like linear equations are the fundamental form ``non-interacting'' PDEs. It turns out that it is possible to naturally obtain these bilinear equations using twist fields and the free-Fermion structure. The trick is to ``double'' the model -- consider two species of Fermions, $\psi_a,\,\psi_b$, which anticommute with each other. Because the state is the exponential of a quadratic form (or, equivalently, because it satisfies Wick's theorem), the factorised state of the replica model, defined by $\bra\bra o_a o_b'\ket\ket_{\rm replica} = \bra o\ket_{\text{copy}\,a}\bra o'\ket_{\text{copy}\, b}$, now has {\em rotation symmetry $O(2)$ amongst the copies $a$ and $b$}. With a complex structure and $U(1)$ conservation, it is extended to a $U(2)$ symmetry -- but this is not essential. One then considers the Ward identities for the copy-rotation symmetry generator $Z$ and its ``descendants'' under space-time symmetry transformations, $[P_a,Z],\,[H_a,Z]$, etc., with the momenta $P_{a,b}$ and Hamiltonian $H_{a,b}$ of each copy. These give a series of equations for two-point functions of twist fields (inserted on copies $a$ and $b$) and their descendants. Because of factorisation of the state, these equations become bilinear equations. In this way, one obtains bilinear Hirota equations for tau functions. I believe the first time such techniques were used was in \cite{perk1980quadratic}, see also \cite{fonseca2003ward,doyon2011correlation}.

\subsection{Large-deviation theory, thermodynamics, hydrodynamics}\label{ssectld}

Thermodynamics and hydrodynamics are powerful emergent theories for the large-scale behaviours of many-body systems, especially in finite-entropy-density states. It turns out that these theories allow us to obtain exact asymptotic behaviours of (at least a certain class of) twist-field correlation functions in such states, not just for free-fermion models but rather generally. This is one of the most general set of techniques: it goes beyond the form factor methods, and applies to both quantum and classical Hamiltonian systems and classical statistical models.

Here, the principal tool are the locality and extensivity properties based on the state, Subsec.~\ref{ssectstate} and \ref{ssectextstate}. Consider a finite-entropy-density state $\bra\cdots\ket$, with density matrix $\rho = \frc1Ze^{-W}$ for some extensive observable $W$,
\beq
	\bra \cdots\ket = \frc1Z\Tr \big(e^{-W} \cdots\big),\quad Z = \Tr\big(e^{-W}\big),\quad W = \int \dd x\,w(x).
\eeq
I will concentrate on two-point functions of a twist field with its conjugate,
\beq\label{ttcorr}
	\bra \mathcal T(x,t)\b{\mathcal T}(x',t')\ket,
\eeq
but similar ideas can be used more generally. For the path integral formulation, recall that such states are represented using a path integral with a finite imaginary time direction -- on the cylinder.

\subsubsection*{Asymptotic exponential behaviour in space: thermodynamics}

Consider the twist field construction based on ultra-local symmetries with spatially factorised Hilbert space, Subsec.~\ref{ssectultra}. Let us look at the equal-time two-point function of a twist field $\mathcal T_U$ and its conjugate $\b{\mathcal T}_U = \mathcal T_{U^{-1}}$, with $x'>x$:
\beq
	\bra \mathcal T_U(x) \b{\mathcal T}_{U}(x')\ket
	=
	\Big\bra \prod_{x''\in[x,x')} U(x'')\Big\ket.
\eeq
The main idea is the factorisation of partition functions, Eq.~\eqref{factZ}. This suggests that
\beq
	\bra \mathcal T_U(x) \b{\mathcal T}_{U}(x')\ket
	\asymp
	\frc{\Tr\Big(e^{-\int_x^{x'}\dd y\, w(y)}\prod_{y\in[x,x')}U(y)\Big)}
	{\Tr\Big(e^{-\int_x^{x'}\dd y\, w(y)}\Big)}
	\asymp e^{-|x-x'|(f_U-f_{\1})}
	\qquad (|x-x'|\to\infty)
\eeq
where $f_U$ is {\em the specific free energy for the system with Boltzman weight modifed by the symmetry operator}:
\beq
	f_U = -\lim_{L\to\infty} \frc1L\log \Tr \Big(e^{-W_L}
	\prod_{y\in[0,L)} U(y)\Big)
\eeq
where $W_L=\int_0^L\dd x\,w(x)$ is the total Boltzmann charge on the finite-volume system. That is, it is possible to recast the large-distance asymptotic of the correlation function on infinite volume, into the large-volume asymptotic of a ratio of partition functions on finite volume, with that on the numerator modified by the insertion of the total symmetry on the system's volume.

Note that if $U(y)$ is unitary, then $\prod_{y\in[0,L)} U(y)$ is bounded and $f_U\geq f_\1$. These are ``bounded'' twist fields. On the other hand, if $U(y)$ is not bounded, then we may have $f_U<f_\1$, hence a diverging exponential -- these are ``unbounded'' twist fields.

A similar notion is expected to hold for twist fields in the standard exponential form \eqref{defTUgen}, Subsec.~\ref{ssectexp}.
\beq
	\bra \mathcal T_q(x) \b{\mathcal T}_{q}(x')\ket
	\asymp e^{-|x-x'|(f_Q-f_{0})}
	\qquad (|x-x'|\to\infty)
\eeq
where
\beq
	f_Q = -\lim_{L\to\infty} \frc1L\log \Tr \Big(e^{-W_L}
	e^{Q_L}\Big).
\eeq
There, if $q$ is anti-Hermitian, the twist field is bounded (such as \eqref{tlambdaspin} and \eqref{defTUgencont}), while if it is Hermitian, it is unbounded..

A similar notion also holds for twist fields in the path integral formulation, Subsec.~\ref{ssecttwistpath},
\beq
	\bra \mathcal T_A(x) \b{\mathcal T}_{A}(x')\ket
	\asymp e^{-|x-x'|(f_A-f_{\1})}
	\qquad (|x-x'|\to\infty).
\eeq
There, the Boltzmann weight for $f_A$ is modified by adding a branch cut, within the cylinder on which the path integral is evaluated. This, then, applies to classical statistical models, where one must evaluate partition functions with ``twisted'' periodicity conditions.

In fact, if the ultra-local symmetry is part of a continuous ultra-local symmetry group, Subsec.~\ref{ssectsymmetry}, with twist field $\mathcal T_{\vec\lambda}$, Eq.~\eqref{defTUgencont}, then the derivation is stronger: we can use the result \eqref{Fff} established in \cite{doyon2020fluctuations}, to obtain essentially the same formula. The main relation is
\beq\label{twistfreenergy}
	\bra \mathcal T_{\vec \lambda}(x) \b{\mathcal T}_{\vec\lambda}(x')\ket \stackrel{\text{(equality in classical systems)}}\asymp \Big\bra e^{\ri \int_x^{x'} \dd y\,\lambda\cdot \vec q(y)}\Big\ket \asymp e^{-|x-x'|(f(\vec \lambda)-f(\vec 0))}
\eeq
where $f(\vec\lambda)$ is the {\em specific free energy for the density matrix $e^{-W + \ri\vec\lambda \cdot \vec Q}$}. An important, if extremely simple, observation is that {\em the two-point function of such twist fields is related to the full counting statistics $F(\lambda)$ of the charge $Q_{[x,x']} = \int_x^{x'}\dd y\,q(y)$}, Eq.~\eqref{deltaf}. For the full counting statistics, one normally replaces $\ri\vec\lambda\to \vec\lambda$, using unbounded twist fields.

The second asypmtotic equality in \eqref{twistfreenergy} follows from \eqref{Fff}. The first equality in \eqref{twistfreenergy} follows in classical systems simply from $e^{\ri  \int_x^\infty \dd y\,\vec\lambda\cdot \vec q(y)} e^{-\ri  \int_{x'}^\infty \dd y\,\vec\lambda\cdot \vec q(y)}= e^{\ri\int_x^{x'}\dd y\,\vec\lambda\cdot\vec q(y)}$. In quantum systems, it is only an asymptotic equality, which is obtained from
\beq\label{expoquant}
	e^{\ri \int_x^\infty \dd y\,\vec\lambda\cdot \vec q(y)} e^{-\ri  \int_{x'}^\infty \dd y\,\vec\lambda\cdot \vec q(y)}= e^{\ri\int_x^{x'}\dd y\,\vec\lambda\cdot\vec q(y) + o(x')}
\eeq
where $o$ is a local observable. Eq.~\eqref{expoquant} is found as follows. Write $Q = \vec\lambda\cdot\vec Q$, where $\vec Q = \int \dd x\,\vec q(x)$. Then as $[Q,Q]=0$, we must have (see Eq.~\eqref{oop}) $[Q,\vec q(x)] = \p_x \vec o(x)$ for some local $\vec o(x)$. Write $\int_x^\infty \dd y\,\vec\lambda \cdot \vec q(y) = Q - \int_{-\infty}^x \dd y\,\vec\lambda \cdot \vec q(y)$. Then
\beqa
	\lefteqn{\Big[\int_x^\infty \dd y\,\vec\lambda\cdot \vec q(y), \int_{x'}^\infty \dd y\,\vec\lambda\cdot \vec q(y)\Big]}&& \n
	&=&
	\Big[
	Q - \int_{-\infty}^x \dd y\,\vec\lambda \cdot \vec q(y),
	\int_{x'}^\infty \dd y\,\vec\lambda\cdot \vec q(y)\Big]\n
	&=&
	\vec\lambda\cdot \vec o(x')
	-\int_{y\sim x,\,y<x} \dd y\int_{y'\sim x',\,y'>x'}\dd y'\,
	[\vec\lambda\cdot\vec q(y),\vec\lambda\cdot\vec q(y')]
\eeqa
where the right-hand side is a local observable supported around $x$ (a different but similar calculation would give a local observable supported around $x$). Because $\ad \int_x^\infty \dd y\,\vec\lambda\cdot \vec q(y)$ preserves locality (as an extension of the result \eqref{adintlocal}), the Baker-Campbell-Hausdorff formula shows \eqref{expoquant}.

That is:
\st{
The equal-time correlation functions of a twist field and its conjugate behaves exponentially, with growing exponential (for ``unbounded'' twist fields) or decaying exponential (for ``bounded'' twist fields), with a rate determined by thermodynamic specific free energies. For continuous ultra-local symmetry group, they generate the full counting statistics of  total charges on a domain.
}

\subsubsection*{Asymptotic exponential behaviour in time: hydrodynamics}

Here I concentrate on the case of a continuous ultra-local symmetry group; similar ideas are expected to hold in other contexts, albeit less clearly. I look at \eqref{ttcorr} for $x,t$ generic. The main technique is the use of topological invariance, the deformation \eqref{defTUgencontpath}. We deform the tail of the twist field $\mathcal T_{\vec\lambda}(x,t)$ to make it a path $(x,t)\to(x',t')\to(\infty,t')$. Then, in a way that is similar to the first relation in \eqref{twistfreenergy}, the parts $(x',t')\to(\infty,t')$ of $\mathcal T_{\vec\lambda}(x,t)$ and of $\b{\mathcal T}_{\vec\lambda}(x',t')$ cancel each other, and we obtain
\beq\label{ttdefTUgencontpath}
	\bra\mathcal T_{\vec\lambda}(x,t) \b{\mathcal T}_{\vec\lambda}(x',t')\ket
	\asymp\Big\bra\exp\Big[-\ri\vec\lambda\cdot\int_{(x,t)\to(x',t')} \dd s^\mu\ep_{\mu\nu}\,\vec j^\nu(s)\Big]
	\Big\ket.
\eeq
The cancellation is less straightforward to argue for in generality, see the analysis in the recent work \cite{horvath2025hydrodynamictheorynonequilibriumcounting}.

Once \eqref{ttdefTUgencontpath} is accepted, there are a number of strong techniques to evaluate the right-hand side. In particular
\beq\label{ttdefTUgencontpathtime}
	\bra\mathcal T_{\vec\lambda}(0,0) \b{\mathcal T}_{\vec\lambda}(0,t)\ket
	\asymp\Big\bra\exp\Big[-\ri\vec\lambda\cdot\int_0^{t} \dd s\,\vec j(0,s)\Big]
	\Big\ket
\eeq
has the interpretation as the generating function for the cumulants of the total amount of charge passing through the point $x=0$ in the time period $[t,t']$. This is an important quantitiy in the context of non-equilibrium physics, see e.g.~\cite{esposito2009nonequilibrium,bernard2016conformal}. For such dynamical quantities, and also for the equivalent quantity along the straight ray $(0,0)\to(x,t)$ in Eq.~\eqref{ttdefTUgencontpath}, hydrodynamic techniques have been established. The results depends strongly on the hydrodynamic phenomenology, and one must know in particular the velocities of hydrodynamic modes and their dependence on $\vec\lambda$.

If the system is purely diffusive -- more precisely, the hydrodynamic mode with vanishing velocity does not depend on the maximal entropy state -- then
\beq
	\Big\bra\exp\Big[-\ri\vec\lambda\cdot\int_0^{t} \dd s\,\vec j(0,s)\Big]
	\Big\ket
	\asymp e^{\sqrt{t} F(\lambda)}
\eeq
for some $F(\lambda)$. This is the realm of the macroscropic fluctuation theory (MFT) \cite{bertini2015macroscopic}, and applies, for instance, to any non-integrable Hamiltonian quantum spin chain, as well as to many classical models, including those with stochastic dynamics such as the single exclusion process. If the hydrodynamic velocities depend on the state, then they depend on $\vec\lambda$ as shown in \cite{doyon2020fluctuations}. If the specific value of $\vec\lambda$ is such that there are no zero-velocity modes, one has
\beq
	\Big\bra\exp\Big[-\ri\vec\lambda\cdot\int_0^{t} \dd s\,\vec j(0,s)\Big]
	\Big\ket
	\asymp e^{t F(\lambda)}.
\eeq
This is the realm of general methods for describing fluctuations at the ballisitc scale. The most important are ballistic fluctuation theory (BFT) \cite{doyon2020fluctuations,myers2020transport}, generalising the CFT techniques of \cite{bernard2016conformal} to generic Euler equations; and the (similarly named) ballistic macroscopic fluctuation theory (BMFT) \cite{doyon2023emergence,doyon2023ballistic}, which in its current form generalises MFT \cite{bertini2015macroscopic} to ballistic transport in integrable systems \cite{doyon2020lecture} and more generally in ``linearly degenerate systems'' \cite{bressan2012hyperbolic}. These theories apply, for instance, to quantum and classical Hamiltonian models with momentum conservation, to many-body integrable systems, and to certain stochastic models such as the totally asymmetric single exclusion process \cite{myersthesis}.

\st{
Dynamical correlation functions of a twist field and its conjugate behave exponentially in a way that is determined by the emergent hydrodynamic theory for the model under study and the large deviation theory for total currents.
}

\subsubsection*{Descendants}

One of the most important observation of \cite{del2022hydrodynamic} is that such hydrodynamic techniques {\em can be applied as well to descendant twist fields}, Subsec.~\ref{ssectalg} and \ref{ssectqft}. For definiteness consider again twist fields for continuous ultra-local symmetry groups, Eq.~\eqref{defTUgencont} -- but the concepts are more general. The main conjecture is that, if $o,o'\in\mathfrak L_0$ are local observables, then, along any ray $(x,t)\to(x',t')$ in space-time (including purely spatial and purely temporal rays), the exponential asymptotics \eqref{ttdefTUgencontpath} is replaced by the {\em factorised expression}
\beq\label{ttdefTUgencontpatho}
	\bra o(x,t)\mathcal T_{\vec\lambda}(x,t) o'(x',t')\b{\mathcal T}_{\vec\lambda}(x',t')\ket
	\asymp
	\bra o(x,t) o'(x',t')\ket_{\vec\lambda}
	\Big\bra\exp\Big[-\ri\vec\lambda\cdot\int_{(x,t)\to(x',t')} \dd s^\mu\ep_{\mu\nu}\,\vec j^\nu(s)\Big]
	\Big\ket.
\eeq
Here, the state $\bra\cdots\ket_{\vec\lambda}$ is the biasing of $\bra\cdots\ket$ by $\exp\Big[-\ri\vec\lambda\cdot\int_{(x,t)\to(x',t')} \dd s^\mu\ep_{\mu\nu}\,\vec j^\nu(s)\Big]$, which is a maximal entropy state (Gibbs or generalised Gibbs state) determined by the flow equation of BFT \cite{doyon2020fluctuations}. In QFT, because of short-distance singularities, $o(x,t)\mathcal T_{\vec\lambda}(x,t)$ is replaced by $(o\mathcal T_{\vec\lambda})(x,t)$ (Eq.~\eqref{oTdesc}), etc.

In \eqref{ttdefTUgencontpatho}, the two-point function of local observables $\bra o(x,t) o'(x',t')\ket_{\vec\lambda}$ may have exponential decay, with a correlation length that depends on $\vec\lambda$. This, therefore, affects the leading exponential decay. This is what is observed in \cite{del2022hydrodynamic} for the spin-spin correlation function in the XX model, where $o,o'$ are local Fermions, such as in \eqref{Tminusfermion}.

However, because of ambiguities such as \eqref{Tambiguity}, or the finite-translation ambituity seen on the right-hand side of \eqref{TUprod} with $U'=\1$, it is not necessarily clear how to extract a local observable contribution from a generic twist field. The theory is still under development.

\subsubsection*{Applications}

Successful applications of these techniques include: the simple derivation of the asymptotics of the spin-spin correlation function in the quantum XX model \cite{del2022hydrodynamic}, originally obtained in \cite{its1993temperature} only in a restricted region of parameters by an intricate analysis of its tau-function representation (Subsec.~\ref{ssectfree}); the large-time analysis of the particle-number full counting statistics \cite{horvath2024full,horvath2025hydrodynamictheorynonequilibriumcounting} and of the R\'enyi entanglement entropy  (Subsec.~\ref{ssectentanglement} below) \cite{del2024entanglement} after quenches; and the analysis of correlation functions of sine-Gordon vertex operators, seen as twist fields (Subsec.~\ref{ssectjordan}) \cite{del2023exact}.

\subsection{Branch-point twist fields and quantum entanglement}\label{ssectentanglement}

One of the latest -- and most successful -- applications of twist fields is to the study of quantum entanglement in many-body quantum systems. Many measures of quantum entanglement can be directly related to twist fields, such as the entanglement entropy, the logarithmic negativity, and their symmetry-resolved versions. See the reviews \cite{Pasquale_Calabrese_2009,laflorencie2016quantum,castro2025symmetry}.

The first time twist fields were introduced in the area of many-body quantum entanglement is in the work \cite{cardy2008form}. There, {\em branch-point twist fields} were defined in order to study the entanglement entropy, as twist fields associated to the $\Z_n$ subgroup of the permutation symmetry group in replica models. The focus was on form factors in integrable QFT and applications of the angular quantisation techniques (Subsec.~\ref{ssectqft}), with some aspects applying as well to non-integrable QFT  \cite{doyon2009bipartite} and the twist field construction being valid also in quantum spin chain \cite{castro2011permutation} (Subsec.~\ref{ssectultra}). Techniques based on conformal mapping of Riemann surfaces in CFT had already been developed earlier \cite{holzhey1994geometric,calabrese2004entanglement,calabrese2009entanglement}. The relation between the replica method (see below) and Riemann surfaces was estbalished in QFT in \cite{calabrese2004entanglement} (extending and clarifying \cite{holzhey1994geometric}), which inspired the twist field construction \cite{cardy2008form}. Free-particle techniques had also been developed \cite{peschel2003calculation} (see the reviews \cite{peschel2009reduced,casini2009entanglement,bernard2401entanglement}). But without the general context of twist fields, technical progress was limited to CFT and free particles. As explained in \cite{cardy2008form}, branch-point twist fields immediately connect with Riemann surface techniques (Subsec.~\ref{ssectriemann}, and see below), but allow us to go further (e.g.~Subsec.~\ref{ssectqft}). In free-particle models, connections to Painlev\'e equations have been established \cite{casini2009entanglement}, which, a posteriori, is expected from the  twist field description (Subsec.~\ref{ssectfree}). Twist fields have been used to study other measures \cite{Pasquale_Calabrese_2009,laflorencie2016quantum}, in particular those associated to combined permutation-$\Z_n$ and -$U(1)$ symmetries in QFT (composite twist fields, Subsec.~\ref{ssectqft}), for symmetry-resolved entanglement measures and entanglement asymetry \cite{horvath2020symmetry,Horv_th_2021,capizzi2023entanglementasymmetryorderedphase,castro2025symmetry}. Very recently, techniques from large-deviation theory and hydrodynamics (Subsec.~\ref{ssectld})  have been applied \cite{del2024entanglement}.

\subsubsection*{Branch-point twist fields}

For any quantum model with Hilbert space $\mathcal H$ and Hamiltonian $H = \int \dd x\,h(x)$, one may construct the $n$-copy model with Hilbert space ${\mathcal H}^{\otimes n}$ and Hamiltonian
\beq
	H^{(n)} = \sum_i H_i,\quad H_i = \1\otimes \cdots\otimes \1\otimes \underbrace{H}_{\text{copy}\ i} \otimes \1 \otimes \cdots \otimes \1.
\eeq
This new, replica model has a natural group of internal symmetries: the {\em permutations of copies}. For any $s\in S(n)$ in the permutation group on $n$ elements, and operator $o_i = \1\otimes \cdots\otimes \1\otimes \underbrace{o}_{\text{copy}\ i} \otimes \1 \otimes \cdots \otimes \1\in\Aut(\mathcal H^{\otimes n})$ acting on the $i$th copy, we define
\beq
	\sigma_s (o_i) = o_{s(i)}.
\eeq
As $H^{(n)} = \int \dd x\,h^{(n)}(x)$ with $h^{(n)} = \sum_{i=1}^n h_i(x)$, it is clear that
\beq
	\sigma_s(h^{(n)}) = h^{(n)}\quad \forall\ s\in  S(n),
\eeq
hence this is an internal symmetry. A {\em branch-point twist field} is a twist field associated to this symmetry:
\beq
	\mathcal T_s \mbox{ has twist } \sigma_s\mbox{ for some $s\in S(n)$}\quad
	\mbox{(branch-point twist field)}.
\eeq
We will see why this name is appropriate below. Often, in the literature one chooses the cyclic permutation symmetry
\beq\label{cycl}
	s_{\rm cycl}(i) = i+1\mod n
\eeq
and we will simplify to this below; any permutation of order $n$ would do. In this case, the branch-point twist field, and its conjugate, is denoted
\beq
	\mathcal T := \mathcal T_{s_{\rm cycl}},\quad \b{\mathcal T} := \mathcal T^\dag = \mathcal T_{s_{\rm cycl}^{-1}}.
\eeq

If the original Hilbert space factorises as \eqref{Htensor}, then so does the new Hilbert space $\mathcal H^{\otimes n} = \bigotimes_x {\mathcal H}_x^{\otimes n}$, and $\sigma_s$ is an ultra-local symmetry as per Subsec.~\ref{ssectultra}:
\beq
	U_s = \prod_x P_s(x),\quad \sigma_s(o) =  U_soU^{-1}_s.
\eeq
Here the linear operator $U_s$ implementing the permutation $s$ is defined as a product of local permutations $P_s(x)$ on site $x$: for any vectors $|v_1\ket_x \otimes\cdots\otimes |v_n\ket_x \in \mathcal H_x^{\otimes n}$ (where $|v_i\ket_x \in \mathcal H_x$ are different vectors for different $i,x$), 
\beqa
	\lefteqn{P_s(x) \bigotimes_{x'} \big(|v_1\ket_{x'}\otimes \cdots \otimes|v_n\ket_{x'}\big)} &&\\
	&=&
	\bigotimes_{x'<x} \big(|v_1\ket_{x'} \otimes\cdots \otimes|v_n\ket_{x'}\big)
	\otimes
	\big(|v_{s^{-1}(1)}\ket_x \otimes\cdots\otimes |v_{s^{-1}(n)}\ket_x\big)
	\otimes
	\bigotimes_{x'>x} \big(|v_1\ket_{x'} \otimes\cdots \otimes|v_n\ket_{x'}\big).\no
\eeqa
In this case, we can construct the associated twist field \eqref{defTU},
\beq
	\mathcal T_s(x) = \prod_{x'\geq x} P_s(x').
\eeq
This is the string representation of branch-point twist fields \cite{castro2011permutation}.

It is also possible to construct the branch-point twist field $\mathcal T_s$ from Hilbert-space generating observables as per App.~\ref{appgen}.

The branch-point twist field also has a representation in the path integral formulation, Subsec.~\ref{ssecttwistpath}. In this formulation, the fundamental field has an additional copy number $\psi_i$ (and may also have internal vector structure), and the cut \eqref{twistcut}, for the branch-point twist field $\mathcal T_s$, becomes
\beq\label{twistcutbp}
	\mathcal C_{x,t}^s : \Psi_i(x',t+ 0^+) = \lt\{
	\ba{ll}
	\Psi_{s^{-1}(i)}(x',t - 0^+) & (x'> x)\\
	\Psi_i(x',t - 0^+) & (x'< x),
	\ea\rt.
\eeq
Clearly, for any $s$, the group generated by $\sigma_s$ is idempotent, $\sigma_s^N=\id$ for some $1\leq N\leq n$. Consider again the case of the cyclic permutation symmetry \eqref{cycl}, which has $N=n$. Then \eqref{twistcutbp} becomes a continuity condition connecting copy $i$ and $i+1\;{\rm mod}\; n$ on the tail $([x,\infty),t)$,
\beq\label{twistcutbp}
	\mathcal C_{x,t} : \Psi_i(x',t+ 0^+) = \lt\{
	\ba{ll}
	\Psi_{i-1\;{\rm mod}\;n}(x',t - 0^+) & (x'> x)\\
	\Psi_i(x',t - 0^+) & (x'< x)
	\ea\rt.\quad \mbox{(branch-point twist field $\mathcal T$).}
\eeq
The path integral formulation \eqref{Tpath} is therefore a {\em path integral for the original model, but extended to lie on an $n$-sheeted Riemann surface} $\mathcal R_{x,t}^{(n)}$. See Fig.~\ref{figriemann} for the case of $\mathcal T(x,t) \b{\mathcal T}(x',t)$, with Riemann surface $\mathcal R_{x\to x',t}^{(n)}$: in this case,
\beq
	\frc{{}^{\otimes n}\bra \vac|\mathsf T\big[ \mathcal T(x,t) \b{\mathcal T}(x',t) o_{k_1}(x_1,t_1)\cdots \big]|\vac\ket^{\otimes n}}{{}^{\otimes n}\bra \vac| \mathcal T(x,t)\b{\mathcal T}(x',t)|\vac\ket^{\otimes n}}
	=\bra o(x_1,t_1,k_1)\cdots \ket_{\mathcal R_{x\to x',t}^{(n)}}
\eeq
where $o_{k_1}$ lies on copy $k_1$, and $(x_1,t_1,k_1)\in\mathcal R_{x\to x',t}^{(n)}$ is a position on sheet $k_1$ of the Riemann surface,
and the two-point function is the partition function
\beq
	{}^{\otimes n}\bra \vac| \mathcal T(x,t) \b{\mathcal T}(x',t)  |\vac\ket^{\otimes n}
	\propto Z(\mathcal R_{x\to x',t}^{(n)}).
\eeq
\begin{figure}
\bc\includegraphics[width=0.5\textwidth]{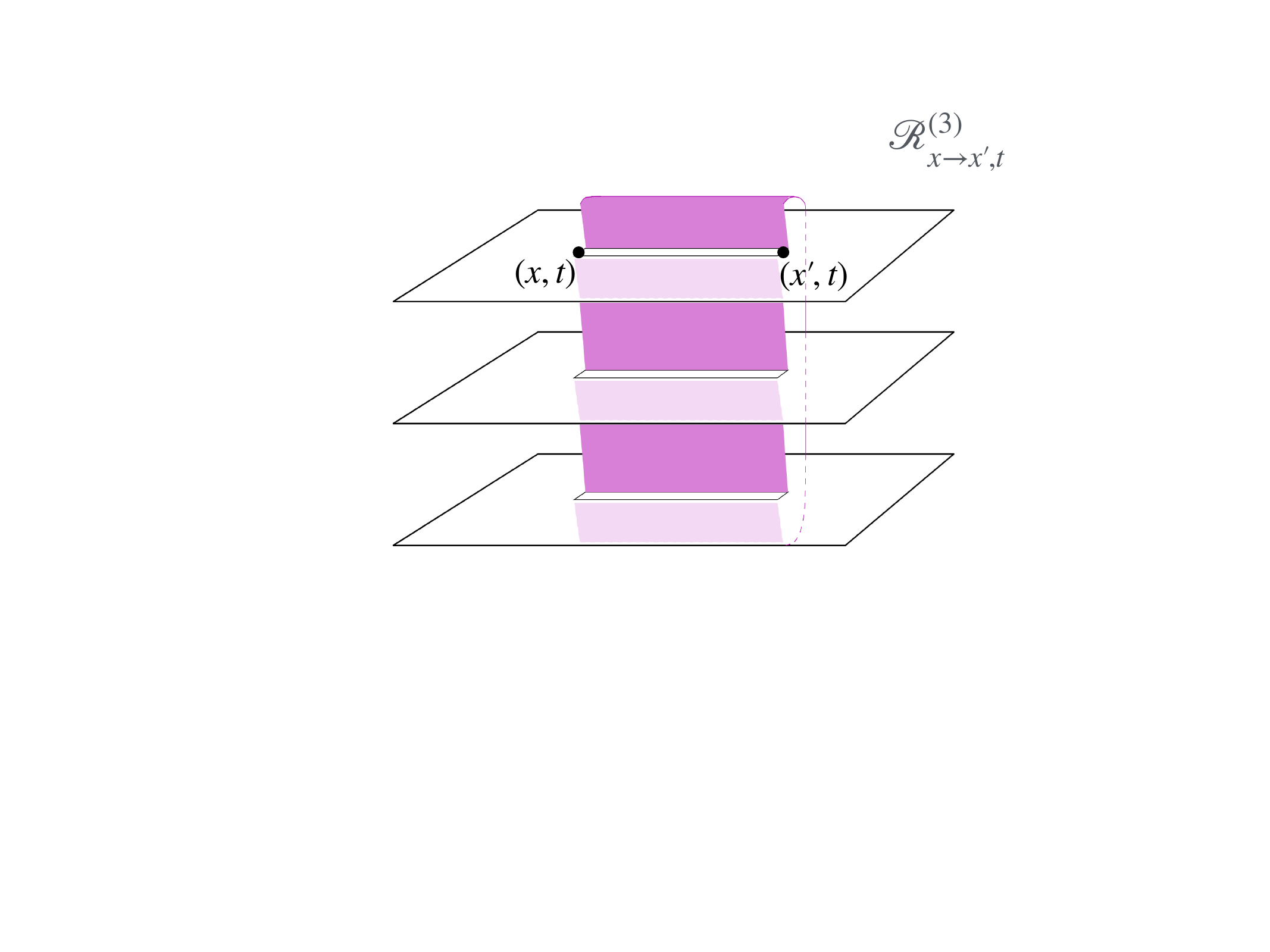}\ec
\caption{The branch-point twist field insertion $\mathcal T(x,t) \b{\mathcal T}(x',t)$ in the $n$-copy model changes the path integral to that of the original 1-copy model, but extended to space-time being the $n$-sheeted Riemann surface $\mathcal R_{x\to x',t}^{(n)}$. Here the case $n=3$. Picture adapted from \cite{cardy2008form}.}
\label{figriemann}
\end{figure}

How does this connect to the Riemann surface  construction of Subsec.~\ref{ssectriemann}? Recall that in this construction, the Riemann sheets are {\em resonant}: the condition \eqref{resonant} holds. Here, we have two indices: the copy index $i$, and the Riemann sheet index, which I will call $a$ (instead of $n$ in \eqref{resonant}). The resonant condition becomes
\beq
	\Psi_{i,\ a+1\;{\rm mod}\; n}(x',t') = \Psi_{i+1\;{\rm mod}\; n,\ a}(x',t').
\eeq
Therefore, we may concentrate on
\beq
	\Psi_{i,1}(x',t') = \Psi_{1,i}(x',t') = : \Psi_{i}(x',t')
\eeq
which are now independent -- that is, copy 1 of the Riemann sheets are all independent, and connected cyclically, Eq.~\eqref{riemannwinding}.

This is the source of the name ``branch-point twist field'': the twist feld $\mathcal T_s(x,t)$ {\em introduces a branch point at $(x,t)$} on formerly unconnected sheets. In this way, by insertion of branch-point twist fields at various positions and with various permutation elements, and by the use of topological invariance \eqref{pathgammagen}, one may construct {\em general Riemann surfaces, with various connectivities}. In particular, multi-point correlations of branch-twist fields are partition functions on such Riemann surfaces, for instance as per \eqref{twistpartitiondeformed}.

Note that although branch points in Riemann surfaces are certain types of conical singularities of integer multiples of $2\pi$, the branch-point twist field $\mathcal T_s$ is {\em different} from the conical twist field $\mathcal T_{\theta= 2\pi \text{order}(s)}$, Eq.~\eqref{Tpathconical}, except in certain situations (such as vacuum two-point functions in CFT). See \cite{castro2018conical}.

As mentioned, the connection between the replica model and the original model on Riemann surfaces in QFT was first established in \cite{calabrese2004entanglement} based on \cite{holzhey1994geometric}, while the connection to branch-point twist fields was established in\footnote{In \cite{calabrese2004entanglement}, a certain primary field $\Phi_n$ was introduced -- however it was not defined as a branch-point twist fields, and some of its properties, as evaluated there, do not agree with those of branch-point twist fields. It is more akin to a conical twist field, Subsec.~\ref{ssectcon}.} \cite{cardy2008form} based on \cite{calabrese2004entanglement}. In massive (integrable) QFT, the copies become additional indices for asymptotic particles, and the formulation \eqref{twistcutbp} for branch-point twist fields is particularly useful as it leads to simple equations for their form factors, including \eqref{kinematicpole} and \eqref{periodicityff}. This was first exploited in \cite{cardy2008form,doyon2009bipartite}, see the reviews \cite{castro2009bi,castro2025symmetry}.

In CFT, the Riemann surface formulation is particularly useful, through the use of conformal maps. For instance, with two branch-point twist field insertions $\mathcal T(x,t)\b{\mathcal T}(x',t')$, we obtain a Riemann surface with two branch points, which can be mapped onto the sphere with two singularities, on which the partition function is known. Thus, two-point functions of branch-point twist fields are easily evaluated, and take the form of two-point functions of primary fields in CFT, e.g.~\cite{calabrese2004entanglement,cardy2008form}
\beq
	\bra \mathcal T(z)\b{\mathcal T}(z')\ket
	=
	|z|^{-2d_n},\quad
	d_n = \frc{c}6\Big(n-\frc1n\Big)\quad \mbox{(CFT on the plane with central charge $c$),}
\eeq
here in the Euclidean formulation.

\st{A branch-point twist field for a quantum model, is a twist field associted to an internal permutation symmetry of the replica $n$-copy version of the model. In the path integrale formulation, this connects the $n$ copies into a Riemann surface, where the positions of twist fields are branch points.}

\subsubsection*{Connection to entanglement entropy}

The von Neumann entanglement entropy is a measure of entanglement in many-body quantum systems. One constructs the reduced density matrix on a subset $A\subset \R$ of space,
\beq
	\rho_A = \Tr_{\otimes_{x\in \b A}\mathcal H_x} \,\rho
\eeq
where $\b A$ is the complement of $A$, that is $\b A = \R\setminus A$ (continuous models) or $\Z\setminus A$ (chains). The von Neumann entropy is then defined as
\beq
	S_A = -\Tr \rho_A\log\rho_A = \lim_{n\to 1} \frc1{1-n}\Tr \rho_A^n.
\eeq
The quantity $\frc1{1-n}\Tr \rho_A^n$ is called the $n^{\rm th}$ R\'enyi entanglement entropy. The trace $\Tr\rho_A^n$, for positive integer $n$, is simply related to branch-point twist fields in the $n$-copy model. Indeed, a simple computation, which as far as I am aware was first produced in \cite{castro2011permutation}, shows that
\beq
	\Tr \Big(\rho_A^n\Big) = \Tr \Big(\rho^{\otimes n} \prod_{x\in A}P_s(x)\Big)
\eeq
for any $s$ of order $n$. For instance with $A=[x,x')$,
\beq\label{EErho}
	\Tr \Big(\rho_A^n\Big) = \Tr \Big(\rho^{\otimes n} \mathcal T(x)\b{\mathcal T}(x')\Big)
\eeq
and in particular for the vacuum state,
\beq\label{EEvac}
	\Tr \Big(\rho_A^n\Big) = {}^{\otimes n}\bra\vac| \mathcal T(x)\b{\mathcal T}(x')|\vac\ket^{\otimes n}.
\eeq
Such two-point functions can be evaluated by a number of techniques as described above. The von Neumann entanglement entropy requires an ``analytic continuation'' to $n\in\R$, with the limit $n\to 1$. Such an analytic continuation is not unique, but there are ways of guessing the right one, depending on the techniques used, see the reviews mentioned above.

\st{Entanglement entropy in CFT, QFT and quantum chains can be evaluated via correlation functions of branch-point twist fields.}

\subsubsection*{The special case of free-particle models}

In free-particle models, where the Hamiltonian is quadratic, then $H^{(n)}$ not only has permutation symmetry, but the {\em full symmetry under rotation of the copies}, of which permutations form a finite subgroup. This is because $\sum_i H_i$ is a quadratic form on the fundamental fields of each copy (the observables in terms of which each $H_i$ is quadratic). If there is a compex structure, then this is often extended to a $U(n)$ symmetry. Concentrating on $s_{\rm cycl}$, the $\Z_n$ group it generates is a subgroup of a $O(2)$ or $U(1)$ symmetry group of the $n$-copy model. As first noted in \cite{cardy2008form}, expressions \eqref{EEvac}, and \eqref{EErho}, in free-particle models, can then be recast into products of $U(1)$ twist fields, which act diagonally on linear combinations (Fourier transforms) of the original copies.

This observation is at the basis of much of the developments in free-particle models. It is, in a disguised way, at the root of the derivation of the Painlev\'e V representation \cite{casini2005entanglement}. It gives rise to vacuum expectation values of branch-point twist fields \cite{cardy2008form}, related to the universal UV-to-IR entropy saturation of the entanglement entropy, for instance, in the quantum Ising model (and to the connection problem of the associated tau function for the Painlev\'e V equation).

Because we map branch-point twist fields to twist fields associated to continuous ultra-local symmetry groups, the above observation also implies that the first relation in \eqref{twistfreenergy}, holds. It turns out that this, in fact, holds as an equatlity, not just an asymptotic equality, in free Fermion models, because the $U(1)$ symmetry is implemented in terms of Fermion bilinears without any spatial structures (without derivatives), and in this case $[\vec\lambda\cdot\vec q(x),\vec\lambda\cdot\vec q(x')]=0$ for all $x,x'$. Thus, {\em entanglement entropy is related to the full counting statistics of the $U(1)$ charge, which has the physical interpretation as the charge counting the number of Fermions}. This is the basis for the relation between entanglement entropy and Fermion number full counting statistics in free Fermion models \cite{klich2009quantum,song2011entanglement,calabrese2012exact,del2024entanglement}. This does not hold, however, beyond free particle models -- it is a special property of the lack of interactions; but see \cite{hsu2009quantum,song2012bipartite}.

\st{Branch-point twist fields in free-particle models can be diagonalised into twist fields associated to $U(1)$ continuous ultra-local symmetry groups. As a consequence, entanglement measures are related to particle-number full counting statistics.}

\section{Conclusion}\label{sectconclu}

In this paper I have pedagogically reviewed notions of locality end extensivity, and the concept of twist field, explaining its various constructions and properties.

Locality may be seen under different lights:  I have explained how it is a property of the algebra of observables and their dynamics (independence at large distances), or a property of the state in which observables are evaluated (decorrelation at large distances). Twist fields require us to extend this notion to semi-locality: twist fields are independent from the energy density at large distances, thus are local observables in its widest definition, however they are not independent from other observables at large distances, because of their non-trivial exchange relations. Nevertheless, most often they still decorrelate in physical states.

Twist fields are based on symmetries, and in many cases can be seen as exponentials of height fields, half-line integrals or sums of conserved densities -- although they are more general. I have given explicit constructions of twist fields based on ultra-local symmetries, symmetries that preserve the energy density and are local in an appropriate sense. I have explained how twist fields have a ``tail'' emanating from their positions in space-time, and how this tail can ``wiggle'', i.e.~there is topological invariance, the shape of the tail can be changed without changing the result, or, in the path integral formulation, changing it in a predictable way.

I have given examples of twist fields whose correlation functions have important physical applications such as those arising from the Jordan-Wigner transformation, bosonisation, or height fields in dynamical systems, and the branch-point twist field to evaluate entanglement measures. I have also reviewed techniques to calculate twist fields correlation functions, such as form factor techniques in QFT, integrable PDE in free Fermion models, and thermodynamics and hydrodynamics techniques in states of finite entropty density.

There is still much to be developed with the techniques I have presented, for instance the use of hydrodynamic methods to obtain asymptotics of entanglement entropy in quantum models, or to study stretch statistics in classical chains. There are also many more extensions with modern applications, such as twist fields in higher dimensions \cite{hung2014twist}, and twist fields based on non-invertible symmetries (certain types of topological defects) \cite{shao2023s}. In the first paragraph of Section \ref{sectmodels} I used the images of dislocations and vortices to explain, intuitively, the need for twist fields. However, these do not manifestly form part of any of the examples I discussed -- except in a sense the disorder parameter, which observes spin dislocations (Subsec.~\ref{ssectcon}). I believe twist fields can be useful to study various topological objects in classical and quantum statistical and dynamical systems beyond what has been done until now, including vortices in many-body wave functions of two-dimensional quantum models, and dislocations in crystal structures and other materials. I do hope that these notes will inspire further research on this subject.

\medskip

{\bf Acknowledgment.} I am grateful for collaborations on this subject over many years with O.A. Castro-Alvaredo, and also for more recent collaborations with G. Del Vecchio Del Vecchio, D. X. Horv\'ath, and P. Ruggiero. This work was supported by the Engineering and Physical Sciences Research Council (EPSRC) under grant number EP/Z534304/1.

\appendix
\section{Path integral representation in quantum and statistical field theory}\label{apppath}

Feynman's path integral formalism for quantum theory is well reviewed in textbooks \cite{zinn2021quantum}. I give here only a brief overview, to fix notation and ideas. Although the path integral formulation exists for spin chains, I concentrate on field theory models for simplicity -- either second quantisation of Galilean gases, or relativistic QFT.


I will only discuss either ground states of local Hamiltonian systems, or finite-temperature states:  $\bra\cdots\ket = \bra \vac|\cdots|\vac\ket$ or $\bra\cdots\ket = \bra\cdots\ket_\beta:=\Tr (\rho \cdots)$, with  
\beq\label{statesquantum}
	|\vac\ket = \mbox{ground state of $H$},\quad \rho = \frc{e^{-\beta H}}{\Tr\big(e^{-\beta H}\big)},\quad H = \int \dd x\,h(x),\quad h(x)\ \mbox{local,}
\eeq
noting that $\bra\vac|\cdots|\vac\ket = \lim_{\beta\to\infty} \bra\cdots\ket_\beta$. I choose the Hamiltonian, shifting by the identity operator if necessary, in such a way that
\beq
	H|\vac\ket = 0,\quad \bra {\rm state}|H| {\rm state}\ket >0\ \forall\;|{\rm state}\ket \neq |\vac\ket.
\eeq
Note that excitations above the vacuum, such as asymptotic states in QFT, can be obtained by inserting appropriate local observables within the vacuum expectation value, by the LSZ formula \cite{peskin2018introduction}.

The foundation of the path integral representation is the choice of a function space $\mathcal F$, over which the path integral will be integrating. This function space represents a {\em spanning set} for $\mathcal H$, so that $\mathcal H = \spn(|\Psi\ket:\Psi\in\mathcal F)$. It may or may not be a basis. This is set of the field configurations in a relativistic QFT, or the spanning set representing coherent states in the second-quantisation of a quantum gas. I call $\mathcal F = F(\R) := \{\R\to\mathcal V\}$ the space of ``field configurations'', for some vector space $\mathcal V$ in which $\Psi(x)$ takes value (which may be one- or higher-dimensional, and over $\R$ or $\C$).

The states $|\Psi\ket$ are eigenstates for some local observables, which are ``fundamental fields''. Fundamental fields are either a single local observable (and its spatial translates), or commuting local observables organised into a vector lying in $\mathcal V$. I will denote the $\mathcal V$-valued fundamental field $\psi(x)$:
\beq\label{fundeq}
	\psi(x)|\Psi\ket = \Psi(x)|\Psi\ket.
\eeq
In a QFT, it can be taken as the real, complex- or vector-valued field from which the QFT is constructed, $\psi(x) = \phi(x)$ or $\psi(x) = \vec\phi(x)$. We represent the sum over all field configurations as $\int [\dd\Psi]$, with appropriate measure such that $\int [\dd\Psi]\, |\Psi\ket\bra \Psi| = \1_{\mathcal H}$. Other local observables will act, in general, as
\beq\label{olocalpath}
	o(x)|\Psi\ket = \Big(O_0(\Psi(x),\p_x \Psi(x),\ldots)
	+
	O_1(\Psi(x),\p_x \Psi(x),\ldots) \frc{\delta}{\delta \Psi(x)}
	+\ldots\Big)|\Psi\ket.
\eeq

Then using the decomposition into such a spanning set, Feynman's idea is to separate time evolution into thin time slices and insert $\1_{\mathcal H}=\int [\dd\Psi(t)]\, |\Psi(t)\ket\bra \Psi(t)|$ at the times $t$ between every two consecutive slices. If the product of operators we average is time-ordered, we use $o(x,t) = e^{\ri H t} o(x)e^{-\ri H t}$ and thereby obtain, in the limit of infinitesimal slices $e^{-\ri H\delta t}$, a representation where we integrate over the spanning set of field configurations, with one for each value of time. The space over which we integrate is therefore $F(\R\times T)$ where $T=[t_{\rm ini},t_{\rm fin}]$ is the total interval of time between the latest and earliest operator.

As a result of this construction, for most physically sensible Hamiltonians in many-body systems, it turns out that functions in $F(\R\times T)$ are continuous\footnote{\label{ft}That is, the measure in the path integral formulation is such that fields are almost surely continuous, much like the Brownian motion. However, fields are not necessarily almost surely differentiable; nevertheless derivatives of fields and their products at different points usually have well-defined averages.}; unless otherwise specified this is implicitly assumed below.

In the case of a ground state, we use $e^{-\ri H t}|\rm state\ket \propto |\vac\ket$ as $t \to (1-\ri \ep)\infty$, where $\ep=0^+$ parametrises the infinitesimal Wick rotation of time. Then, we can extend the time interval to $T = \R$ (infinitesimally Wick rotated), and obtain a path integral with asymptotic conditions in time,
\beq\label{vacpi}
	\bra \vac|o_1(x_1,t_1)\cdots
	o_n(x_n,t_n)|\vac\ket
	=
	\int [\dd\Psi] \,e^{\ri S[\Psi]}
	o_1[\Psi](x_1,t_1)\cdots o_n[\Psi](x_n,t_n)
	\quad
	(t_1>\ldots >t_n).
\eeq
Here
\beq
	S[\Psi] = \int \dd x\dd t\,\mathcal L[\Psi](x,t)
\eeq
is the ``action'' representing the Hamiltonian evolution, with
\beq\label{lagrangiandensity}
	\mathcal L[\Psi](x,t) = \mathcal L(\Psi(x,t),\p_x\Psi(x,t),\p_t\Psi(x,t),\ldots)
\eeq
the Lagrangian density. In the case of Galilean quantum gases or relativistic QFT, $S[\Psi]$ is, indeed, the classical action associated with the classical version of the field-theoretical Hamiltonian of the system. In this representation, every local observables are  functionals of the field configuration and its space and time derivatives  at $x,t$,
\beq\label{olocalpathrep}
	o[\Psi](x,t) = o(\Psi(x,t),\p_x \Psi(x,t),\p_t \Psi(x,t),\ldots),
\eeq
obtained from \eqref{olocalpath} (in a way that I will not explain here). In particular, of course,
\beq
	\psi[\Psi](x,t) = \Psi(x,t).
\eeq
In \eqref{vacpi}, we integrate over the space of functions on $\R^2$ with the asymptotic conditions on $\Psi(x,t\to\pm\infty)$ representing the vacuum condition. With an appropriate choice of fundamental field $\psi(x)$, in particular with $\bra \vac|\psi(x)|\vac\ket = 0$, this is
\beq\label{condvac}
	\int [\dd\Psi] \quad : \quad \lim_{t\to\pm(1-\ri\ep)\infty}\Psi(x,t)= 0.
\eeq

At finite temperatures, we see $e^{-\beta H}$ as an imaginary time evolution, by time $t = -\ri \beta$. We use this, instead of $e^{-\ri H t}|\rm state\ket \propto |\vac\ket$ as $t \to (1-\ri \ep)\infty$, to determine the time-boundary condition. The result is easier to express if operators are time-evolved in purely imaginary time, $o(x,-\ri \tau) = e^{ \tau H} o(x)e^{- \tau H}$, with time-slicing with $\1_{\mathcal H}=\int [\dd\Psi(\tau)]\, |\Psi(\tau)\ket\bra \Psi(\tau)|$. We get
\beq\label{betapi}\begin{aligned}
	\bra o_1(x_1,-\ri \tau_1)\cdots
	o_n(x_n,-\ri \tau_n)\ket_\beta
	&=
	\int_\beta [\dd\Psi]\,e^{- S_{\rm E}^\beta[\Psi]}
	o_1[\Psi](x_1,\tau_1)\cdots o_n[\Psi](x_n,\tau_n)
	\\ & \hspace{7cm}
	(\tau_1>\ldots >\tau_n)
	\end{aligned}
\eeq
where $ S_{\rm E}^\beta(\Psi)$ is the ``Euclidean'' version of $S(\Psi)$ on a cylinder of circumference $\beta$, formally obtained by the analytical continuation $t \to -\ri \tau$,
\beq
	S_{\rm E}^\beta[\Psi] = \int \dd x\int_0^\beta\dd \tau\,\mathcal L_{\rm E}(\Psi(x,\tau),\p_x\Psi(x,\tau),\p_\tau\Psi(x,\tau),\ldots)
\eeq
with $\mathcal L_{\rm E}(\Psi(x,\tau),\p_x\Psi(x,\tau),\p_\tau\Psi(x,\tau),\ldots) = -\mathcal L(\Psi(x,t),\p_x\Psi(x,t),\p_t\Psi(x,t),\ldots)|_{t=-\ri \tau}$. Now, the space over which we integrate represents periodic conditions, which amount to implementing the so-called Kubo-Martin-Schwinger condition\footnote{For Fermionic fields, it is an anti-periodic condition, with a minus sign.} \cite{brattelioperator1,brattelioperator2}:
\beq\label{condtemp}
	\int_\beta [\dd\Psi]\quad : \quad \Psi(x,\beta) = \Psi(x,0).
\eeq
One may also take the limit $\beta\to0$, which reproduces vacuum expectation values of imaginary time-ordered products of observables:
\beq\label{inftypi}\begin{aligned}
	\bra\vac| o_1(x_1,-\ri \tau_1)\cdots
	o_n(x_n,-\ri \tau_n)|\vac\ket
	&=
	\int_{\rm E} [\dd\Psi]\,e^{- S_{\rm E}[\Psi]}
	o_1[\Psi](x_1,\tau_1)\cdots o_n[\Psi](x_n,\tau_n)
	\\ & \hspace{7cm}
	(\tau_1>\ldots >\tau_n)
	\end{aligned}
\eeq
where
\beq
	S_{\rm E}[\Psi] = \int \dd x\dd \tau\,\mathcal L_{\rm E}(\Psi(x,\tau),\p_x\Psi(x,\tau),\p_\tau\Psi(x,\tau),\ldots)
\eeq
and
\beq\label{condvaceucl}
	\int_{\rm E} [\dd\Psi] \quad : \quad \lim_{\tau\to\infty}\Psi(x,\tau)= 0.
\eeq

We note that if $o(x)$ has vanishing vacuum expectation value $\bra \vac|o(x)|\vac\ket = 0$, then imaginary time-evolution of local operators can be used to {\em define} the vacuum state, via
\beq
	\lim_{\tau\to-\infty}o(x,-\ri \tau)|\vac\ket = 0,\quad
	\lim_{\tau\to\infty}\bra\vac|o(x,-\ri \tau) = 0.
\eeq
This is reminescent of \eqref{condvac}: the field configurations $\Psi(x)$ are eigenvalues on coherent states $|\Psi\ket$ of the local operator $\psi(x)$, which, as mentioned, must have zero vacuum expectation value for \eqref{condvac} to hold.

The Euclidean path integral \eqref{betapi}, and its limit $\beta\to\infty$ \eqref{inftypi}, make the relation with statistical field theory clear: the quantity $e^{-S_{\rm E}^\beta[\Psi]}$, or $e^{-S_{\rm E}[\Psi]}$, can be seen as a Boltzmann weight, and the Euclidean space-time, $(x,\tau)\in \R\times[0,\beta]$ or $\R^2$, as the two-dimensional space on which the statistical degrees of freedom $\Psi(x,\tau)$ lie, either a cylinder or the full plane, with $o(x,-\ri \tau)$ local statistical observables at $(x,\tau)$. Other two-dimensional topologies and shapes can be represented with the path integral in similar ways: for $ z_i\in\mathcal S\subset\R^2$, and $o_i( z_i)$ representing commuting classical, local  random variables in the statistical model within the state $\bra\cdots\ket$,
\beq\label{statpi}
	\bra o_1( z_1)\cdots
	o_n( z_n)\ket
	=
	\int_{\mathcal S} [\dd\Psi]\,e^{- S_{\rm E}^{\mathcal S}[\Psi]}
	o_1[\Psi]( z_1)\cdots o_n[\Psi]( z_n)
\eeq
where
\beq
	S_{\rm E}^{\mathcal S}[\Psi] = \int_{\mathcal S} \dd^2 z\,\mathcal L_{\rm E}(\Psi(x,\tau),\p_x\Psi(x,\tau),\p_\tau\Psi(x,\tau),\ldots)
\eeq
and $\int_{\mathcal S}$ implements the boudary, periodicity or asymptotic conditions on $\p \mathcal S$ that are appropriate for the state $\bra\cdots\ket$. Note how, in \eqref{statpi}, {\em there is no need for time ordering}, as this is a path-integral representation of an expectation value of classical observables.

Below, in the quantum context, I will use the standard time-ordering symbolism
\beq
	\mathsf T\big[o_1(x_1,t_1)\cdots
	o_n(x_n,t_n)] =
	o_{\mu(1)}(x_{\mu(1)},t_{\mu(1)})
	\cdots
	o_{\mu(n)}(x_{\mu(n)},t_{\mu(n)}),\quad
	t_{\mu(1)}>\cdots>t_{\mu(n)}
\eeq
if all times are different, where $\mu$ is the unique permutation of $\{1,\ldots,n\}$ implementing the ordering, and similarly for imaginary times.

\section{Twist fields from Hilbert-space generating observables}\label{appgen}

In Subsec.~\ref{ssectultra} and \ref{ssectexp} we constructed twist fields, and found that they are naturally associated to groups of symmetries. Further, they had certain, natural transformation properties under the symmetries: both for twist fields associated with ultra-local symmetries in spatially factorised Hilbert spaces, and for those associated with continuous ultra-local symmetry groups, their transformation properties were given by inner automorphisms, or equivalently, by adjoint group action on Lie algebra elements.

Is it possible to start with a symmetry group -- automorphisms of the algebra of mutually local observables -- and construct twist fields out of this, without requiring a specific form of these automorphisms as we did when defining ultra-local symmetries in \eqref{Uultra} and \eqref{Uultragen}? Do we again get this natural transformation property?

It turns out that the answers to these question is yes, if we have: (1) a group $G=\{g\}$, (2) a representation $\sigma_g$ as internal symmetries on the algebra of mutually local observables $\mathfrak L_0$, and (3) a $G$-invariant subspace of {\em Hilbert-space generating observables $\mathfrak L_0^\sharp\subset \mathfrak L_0$}.

Hilbert-space generating observables arise if the Hilbert space can be formed from a vector $|V\ket$ by application of local observables $\mathfrak L_0$,
\beq
	\mathcal H = \mathfrak L_0 |V\ket.
\eeq
It is a vector subspace $\mathfrak L_0^\sharp\subset \mathfrak L_0$ of mutually local observables, such that the set
\beq
	o_1(x_1)\cdots o_n(x_n) |V\ket\quad \forall\ o_1,\ldots,o_n \in \mathfrak L_0^\sharp,
\eeq
perhaps with some ordering imposed on the positions $x_1>\cdots>x_n\in\R$, is a basis for $\mathcal H$. $G$-invariance means that
\beq\label{Ginv}
	\sigma_g(o)\in\mathfrak L_0^\sharp\quad\forall\ o\in\mathfrak L_0^{\sharp},\ g\in G.
\eeq

In this case, for every $g\in G$ we simply have to define $\mathcal T_g(x)|V\ket = |V\ket$, and the operators $\mathcal T_g$ are defined on all of $\mathcal H$ by the exchange relations: we set
\beq\label{twistg}
	\mathcal T_g(x) o_1(x_1)\cdots o_n(x_n) |v\ket =
	\sigma_g^{\chi(x\geq x_1)}(o_1(x_1))\cdots \sigma_g^{\chi(x\geq x_n)}(o_n(x_n))
	|v\ket
	\quad\forall\ o_1,\ldots,o_n \in \mathfrak L_0^\sharp
\eeq
where $\chi(x\geq x_1) = 1$ if $x\geq x_1$ and $0$ otherwise. Naturally, the conjugate twist field is
\beq
	\b{\mathcal T}_g = \mathcal T_{g^{-1}}.
\eeq

This clearly satisfies \eqref{exch2} for the observables obtained by products of translates of $\mathfrak L_0^\sharp$, by the explicit definition and because $\mathfrak L_0^\sharp$ are Hilbert-space generating observables. It also satisfies \eqref{exch}, {\em with the inner automorphism action}:
\beq\label{gtransfo}
	\sigma_g(\mathcal T_{g'}(x)) = \mathcal T_{gg'g^{-1}}(x).
\eeq
Let us check this. Assume that $x<x'$. Then
\beqa
	\mathcal T_g(x)\mathcal T_{g'}(x')o(y)|V\ket
	&=&
	\mathcal T_g(x)\sigma_{g'}^{x'\leq y}(o(y))|V\ket
	=
	(\sigma_g^{x\leq y}\circ\sigma_{g'}^{x'\leq y})(o(y))|V\ket\n
	&=&
	(\sigma_{gg'g^{-1}}^{x'\leq y}\circ\sigma_g^{x\leq y})(o(y))|V\ket
	= \mathcal T_{gg'g^{-1}}(x')\mathcal T_g(x) o(y)|V\ket
\eeqa
where in the second equality I used $G$-invariance of $\mathfrak L_0^\sharp$. In the third equality, I used the fact that $\sigma_g^{\chi(x\leq y)}\circ\sigma_{g'}^{\chi(x'\leq y)}\circ \sigma_g^{-\chi(x\leq y)} = \sigma_{gg'g^{-1}}^{\chi(x'\leq y)}$ holds true in the three cases that may arise if $x<x'$: either $x\leq y,\,x'\leq y$, or $x\leq y,\,x'>y$, or $x>y,\,x'>y$. A similar calculation can be done for $x>x'$, and with more observables $o_1(x_1),\ldots, o_n(x_n)\in\mathfrak L_0^\sharp$, and we see that this is in agreement with \eqref{exch}, \eqref{gtransfo}.

More work is needed to verify that, in this construction, \eqref{exch2} holds for all mutually local observables $\mathfrak L_0$; I will omit this calculation, but I just mention that this follows from mutual locality. Naturally, one may extend this construction to a subspace $\mathcal V$ instead of $\C|V\ket$.

An example of Hilbert-space generating observables $\mathfrak L_0^\sharp$ is that of a quantum model with all local observables generated by the canonical fields $\Psi(x),\,\Psi^\dag(x)$, $[\Psi(x),\Psi^\dag(x')] = \delta(x-x')$, $[\Psi(x),\Psi(x')] = [\Psi^\dag(x),\Psi^\dag(x')]=0$, and with Hilbert space being the span of $\Psi^\dag(x_1)\cdots\Psi^\dag(x_n)|0\ket$. In this case $|v\ket = |0\ket$ is the pseudovacuum, annihilated by $\Psi(x)$, and $\mathfrak L_0^\sharp = \C\Psi^\dag(0)$.

\medskip

Therefore, here we conclude that:
\st{
For a group $G= \{g\}$, a representation as internal symmetries $\sigma_g$ on $\mathfrak L_0$, and Hilbert-space generating observables $\mathfrak L_0^\sharp\subset \mathfrak L_0$ with $g\mathfrak L_0^\sharp = \mathfrak L_0^\sharp\ \forall g\in G$, there is a family of twist fields $\mathcal T_g$ with twist $\sigma_{\mathcal T_g} = \sigma_g$ as defined in \eqref{twistg}, which transform by inner automorphisms of the group, Eq.~\eqref{gtransfo}.
}

\section{Unwinding gauge transformation for the complex relativistic Boson}\label{appunwinding}

I illustrate the technique of Subsec.~\ref{ssectriemann}, concerning the unwinding gauge transformation, with the complex relativistic boson
\beq
	H = \int \dd x\,\Big(\p_x\phi^\dag(x)\p_x\phi(x) + \pi^\dag(x)\pi(x)
	+ V(\phi^\dag(x)\phi(x))\Big),
\eeq
with
\beq
	[\phi(x),\pi^\dag(x')] = \ri \delta(x-x'),\quad
	[\phi(x),\phi(x')] = [\pi(x),\pi(x')] = [\phi(x),\pi(x')] = 0.
\eeq
We can take the fundamental field to be the complex bosonic field itself,
\beq
	\psi(x) = \phi(x),\quad
	\phi(x)|\Psi\ket = \Psi(x)|\Psi\ket,
\eeq
and we have $U(1)$ symmetry group represented as $\sigma_\lambda(\phi(x)) = e^{-\ri \lambda}\phi(x)$, $\sigma_\lambda(\pi(x)) = e^{-\ri \lambda}\pi(x)$, which is an ultra-local symmetry for the path integral representation with 1 by 1 matrix
\beq
	A_\lambda = e^{\ri \lambda}
\eeq
and obtained from the unitary operator $U_\lambda$
\beq
	U_\lambda = e^{\lambda \int\dd x\,(\pi^\dag(x)\phi(x))-\phi^\dag(x)\pi(x)},\quad
	U_\lambda\phi(x)U_\lambda^{-1} = e^{-\ri \lambda}\phi(x),\quad 
	U_\lambda\pi(x)U_\lambda^{-1} = e^{-\ri \lambda}\pi(x).
\eeq
This is also, of course, an ultra-local symmetry, and forms a continuous ultra-local symmetry group, with
\beq
	\sigma_\lambda = e^{\ri \lambda \ad\int q},\quad q(x) =
	\ri(\phi^\dag(x)\pi(x) - \pi^\dag(x)\phi(x)),
\eeq
in the definitions of Subsec.~\ref{ssectultra}, \ref{ssectexp}. The path-integral representation gives
\beq
	S[\Psi] = \int \dd x\dd t\,\Big(\p_t\Psi^\dag(x,t)\p_t\Psi(x,t)
	-
	\p_x\Psi^\dag(x,t)\p_x\Psi(x,t)
	- V(\Psi^\dag(x,t)\Psi(x,t))\Big).
\eeq
Hence we can define the twist fields $\mathcal T_\lambda(x,t)$, associated to the transformation with $A_\lambda = e^{\ri\lambda}$, and here we consider the Riemann-surface formulation \eqref{Tpathriemann}.

Now consider the following smooth function on the Riemann surface $\mathcal R_{x,t}$:
\beq\label{functionf}
	f(z') = \frc1{2\pi\ri}\Big(\log_{\rm princ}^{(\ep)}(x-x' + \ri (t'-t))
	- \log_{\rm cont}(x-x' + \ri (t'-t),n)\Big).
\eeq
Here $\log_{\rm princ}^{(\ep)}(\rm z)$, $\rm z\in\C$, for small but finite $\ep>0$, is a regularisation of the logarithmic function on its principal branch, with a discontinuity on $\Re(\rm z)\in (-\infty,0)$ that is regularised to a smooth function on an $\ep$-neighbourhood $\Re(\rm z)\in (-\infty,0),\,\Im{(\rm z})\in[-\ep,\ep]$; and $\log_{\rm cont}$ is the continuous extension of the logarithmic function, which is a smooth function on the Riemann surface $\mathcal R_{0,0}$. The function $f(z')$ is constant on $\mathcal R_{x,t}$, except for smooth jumps through $\ep$-neighbourhoods of the half-lines $\{(x',t,n):x'>x\}\subset\mathcal R_{x,t}$ for all $n$, where its non-trivial the winding is accumulated.
On the Riemann surface, the jumps through the $\ep$-neighbourhood come from $\log_{\rm princ}^{(\ep)}$:
\beq
	f(x',t+\ep,n+1) - f(x',t-\ep,n) = 1,\quad x'>x.
\eeq
Because it is constant on every sheet away from the half-line, then this implies
\beq
	f(x',t',n+1) - f(x',t',n) = 1,\quad (x',t')\not\in [x,\infty)\times [-\ep,\ep]
\eeq
and clearly, on the principal branch, $f(x',t',0)=0,\,(x',t')\not\in [x,\infty)\times [-\ep,\ep]$.
On each sheet, because of the winding due to $\log_{\rm cont}$, there are sharp jumps,
\beq
	f(x',t+0^+,n) - f(x',t-0^+,n) = -1,\quad x'>x.
\eeq
With an appropriate choice of $\log^{(\ep)}_{\rm princ}(\rm z)$ (not analytic in an $\ep$-neighbourhood of $(-\infty,0]$), it is possible to choose $f(z')\in \R$.

On the right-hand side of \eqref{Tpathriemann}, make the following change of integration variable:
\beq
	\t\Psi(z') = e^{\ri\lambda f(z')}\h\Psi(z'),\quad
	\t\Psi^\dag(z') = e^{-\ri\lambda f(z')}\h\Psi^\dag(z').
\eeq
As this is a linear transformation of the integration variable, the Jacobian does not depend on the variable itself, hence the path integral measure is only affected by an overall normalisation, which is cancelled in the ratio \eqref{Tpathriemann}.

Because of the ``unwinding" due to the function $f$, the resulting field configuration is continuous on every sheet:
\beq
	\t \Psi(x',t+0^+,n) =  e^{\ri\lambda f(x',t+0^+,n)}\Psi_{n}(x',t+0^+)
	= e^{-\ri \lambda}e^{\ri\lambda f(x',t-0^+,n)}e^{\ri\lambda}\Psi_{n}(x',t-0^+) = \t \Psi(x',t-0^+,n).
\eeq
With smooth coordinates $(x',t')$ on the patch $(x,\infty)\times [-\ep,\ep]$ on any branch of the Riemann surface, we have
\beq
	\p_{x'} f(z') = 0,\ \p_{t'} f(z') = \delta^{(\ep)}(t'-t)
\eeq
where $\delta^{(\ep)}(t'-t)$ is a regularisation of the delta-function; while everywhere else, the derivatives vanish. Using this, we obtain
\beqa
	\p_{t'}\h\Psi^\dag(x',t')\p_{t'}\h\Psi(x',t')
	&=&
	\p_{t'}\t\Psi^\dag(x',t')\p_{t'}\t\Psi(x',t')\n
	&&
	+\, \ri\lambda \delta^{(\ep)}(t'-t)\Theta(x'-x)
	\big(\t\Psi^\dag(x',t')\p_{t'}\t\Psi(x',t')
	-
	\p_{t'}\t\Psi^\dag(x',t')\t\Psi(x',t'))\big)\n
	&&
	+\,\mathcal O(\lambda^2)
\eeqa
and therefore
\beq
	\ri S[\h\Psi]
	=
	\ri S[\t\Psi]
	+ \lim_{N\to\infty} \frc1{2N+1}\sum_{n=-N}^N \ri \lambda \int_x^\infty \dd x'\,q[\t\Psi](x',t,n) + \mathcal O(\lambda^2)
\eeq
As the new path-integration variables $\t\Psi(x',t',n)$ are continuous on every sheet, and as every sheet contributes in the same say, we recover the usual, single-sheeted path integral for the now continuous field $\Psi = \t\Psi(\cdot,\cdot,0)$
\beq
	\ri S[\h\Psi]
	=
	\ri S[\Psi] + \ri \lambda \int_x^\infty \dd x'\,q[\Psi](x',t)
	+ \mathcal O(\lambda^2).
\eeq
This is only a leading-order calculation, and we see that there are, indeed, $\mathcal O(\lambda^2)$ correction, because the kinetic term has two derivatives. These come from the transformation of the conjugate momentum $\pi(x,t)$, represented as $\p_t \Psi(x,t)$ in the path integral, under the gauge transformation. However, it can be adapted to show, more generally, the relation \eqref{twistpathexpsmall},
\beq
	\frc{\bra \vac|\mathsf T\big[ \mathcal T_{\lambda+\delta\lambda}(x,t) o_1(x_1,t_1)\cdots\big]|\vac\ket}{\bra \vac| \mathcal T_{\lambda+\delta\lambda}(x,t)|\vac\ket}
	=
	\frc{\bra \vac|\mathsf T\big[ \mathcal T_{\lambda}(x,t) 
	e^{\ri \delta\lambda \int_x^\infty \dd x'\,q(x',t)}
	o_1(x_1,t_1)\cdots \big]|\vac\ket}{\bra \vac| \mathcal T_\lambda(x,t)
	e^{\ri \delta\lambda \int_x^\infty \dd x'\,q(x',t)}|\vac\ket}
\eeq
to leading order in $\delta\lambda$. Because $q(x',t)$ is not invariant under the gauge transformation, it is crucial to move $t\to t+0^+$ in order to continue the argument, and we indeed get the stack \eqref{twistpathexp}, and not $e^{\ri \lambda \int_x^\infty q(x',t)}$.

\end{document}